\documentclass[fleqn,usenatbib]{mnras}
\usepackage{comment}
\usepackage[T1]{fontenc}
\usepackage{ae,aecompl}
\usepackage{graphicx}	
\usepackage{amsmath}	
\usepackage{amssymb}	
\usepackage{multicol}
\usepackage[normalem]{ulem}

\title[Degeneracies between SIDM and SN feedback]{Degeneracies Between Self-interacting Dark Matter and Supernova Feedback as cusp-core transformation mechanisms}

\graphicspath{{./}{figures/}}

\author[Burger et. al.]{Jan D. Burger$^{1,2}$\thanks{E-mail: jburger@ucr.edu (JB)},
Jes\'us Zavala$^2$,
Laura V. Sales$^1$,
Mark Vogelsberger$^3$, 
Federico Marinacci$^4$,\newauthor
and Paul Torrey$^5$
\\ \\
$^1$University of California Riverside, Riverside, CA, US\\
$^2$University of Iceland, Dunhagi 5, 107 Reykjavík, Iceland\\
$^3$MIT Kavli Institute for Astrophysics and Space Research, Ronald McNair Building, 37-611\\
$^4$Department of Physics and Astronomy ``Augusto Righi'', University of Bologna, via Gobetti 93/2, I-40129 Bologna, Italy\\
$^5$Department of Astronomy, University of Florida, 211 Bryant Space Sciences Center, Gainesville, FL 32611, USA
}

\date{Accepted XXX. Received YYY; in original form ZZZ}

\pubyear{2022}

\begin{document}
\label{firstpage}
\pagerange{\pageref{firstpage}--\pageref{lastpage}}
\maketitle

\begin{abstract}
We present a suite of 16 high-resolution hydrodynamic simulations of an isolated dwarf galaxy (gaseous and stellar disk plus a stellar bulge) within an initially cuspy dark matter (DM) halo, including  self-interactions between the DM particles (SIDM); as well as stochastic star formation and subsequent supernova feedback (SNF), implemented using the stellar feedback model \texttt{SMUGGLE}. 
The simulations start from identical initial conditions and we regulate the strength of SIDM and SNF by systematically varying the SIDM momentum transfer cross section and the gas density threshold for star formation. 
The DM halo forms a constant density core of similar size and shape for several combinations of those two parameters. Haloes with cores that are formed due to SIDM (adiabatic cusp-core transformation) have velocity dispersion profiles which are closer to isothermal than those of haloes with cores that are formed due to SNF in simulations with bursty star formation (impulsive cusp-core transformation). Impulsive SNF can generate positive stellar age gradients and increase random motion in the gas at the centre of the galaxy. Simulated galaxies in haloes with cores that were formed adiabatically are spatially more extended, with stellar metallicity gradients that are shallower (at late times) than those of galaxies in other simulations. 
Such observable properties of the gas and the stars, which indicate either an adiabatic or an impulsive evolution of the gravitational potential, may be used to determine whether observed cores in DM haloes are formed through self-interactions between the DM particles or in response to impulsive SNF.

\end{abstract}


\section{Introduction}

Precision measurements of the cosmic microwave background (\citealt{2020A&A...641A...6P}) reveal that the matter distribution in the early Universe was almost completely homogeneous, perturbed only by small density fluctuations. The $\Lambda$CDM concordance model, in which $\sim 80$ per cent of the matter content in the Universe consists of collisionless, cold dark matter (CDM), successfully explains the growth of these small fluctuations into the large scale structure we observe today (\citealt{2006Natur.440.1137S}). Collisionless N-body simulations predict the hierarchical collapse of overdensities into sheets, filaments, and eventually self-gravitating virialized dark matter (DM) haloes. Galaxies, consisting of ordinary baryonic matter, are hosted by such DM haloes. 
Under the assumption that the brightest observed galaxies are hosted by the most massive DM haloes, the clustering and the abundance of observed galaxies are well explained by the spatial distribution of DM haloes in large cosmological simulations (\citealt{1988ApJ...327..507F}, \citealt{2004ApJ...608..663K}, \citealt{2006ApJ...647..201C}, \citealt{2010ApJ...710..903M}, \citealt{2013ApJ...770...57B}).
Moreover, the observed rotation curves of large spiral galaxies are well explained by the combined mass of visible matter and DM (see e.g. \citealt{1985ApJ...295..305V}). 

On the scale of dwarf galaxies, however, the situation is far more uncertain. The dynamical properties of some observed dwarf galaxies appear to be inconsistent with predictions from collisionless $N$-body simulations in regards to the abundance and the inner structure of low-mass CDM haloes. These mismatches between simulations and theory are longstanding issues that have become known as the small-scale challenges to $\Lambda$CDM \citep[see][for a review]{Bullock2017}. To date, it remains unclear whether these challenges are a manifestation of known but uncertain non-gravitational baryonic physics, which is not present 
in DM only N-body simulations, or whether a modification of the $\Lambda$CDM concordance cosmogony is needed to tackle them. 

One of these challenges is the so-called cusp-core problem. Cosmological CDM N-body simulations predict that the spherically-averaged density profiles of DM haloes can be uniquely described by a single two-parameter fitting function, the so-called Navarro-Frenk-White (NFW) profile \citep{1996ApJ...462..563N,1997ApJ...490..493N}. This universality has been demonstrated over 20 orders of magnitude in halo mass \citep{2020Natur.585...39W}. Importantly, the spherically-averaged density of NFW haloes rises inversely proportional with radius close to the halo's centre; NFW haloes are cuspy.
However, the observed rotation curves of some dwarf Irregulars and Low Surface Brightness galaxies in the field \citep[e.g.,][]{Moore1994,deBlok2008,Kuzio2008,2019MNRAS.484.1401R}, and at least two Milky Way dwarf spheroidals (Fornax and Sculptor) \citep{Walker2011}, are seemingly inconsistent with the assumption that these galaxies are hosted by cuspy DM haloes. Instead, the slow-rising nature of their rotation curves suggests that these galaxies 
may be hosted by DM haloes with extended central cores of constant density. A potentially related issue is that some observations suggest that the mass enclosed within the central kiloparsec of 
dwarf galaxies 
may be overpredicted by collisionless CDM $N$-body simulations (\citealt{2002ApJ...572...34A}, \citealt{2015MNRAS.452.3650O}). Currently, there is ongoing debate about whether measurements of the HI-rotation curves of field dwarfs are interpreted correctly (e.g. \citealt{2019MNRAS.482..821O}, \citealt{2020MNRAS.495...58S}), and whether strong deviations of spherical symmetry in the dwarf spheroidals Fornax and Sculptor may have wrongfully led to the conclusion that the inner density profiles of their host haloes are cored (\citealt{2018MNRAS.474.1398G}). However, it is clear that if observations of slow-rising rotation curves in dwarf galaxies stand the test of time, a non-gravitational physical mechanism that transforms central density cusps into cores is needed to reconcile them with the success of $\Lambda$CDM on larger scales. 
Several such mechanisms of cusp-core transformation have been proposed and while some of them invoke baryonic physics to flatten out the central density profile of dwarf-size haloes, others require abandoning $\Lambda$CDM for a different cosmogony that resembles $\Lambda$CDM on large scales.   

Among the mechanisms of cusp-core transformation that work within $\Lambda$CDM, the most viable one is core formation induced by supernova (SN) feedback \citep{Navarro1996, Gnedin2002, 2005MNRAS.356..107R, 2008Sci...319..174M, Pontzen:2011ty, GarrisonKimmel:2013aq, DiCintio2014, Tollet2016, Chan2015, 2017MNRAS.471.3547F, 2020MNRAS.497.2393L, 2021arXiv210301231B}. Repeated energy injection from supernovae in the dwarf galaxy can give rise to galactic-scale gas outflows, causing rapid fluctuations of the enclosed baryonic mass, and hence of  the total gravitational potential within the inner DM halo. As shown in detail by \cite{Pontzen:2011ty}, repeated impulsive changes in the gravitational potential cause a net radial expansion of the orbits of particles that move within. 
In the case of core formation induced by SNF, this means that the strongly fluctuating gravitational potential causes a radial expansion of the orbits of individual DM particles in the halo centre, thus flattening the central density profile. 

To be a feasible mechanism of cusp-core transformation, SNF needs to fulfill a number of conditions. First and foremost, the total energy that is released by supernovae has to be sufficient to unbind the DM halo's central cusp \citep{Penarrubia:2012bb}. 
A secondary condition is that SNF needs to be impulsive, i.e., SN-driven gas outflows need to give rise to sizeable changes of the gravitational potential on timescales which are shorter than the typical dynamical times of DM particles in the inner halo \citep{Pontzen:2011ty, 2021arXiv210301231B}. From the observational side, there is evidence that starbursts in bright dwarfs, and thus, their associated supernova cycles, happen on timescales that are comparable to the typical dynamical times of those galaxies \citep{Kauffmann2014}. However, observations still lack the time resolution required to resolve starburst cycles on the smaller 
dynamical timescales of the low-mass MW dwarf spheroidals \citep{Weisz2014}. 
In general, 
the more energy is injected during a SNF cycle, the shorter the time is over which that energy is injected, and the more concentrated the baryonic mass is to the centre of the DM halo (\citealt{2021arXiv210301231B}), the more efficient the SNF-induced cusp-core transformation will be. In hydrodynamic simulations of galaxy formation, the implementations of SNF are 
calibrated to the resulting structural properties of larger galaxies. 
Recent studies suggest that, in cosmological simulations, the efficiency of SNF at flattening the cusps of dwarf-size DM haloes is 
mainly determined by one model parameter, the gas density threshold for star formation (\citealt{2019MNRAS.488.2387B}, \citealt{2020MNRAS.499.2648D}). In a given dwarf galaxy, larger star formation thresholds lead to more bursty star formation, more concentrated and impulsive feedback, and a stronger contribution of baryons to the central potential, and hence to enhanced core formation \citep{2019MNRAS.486.4790B,2019MNRAS.488.2387B}. 

Among the most viable mechanisms of cusp-core transformation that require changes to the assumed cosmogony is one that was proposed specifically as a possible solution to the cusp-core problem. It proposes that the DM is in fact not collisionless but self-interacting 
(SIDM, \citealt{Spergel2000}, \citealt{Yoshida2000}, \citealt{Dave2001}, \citealt{Colin2002}, \citealt{Vogelsberger2012}, \citealt{Rocha2013}, see \citealt{Tulin2018} for a review). In SIDM, particles can exchange energy and momentum through elastic scattering,
causing an outside-in energy redistribution within the centre of DM haloes, resulting in the formation of an isothermal core. 
The timescale on which an initially cuspy SIDM halo forms a flat and isothermal core is roughly given by the time it takes for each DM particle in the inner halo to scatter at least once \citep{Vogelsberger2012,Rocha2013}. 
The strength of the self-interaction in SIDM models is 
parametrized in terms of the momentum transfer cross section per unit mass, $\sigma_T/m_\chi$. Depending on the specific SIDM model, $\sigma_T/m_\chi$ can either be constant or dependent on the relative velocity between the two scattering DM particles. SIDM is an efficient mechanism of cusp-core transformation in dwarf-size haloes for $\sigma_T/m_\chi \gtrsim 1\,{\rm cm^2g^{-1}}$, whereas SIDM haloes are virtually indistinguishable from CDM haloes if $\sigma_T/m_\chi \lesssim 0.1\,{\rm cm^2g^{-1}}$ (\citealt{Zavala2013}). The most stringent and precise constraints on the self-interaction cross section have been put on the scales of galaxy clusters (e.g. \citealt{Robertson2017}, \citealt{Robertson2018}) and large elliptical galaxies (\citealt{Peter2013}), where observations require that $\sigma_T/m_chi \lesssim 1\,{\rm cm^2g^{-1}}$. On smaller scales, 
\cite{Read2018} concluded that $\sigma_T/m_\chi \lesssim 0.6\,{\rm cm^2g^{-1}}$, based on their findings that the central density profile of the MW dwarf spheroidal galaxy Draco is cuspy (see also the SIDM results of \citealt{2018NatAs...2..907V}). Moreover, based on a DM only analysis of the updated too-big-to-fail problem, \cite{2019PhRvD.100f3007Z} concluded that SIDM models with a constant cross section of $\sigma_T/m_\chi \sim 1\,{\rm cm^2g^{-1}}$ fail to explain the apparently large central densities of the host haloes of the ultra-faint satellites of the MW \citep{Errani2018}. 
It should be pointed out that the constraints on $\sigma_T/m_\chi$ on the scale of dwarf galaxies are affected by significantly larger systematic uncertainties than on the scales of galaxy clusters or elliptical galaxies. Moreover, \cite{2019PhRvD.100f3007Z} demonstrate that SIDM with a strongly velocity-dependent self-interaction cross section may provide a natural explanation for the observed diversity in the rotation curves of the MW dwarf spheroidals (see also \citealt{Correa_2021}). The strong dependence of the self-interaction cross section on the typical DM velocities would create a bimodal distribution of rotation curves in the MW satellites in which the heavier haloes have constant density cores while the lighter haloes have undergone gravothermal collapse and have very steep central cusps as a consequence . The same mechanism of gravothermal collapse might be accelerated by tidal interactions in the environment of the MW leading to an agreement between constant cross section SIDM models with $\sigma_T/m_\chi \sim 3\,{\rm cm^2g^{-1}}$ and the internal kinematics of MW satellites (e.g. \citealt{Kahlhoefer_2019,Sameie_2020}).  

\cite{2019MNRAS.485.1008B} have shown that while both SNF and SIDM can transform cusps into cores in dwarf-size haloes, the two mechanisms leave distinct signatures in the dynamical properties of kinematic tracers. This difference is related to the different timescales on which SNF and SIDM affect the gravitational potential. While SNF is a viable mechanism for cusp-core transformation only if it causes strong and impulsive fluctuations in the central potential, SIDM thermalizes the central region of DM haloes on timescales that are comparable to or larger than the typical dynamical timescales at distances of $\sim 1{\rm kpc}$ from the centre of dwarf galaxies. 
In other words, SIDM haloes form cores adiabatically, while SNF forms them impulsively. Stars, which approximately act as tracers of the gravitational potential, respond differently to impulsively changing potentials than they do to adiabatically changing potentials. In particular, while the 
actions of tracers on regular orbits are conserved in adiabatically evolving potentials (e.g. \citealt{2008gady.book.....B}), this is not the case in impulsively evolving potentials. Moreover, the orbits of tracers in adiabatically changing potentials quickly adapt to the evolution of the potential, while an ensemble of tracer particles can be put out of dynamical equilibrium in impulsively changing potentials. Hence, the dynamical properties of the stars may differ considerably between {\it i}) dwarf galaxies with cuspy haloes, {\it ii}) cored haloes with an adiabatic core formation history, and {\it iii}) cored haloes with an impulsive core formation history. 

In this article, we aim to identify such differences using 16 high-resolution hydrodynamical simulations of an isolated dwarf galaxy with global parameters resembling 
the Small Magellanic Cloud (SMC) embedded within a live halo (similar to \citealt{2012MNRAS.421.3488H}). Starting from idealized initial conditions, we simulate the evolution of the system over roughly half a Hubble time using the moving-mesh code \texttt{AREPO}  (\citealt{Springel:2009aa}) with the interstellar medium (ISM) and stellar evolution model "Stars and MUltiphase Gas in GaLaxiEs" (\texttt{SMUGGLE}) introduced in (\citealt{2019MNRAS.489.4233M}) and the Monte-Carlo code for self-interactions between DM particles described in \cite{Vogelsberger2012}. Core formation within colissionless dark matter haloes in \texttt{SMUGGLE} is investigated in detail in \citet{2021arXiv211000142J}.

All simulations start from identical initial conditions and are carried out using different combinations of the momentum transfer cross section per unit mass $\sigma_T/m_\chi$ and the gas density threshold for star formation $n_{\rm th}$. Within the context of our idealized setup, we identify for which combinations of those two parameters cause the DM halo hosting the SMC-analogue to form a constant-density core, and for which parameter combinations the halo retains its initial central cusp. To determine how we can differentiate between SIDM and SNF as core formation mechanisms, we then look for observable quantities that are characteristically different between simulations in which the DM halo forms a core of similar size.
In other words, we look for ways in which we can break the degeneracy between SNF and SIDM as cusp-core transformation mechanisms. To that end, we compare three observable quantities, which are derived from the dynamical properties of either the stars or the gas: 
i) the spatial extent of the visible galaxy, ii) the amount of random motion in the line-of-sight gas velocity, and iii) the age and metallicity gradients of the stars formed throughout the simulation. 

This article is structured as follows. We describe the simulations and initial conditions in Section \ref{sec:simulations}, present our results in Section \ref{sec:results}, and summarize our findings in Section \ref{sec:sum}. In Appendix \ref{sec:discussion} we discuss several caveats that arise because of the stochastic nature of star formation. In Appendix \ref{app_addfigs} we demonstrate that our main conclusions are independent of how we analyze our results. Appendix \ref{appc} explores how the effectiveness of both SNF and SIDM depends on the ratio of DM to baryons in the centre of the dwarf.

\section{Simulations}\label{sec:simulations}
Our goal is to investigate 
how we can differentiate between cores (in a dwarf-size DM halo) that have been formed adiabatically and cores that have been formed impulsively. To that end, we perform a suite of 16 different hydrodynamical simulations starting from 
the same idealized system. In this suite of 16 simulations, we investigate the impact of different star formation histories -- resulting from different choices for the model parameters of our ISM and feedback model -- and SIDM with different self-interaction cross sections. 
Specifically, we use the ISM and stellar feedback model \texttt{SMUGGLE} \citep{2019MNRAS.489.4233M}
with four different values of the gas density threshold for star formation and the \citet{Vogelsberger2012} SIDM model with four different constant self-interaction cross sections. 

In this Section, we briefly outline how the initial conditions of our simulations are generated and how an orbital family of kinematic tracers is included into the initial conditions. Then, we will discuss the ISM model used in our simulations, as well as the algorithm employed to model DM self-scattering. 

\subsection{Initial conditions}\label{subsec:ICs}

We set up an isolated DM halo in dynamical equilibrium containing a baryonic galaxy consisting of a stellar disk, a gaseous disk and a stellar bulge. The structural parameters of our initial conditions are similar to the SMC-like galaxy presented in table 1 of \citet{2012MNRAS.421.3488H}. 

The DM halo is modeled as a Hernquist sphere whose structural parameters are defined by its circular velocity $v_{200}$ at the virial radius $r_{200}$ and its concentration parameter $c_{200}$\footnote{$r_{200}$ is defined through the equation $M_{200} = 200\times 4\pi/3\rho_{\rm crit}r_{200}^3$, where $M_{200}$ is the halo's virial mass, $\rho_{\rm crit}$ is the critical density of the Universe, and $c_{200} = r_{200}/r_{-2}$ is the halo concentration, with $r_{-2}$ being the radius at which the logarithmic slope of the halo's density profile equals $-2$.}. Here, we use $v_{200} = 36.3\,{\rm km\,s^{-1}}$ and $c_{200} = 18$. Assuming $h = 0.7$, this implies a virial mass of $M_{200} \approx 1.6\times 10^{10}M_\odot$ and a virial radius $r_{200} \approx 51.8 \,{\rm kpc}$. 

\begin{figure}
    \centering
    \includegraphics[trim={0.5cm 0.5cm 0.5cm 0.5cm},clip=true,width=\linewidth]{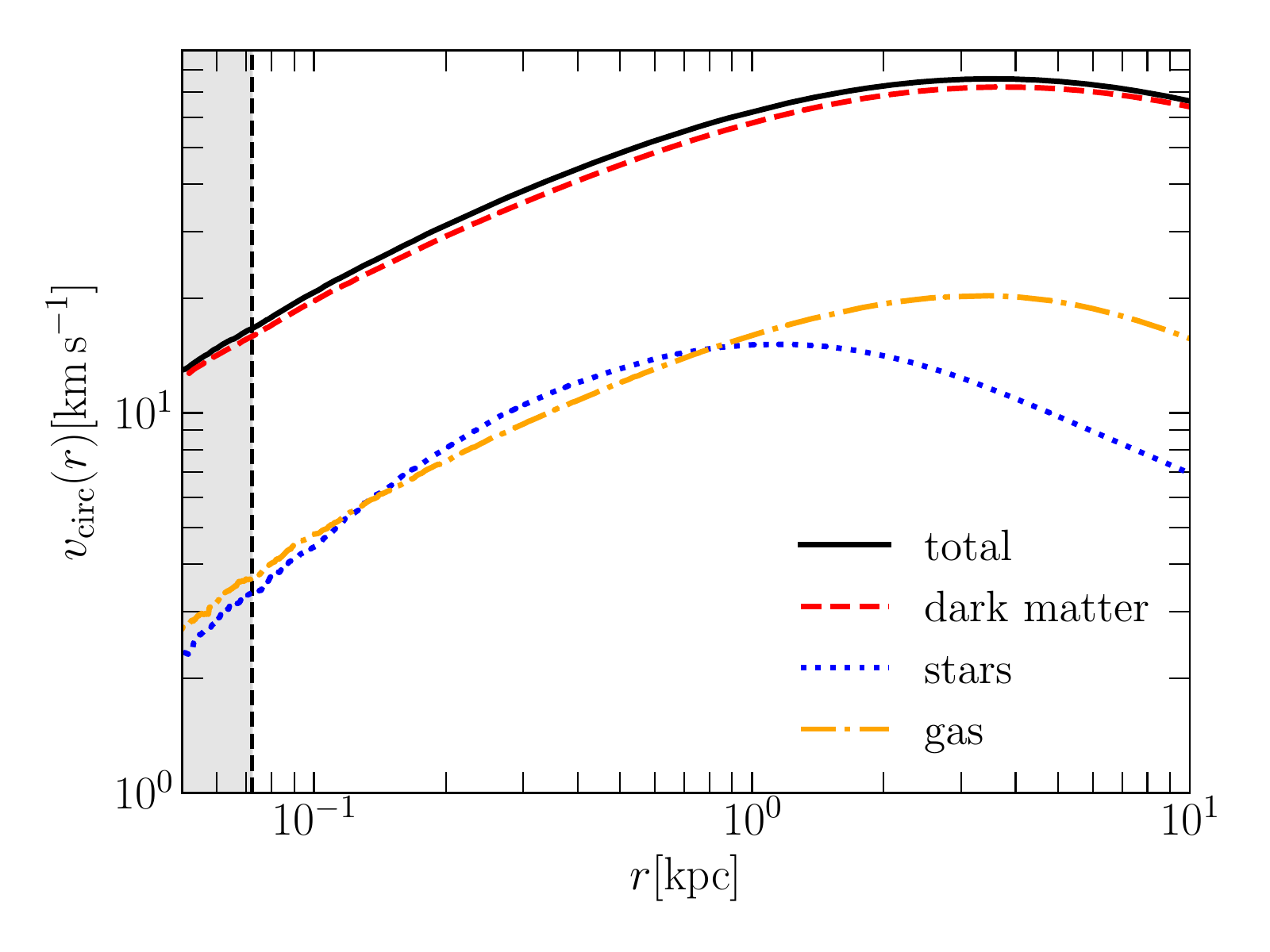}
    \caption{Visualization of the relative mass distribution of our initial conditions. Shown are {\it inferred} circular velocity curves of the dark matter, gas, and stellar (disc and bulge) components. Dark matter dominates the mass distribution in the centre of the galaxy. The grey area denotes the resolution limit of our simulations.}
    \label{fig:icrot}
\end{figure}

The baryonic components are defined by their mass fractions, relative to the DM halo mass, and by their density profiles. 
Both the stellar and gaseous disks combined have a mass of $0.0445\times M_{200}$ ($84\%$ of the disk mass is gas while the rest consists of collisionless star particles). The stellar bulge consists fully of collisionless star particles and makes up for a fraction of $0.005$ of the total mass $M_{200}$. The gas disk has an exponential surface density profile,
\begin{equation}
    \Sigma(R)\propto\exp\left(-\frac{R}{H_{\rm gas}}\right),
\end{equation}
with a scale length $H_{\rm gas} = 2.1\,{\rm kpc}$, where $R$ is the cylindrical radius. As we are interested in a late time dwarf galaxy, we consider a fully ionized gas composition. The gas is initially isothermal with a temperature of $10^4\,{\rm K}$, and has solar metallicity. 
The vertical structure of the gaseous disk is initialized such that the gas is in hydrostatic equilibrium (see \citealt{1993ApJS...86..389H} and \citealt{2005MNRAS.361..776S}). 

The stellar disk also has an exponential surface density profile, but with a smaller scale length of $H_\star = 0.7\,{\rm kpc}$, and a vertical distribution given by: 
\begin{equation}
    \rho(R,z) \propto \Sigma(R)\left[\cosh\left(\frac{z}{z_0}\right)\right]^{-2},
\end{equation}
with a scale height $z_0 = 0.14\,{\rm kpc}$. 
The bulge is modelled as a Hernquist sphere with a scale length $A = 0.233\,{\rm kpc}$.

Due to the spherical symmetry of halo and bulge, we can make use of Eddington's equation \citep{Eddington1916} to calculate the full distribution functions of both the halo and bulge particles. The velocities of different particles are subsequently sampled directly from the distribution function. However, due to the presence of the baryonic disk, the total gravitational potential deviates from spherical symmetry. For definiteness,  we calculate the distribution function by performing Eddington's integral along the 
direction perpendicular to the plane of the disk and note that this procedure introduces a small degree of inaccuracy.

For the stellar disk, we calculate the velocity dispersion tensor on a logarithmic grid of $R,z$ values using the Jeans equation in cylindrical coordinates and the streaming velocity from the enclosed mass profile using the epicyclic approximation \citep[see][for details]{2005MNRAS.361..776S}. The velocities of individual 
disk particles are then comprised of the streaming velocity and an added random component which is calculated using a local Maxwellian velocity distribution based on the calculated velocity dispersion tensor. The velocities of individual gas cells in the gaseous disk are set to the gas' streaming velocities (calculated taking into account both gravity and the gas pressure gradient) at the position of the respective cell. 

We initially set up $1.2 \times 10^7$ DM particles, $4\times 10^5$ gas cells, $8\times 10^4$ collisionless disk particles and $8\times 10^3$ bulge particles. The mass of each particle is then approximately $1.3\times 10^3M_\odot$. The gravitational softening length is $\epsilon_g = 24\,{\rm pc}$ for all particle species. 

Since the algorithm we use to set up our initial conditions relies on 
assuming spherical symmetry to calculate the distribution functions of the bulge and the halo, the resulting distribution of particles is not fully in dynamical equilibrium due to the presence of the axisymmetric disk component. To remedy this, we evolve the system for a time of $1\,{\rm Gyr}$, solving for the dynamical evolution of the gas and the collisionless particles but disabling cooling processes and deactivating 
star formation and stellar feedback.
 
After letting the system relax for $1\,{\rm Gyr}$, we take the final snapshot as our new initial conditions. The resulting initial conditions are presented in Fig.~\ref{fig:icrot}, where we show the calculated rotation curve of our relaxed system, along with a decomposition into the contributions from the DM, gas component, and stellar components. DM initially dominates the central gravitational potential, with nearly equal minor contributions coming from the stellar and the gaseous disc (note, however, that the gaseous disc is more extended). 



\subsection{The stellar evolution model}\label{subsec:sev}
We use the \texttt{SMUGGLE} stellar feedback and ISM model \citep{2019MNRAS.489.4233M} for the moving mesh code \texttt{AREPO} \citep{Springel:2009aa}. We refer the reader to the original paper for details about the model and its implementation. Here, we briefly review two components that are of key importance in our work, namely the stochastic implementation of star formation and the implementation of SNF. 

The formation of star particles proceeds stochastically and is based on the star formation rate within a given gas cell, which is given by 
\citep{2019MNRAS.489.4233M}:
\begin{equation}
    \dot{M}_\star = \displaystyle\left\{ \begin{array}{ll}
        0\qquad &\rho < \rho_{\rm th} \\
        \displaystyle\epsilon \frac{M_{\rm gas}}{t_{\rm dyn}}\qquad &\rho \ge \rho_{\rm th}
    \end{array}  \right.,\label{sfr}
\end{equation}
where $M_{\rm gas}$ is the gas mass in a given gas cell, $t_{\rm dyn}$ is the dynamical time of the gas cell, and $\epsilon$ is the star formation efficiency parameter, set to a value of $0.01$ in all of our runs. 
It is evident from Eq.~(\ref{sfr}) that star formation can only proceed if the gas density in a given gas cell is larger than the threshold density $\rho_{\rm th}$. This in itself is of key importance, as it implies that changing this parameter can significantly impact the distribution of gas densities throughout the simulation. In particular, increasing the threshold will lead to more concentrated gas and therefore to more concentrated star formation. Apart from the density criterion, gas cells are also required to be gravitationally bound, meaning that they cannot overcome their self-gravity through gas motion and thermal energy. 
If both of these criteria are fulfilled, a gas cell is stochastically converted into star particles with a probability of $p = 1-\exp(-\dot{M}_\star \Delta t/M_i)$, where $M_i$ is the mass in the gas cell i and $\Delta t$ here denotes a simulation time step. This probability is then compared to a random number $x$ in the interval $(0,1)$ drawn from a uniform distribution. The gas cell is converted into a star particle if $p \ge x$. The formed star particles represent stellar populations with a \citet{2001ApJ...554.1274C} initial mass function. 

The implementation of SNF is 
explained in great detail in Section 2.3 of \citet{2019MNRAS.489.4233M}. The algorithm differentiates between type II supernovae and type Ia supernovae. The total momentum injected into the ISM is boosted if the cooling radius, the radius at which the SN remnant transitions from an adiabatic Sedov-Taylor phase to a momentum conserving phase, cannot be resolved in the simulation.  This is the case for most simulations, given that the cooling radius is of the order of a few pc, well below the scales that are resolved in galaxy formation simulations. For a given gas cell, the boost factor depends on the ratio between the gas cell's mass and the total (type II and type Ia) ejecta mass, and the fraction of the solid angle covered by the gas cell as seen from the star particle's position (see Eqs. 31-35 of \citealt{2019MNRAS.489.4233M}). The expected values of ejected mass, energy, and the total number of supernovae are self-consistently calculated at each time-step and for each star particle. Time steps are chosen such that the expected number of supernovae is below one at essentially all times. A discrete number of supernovae is then sampled from a Poisson distribution with the expected number of supernovae as the distribution's mean. Once the number of supernovae, the ejected energy, momentum, mass, and metallicity have been determined, these quantities are 
distributed over a fixed number of nearest neighbour gas cells. Fixing the number of nearest neighbours implies defining a search radius $h$ via 
\begin{equation}
    N_{\rm ngb} = \frac{4\pi}{3}h^3\sum_i W(|\mathbf{r}_i-\mathbf{r}_s|,h),
\end{equation}
where $\mathbf{r}_i$ is the position vector of the {\it i}'th neighbouring gas cell, $\mathbf{r}_s$ is the star particle's position vector, and $W$ is the cubic spline kernel. 
If the radius $h$ determined in this way is larger than $R_{\rm SB}$, the typical radius of a super bubble ($\sim 1\,{\rm kpc}$), then the feedback energy and momentum are distributed amongst cells within $R_{\rm SB}$, while mass and metallicity are distributed amongst the $N_{\rm ngb}$ nearest neighbours within the search radius $h$. If there are $N_{\rm ngb}$ nearest neighbouring gas cells within the super bubble radius, no distinction is made. The SN ejecta are divided amongst cells using weights that are proportional to the solid angle covered by the cells as seen from the stellar particle's position (see Eq.~(35) in \citealt{2019MNRAS.489.4233M}). 

A key parameter of the model is $n_{\rm th}$, the number density threshold for star formation. Together with the average mass per gas cell this determines the density threshold $\rho_{\rm th}$ in Eq.~(\ref{sfr}). Its value is therefore directly related to how clustered the stellar populations that form are, and hence, how clustered SNF is. For a fixed initial DM to baryon ratio\footnote{In cosmological simulations, the ratio of DM to baryons in the central regions of dwarf galaxies varies depending on the galaxys' initial configurations and their dynamical histories. We discuss this point in Appendix \ref{appc}, where we modify the amount of DM in the centre of our simulated galaxy by changing the halo concentration parameter.}, a larger star formation threshold 
leads to more bursty star formation, leading in turn to more energetic and impulsive SNF. We thus expect $n_{\rm th}$ to play a key role in determining whether SNF is effective at forming cores in our simulations. 

\subsection{The SIDM algorithm}\label{subsec:sidm_alg}
To model SIDM, we use the algorithm introduced in \citet{Vogelsberger2012} and described in Section 2.2 therein. In this algorithm, the probability for scatter between two DM particles $i$ and $j$ is given by 
\begin{equation}
    P_{ij} = m_i\frac{\sigma_T}{m_\chi}v_{ij}W(r_{ij},h_i)\Delta t_i, \label{eq:prop}
\end{equation}
where $m_i$ is the {\it i}'th DM simulation particle's mass, $\sigma_T/m_\chi$ is the SIDM momentum transfer cross section per unit mass, $v_{ij}$ is the relative velocity between particles {\it i} and {\it j}, and $\Delta t_i$ is the time step of particle {\it i}. The scattering probability is smoothed by the cubic spline kernel $W$, whose arguments are the distance between particles {\it i} and {\it j} and the smoothing length $h_i$, denoting the radius of a sphere around simulation particle {\it i} which contains a predetermined number of nearest neighbours. The total probability for a scatter is given by a sum of the probabilities calculated according to Eq.~(\ref{eq:prop}) over all the nearest neighbours and multiplied by $1/2$. Whether and with which of the nearest neighbours a scatter occurs in a given timestep is determined stochastically as outlined in \citet{Vogelsberger2012}. To model an elastic scatter, the two colliding particles are assigned new velocities in a way that conserves both total momentum and total energy. 
In a halo with a fixed initial density profile, the total number of scattering events over a given time is regulated by 
$\sigma_T/m_\chi$; if each DM simulation particle in the halo centre takes part in $\sim 1$ scattering event, the inner halo forms an isothermal constant density core. The size of this core and the timescale at which it forms depend on the strength of the interaction, and hence on the SIDM transfer cross section. 

\subsection{Simulation suite parameter space}\label{subsec:simparams} 

The goal of our work is to identify those simulations -- out of the 16 runs in our simulation suite -- in which DM cores of a near identical size form, and to then highlight how the observable kinematic properties of the baryons differ between them. Thus, we aim to break the degeneracy in core size between simulations in which cores are predominately formed by either SNF, or SIDM. 
By changing the parameters of both the star formation and stellar feedback model, 
and the SIDM algorithm, we are able to regulate the relative importance of 
the impulsive and adiabatic processes, respectively, in our simulations.  
As outlined above, the key parameter determining the impact of DM self-interactions is the momentum transfer cross section $\sigma_T/m_\chi$,
while we chose to vary the density threshold for star formation, $n_{\rm th}$, in order to regulate the impulsiveness of SNF. 
Our simulation suite consists of 16 simulations, for each of which 
we adopt a different combination of these two model parameters. We thus cover a four by four grid in parameter space, running one simulation for each combination of $\sigma_T/m_\chi = \{0,0.1,1,10\}$ (in units of ${\rm cm^2g^{-1}}$) and $n_{\rm th} = \{0.1,1,10,100\}$ (in units of $\rm cm^{-3}$). 

The numerical values of the other parameters of the 
star formation and stellar feedback model are given in table 3 of \citet{2019MNRAS.489.4233M}. For the SIDM algorithm, we adopt $N_{\rm ngb} = 32\pm 5$ for the nearest neighbour search. 

\section{Results}\label{sec:results}

In this Section, we present the results of our simulations. We start by showing how the numerical value of the star formation threshold affects the star formation history, both in a CDM and in an SIDM halo. Then, we compare the evolution of the density and velocity dispersion profiles of the DM halo for all 16 combinations of $n_{\rm th}$ and $\sigma_T/m_\chi$. Thereafter, we focus on several dynamical quantities that help to break the degeneracy 
between simulations in which the final density profiles (and in turn the galaxy rotation curves) look nearly identical. Our simulations are run for a total of $4\,{\rm Gyr}$. 
Most results presented in this Section are derived from snapshots taken after $3\,{\rm Gyr}$, except for a few relevant cases in which we present results for $4\,{\rm Gyr}$. We have verified that the differences between simulations with different transfer cross sections and star formation thresholds persist at later times. 

We note that the final structural and dynamical properties of the simulated galaxy-halo system do not depend solely on $n_{\rm th}$ and $\sigma_T/m_\chi$ -- but also on the initial DM to baryon ratio. For the initial conditions used here, our selected combinations of $n_{\rm th}$ and $\sigma_T/m_\chi$ allow us to explore cases in which a core forms due to SNF, SIDM, or not at all -- which is desirable for the purpose of identifying signatures of different core formation mechanisms. For an exploration of the dependence of the galaxy-halo systems final structural and dynamical properties on the initial DM to baryon ratio we refer the reader to Appendices \ref{app_ics} and \ref{appc}.

\subsection{Star formation histories}\label{subsec:sfh}
\begin{figure*}
    \centering
    \includegraphics[width=0.49\linewidth,trim={0.5cm 0.5cm 0.5cm 0.5cm},clip=true]{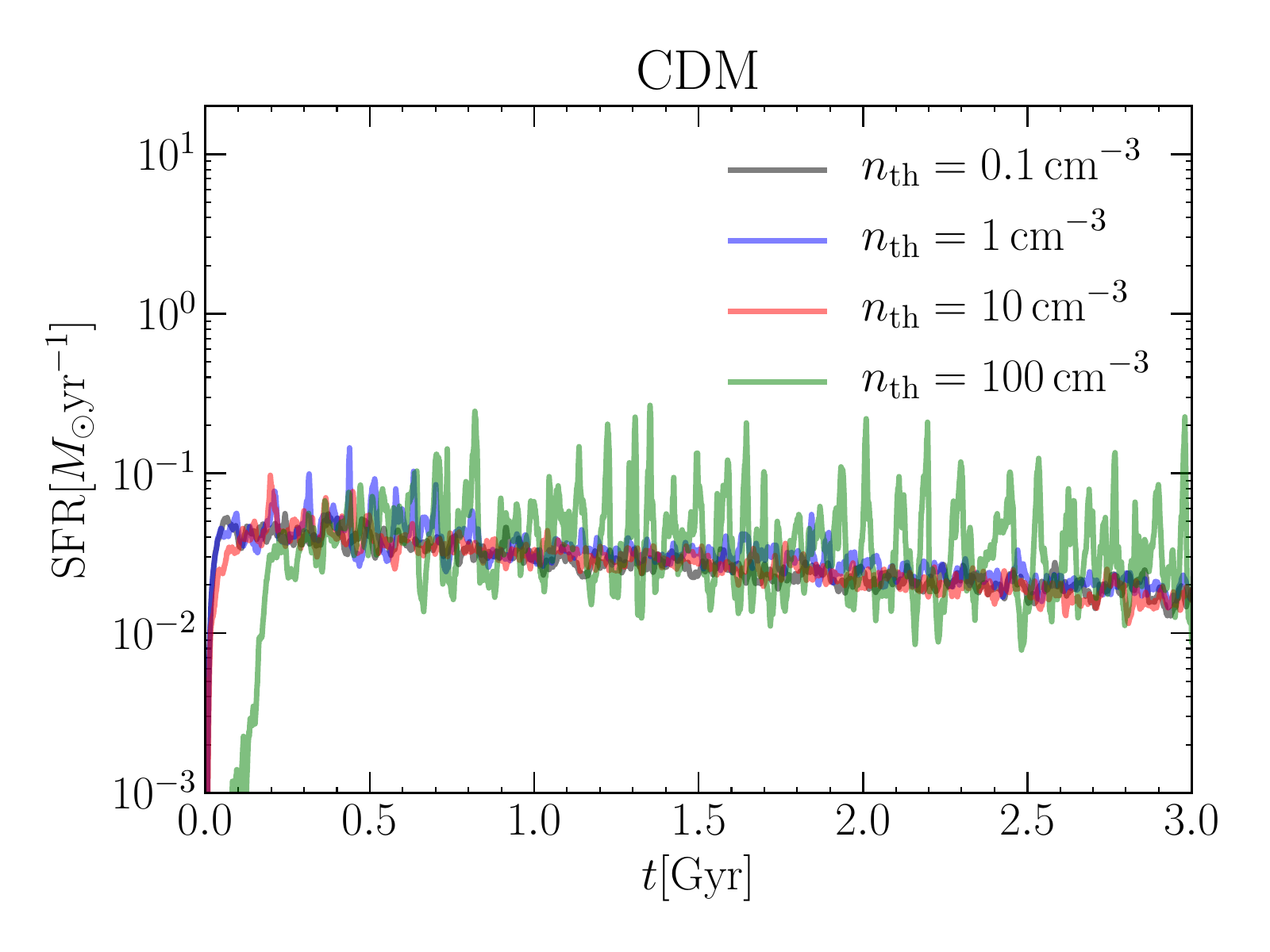}
    \includegraphics[width=0.49\linewidth,trim={0.5cm 0.5cm 0.5cm 0.5cm},clip=true]{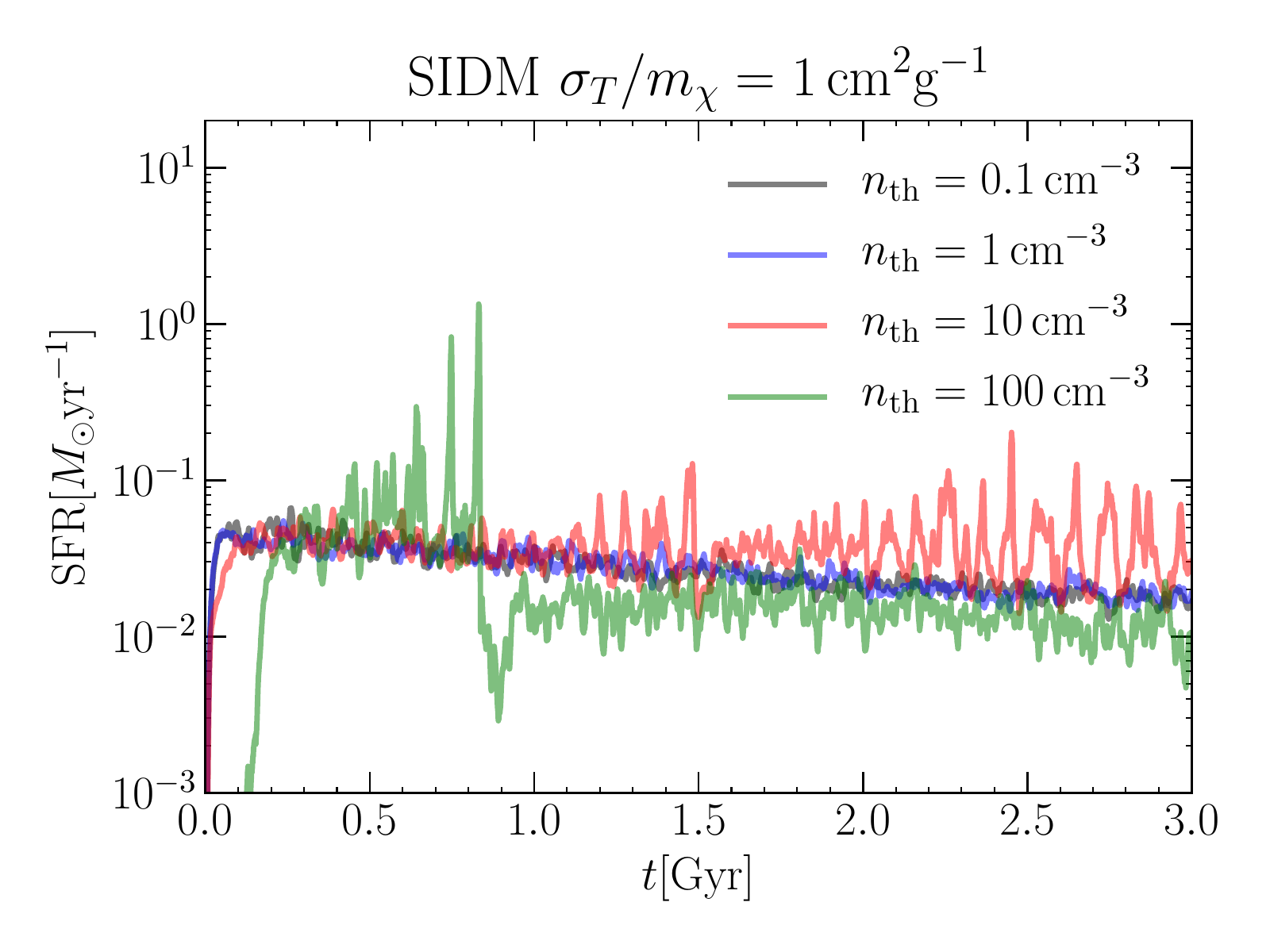}
    \caption{Star formation history for two different DM models over 3 Gyr of simulated time. On the left (right) panel we show the star formation rates in the simulations of the CDM (SIDM, $\sigma_T/m_\chi = 1\,{\rm cm^2g^{-1}}$) halo with four different star formation thresholds as indicated. On average, larger star formation thresholds lead to burstier star formation, independent of the self-interaction cross section. However, star formation can also shut down in simulations with large star formation thresholds if gas is ejected from the galaxy early on (see for instance the green line on the right panel).} 
    \label{fig:sfhs}
\end{figure*}

For our benchmark simulation suite we find that, on average, the burstiness of star formation depends on the numerical value of the star formation threshold $n_{\rm th}$. 
In Fig. \ref{fig:sfhs}, we show the star formation rates measured in eight different simulations over a simulated time of 3 Gyr. The left panel shows the star formation rates of all CDM simulations, while the right panel shows the star formation rates of all SIDM simulations with $\sigma_T/m_\chi = 1\,{\rm cm^2g^{-1}}$. 

On the left panel, we find quasi-periodic bursty star formation cycles only in the CDM simulation with $n_{\rm th} = 100\,{\rm cm^{-3}}$. In all other simulations, star formation decreases monotonously after $\sim 300$ Myr. We can identify a single star burst in the simulation with $n_{\rm th} = 10\,{\rm cm^{-3}}$, after $\sim 250$ Myr. The star formation histories in the two simulations with low star formation thresholds are smooth over the entire simulated time. 

On the right panel, we find bursty star formation in both simulations with larger star formation thresholds, i.e. for $n_{\rm th} = 10\,{\rm cm^{-3}}$ and for $n_{\rm th} = 100\,{\rm cm^{-3}}$. For $n_{\rm th} = 10\,{\rm cm^{-3}}$, bursty episodes of star formation start appearing after $\sim 1$ Gyr. For $n_{\rm th} = 100\,{\rm cm^{-3}}$, on the other hand, we identify massive bursts of star formation only during the first gigayear of simulated time. After a particularly strong burst, the star formation rate drops significantly and does not recover. This drop is directly related to the strong star burst before. The large number of supernovae that occur shortly after this star burst drive a large amount of gas out of the galaxy, effectively shutting off star formation (see Appendix \ref{appendix_3} for a further discussion of this run). In the two simulations with lower star formation thresholds we once again observe a smooth star formation history throughout the simulations, with a steadily decreasing star formation rate as more of the gas is converted into stars. 

Overall, we find that in our default set up, bursty star formation can only occur in our simulations with large star formation thresholds. 
For $n_{\rm th} \lesssim 1\,{\rm cm^{-3}}$, the star formation rate is smooth and monotonously decreases with time. The larger the star formation threshold, the burstier star formation can be. However, star formation also becomes more stochastic in simulations with larger star formation thresholds. In particular, SNF following a massive star burst can result in star formation being completely shut off, due to a large amount of gas being removed from the galaxy in large-scale galactic winds.

We note that a different dependence of the burstiness of star formation on $n_{\rm th}$ arises when adopting a different initial DM to baryon ratio in the centre of the simulated galaxy. In particular, we demonstrate in Appendix \ref{appc} that substantially increasing the initial relative amount of baryons in the galaxy's centre results in bursty star formation for all adopted values of $n_{\rm th}$.
Crucially, the fact that our benchmark simulation suite contains runs with both smooth and bursty star formation is what enables us to compare simulations with impulsive SNF to simulations without impulsive SNF.

\subsection{Density profiles and final galaxy rotation curves}\label{subsec:profiles}

\begin{figure*}
    \centering
    \includegraphics[width=0.48\linewidth,trim={0.5cm 0.5cm 0.5cm 0.5cm},clip=true]{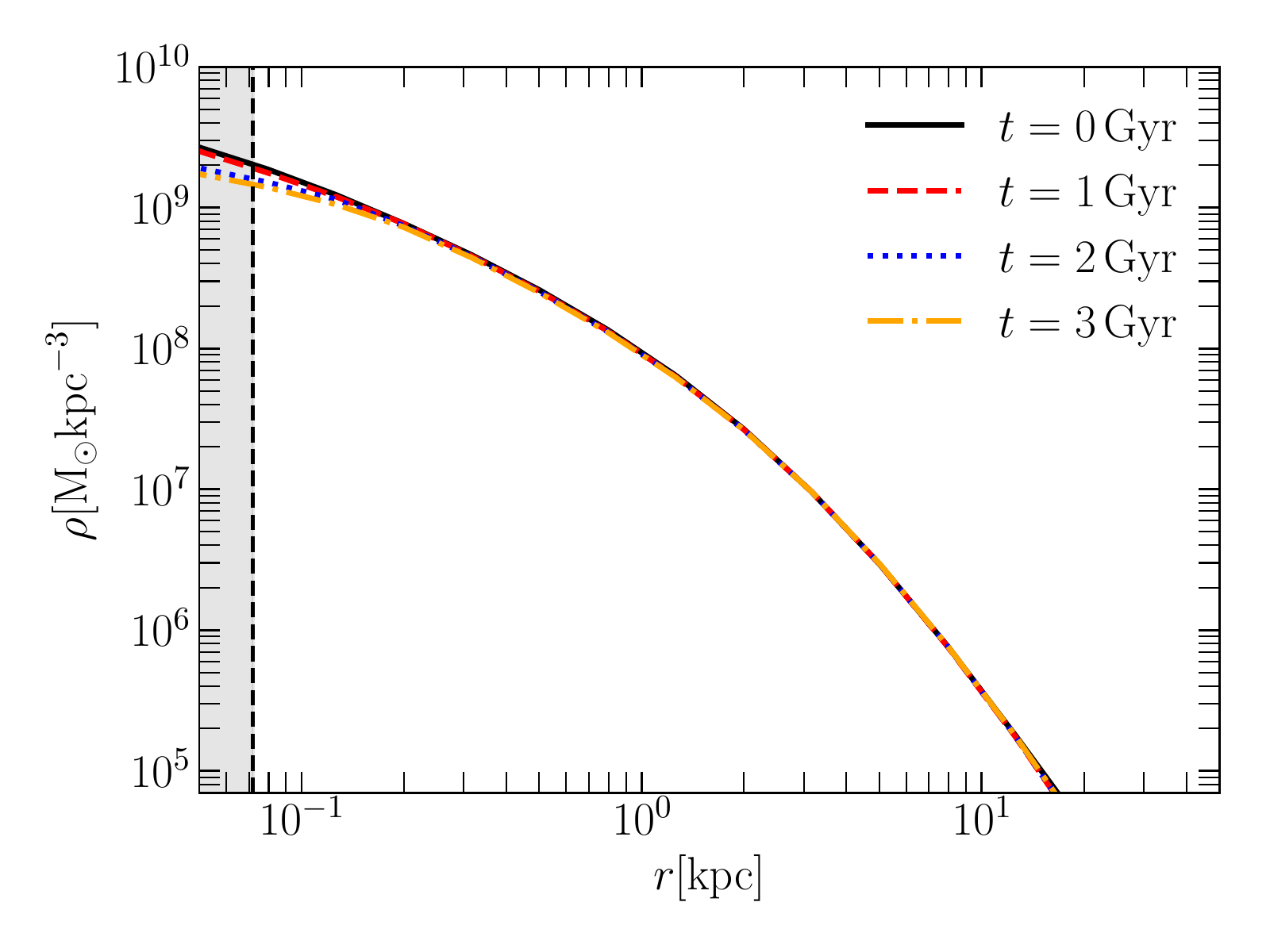}
    \includegraphics[width=0.48\linewidth,trim={0.5cm 0.5cm 0.5cm 0.5cm},clip=true]{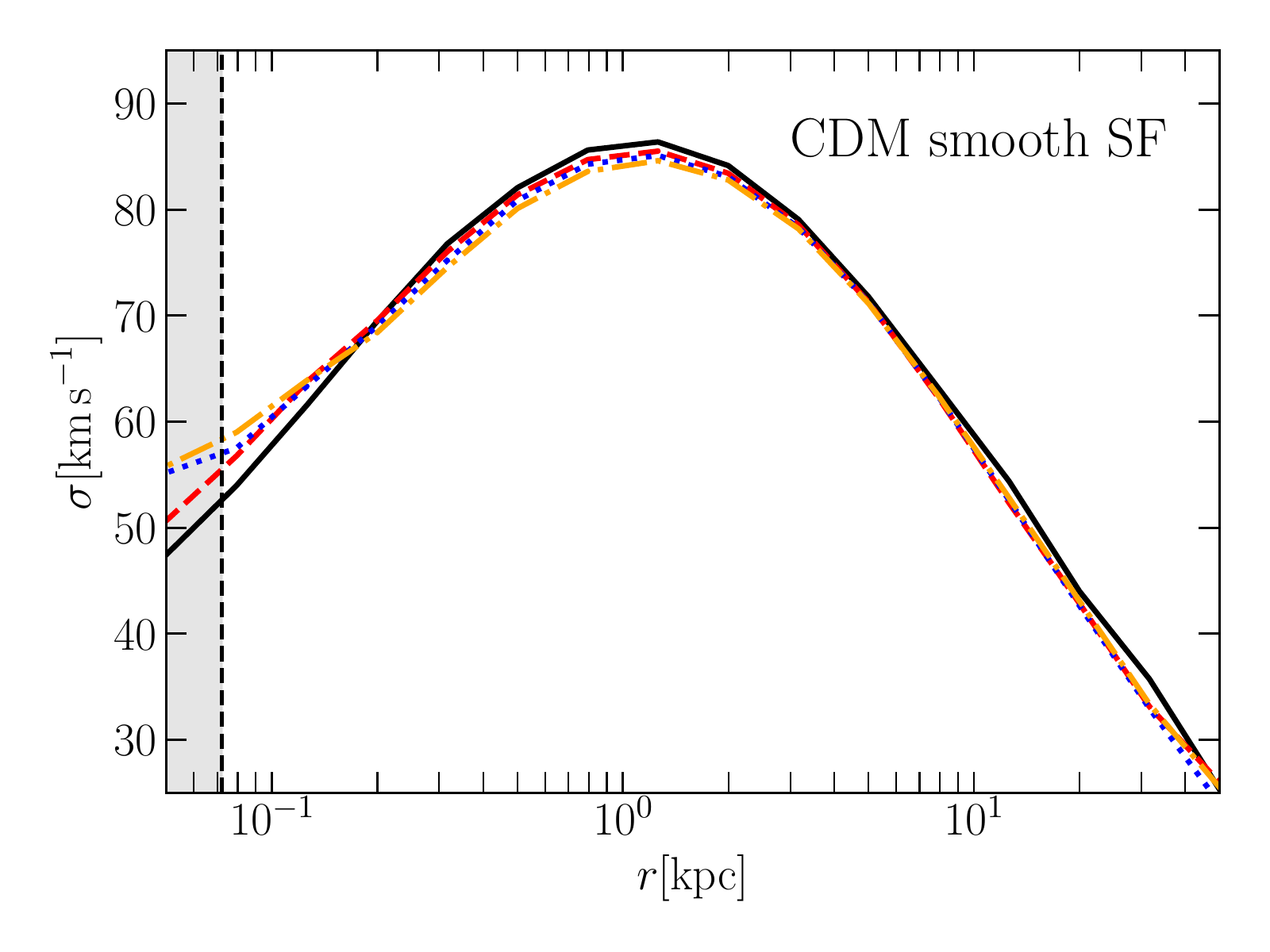}\\
    \includegraphics[width=0.48\linewidth,trim={0.5cm 0.5cm 0.5cm 0.5cm},clip=true]{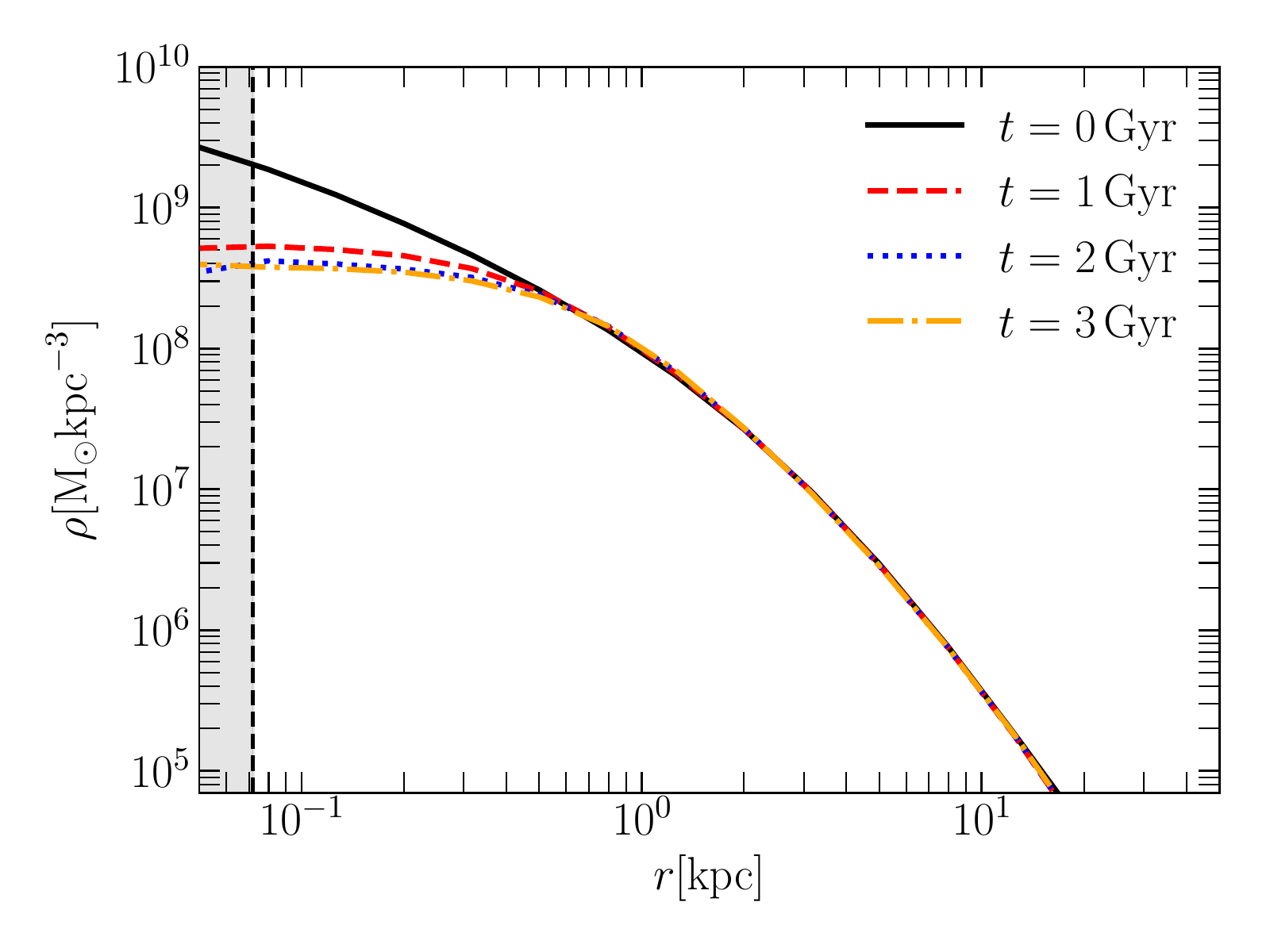}
    \includegraphics[width=0.48\linewidth,trim={0.5cm 0.5cm 0.5cm 0.5cm},clip=true]{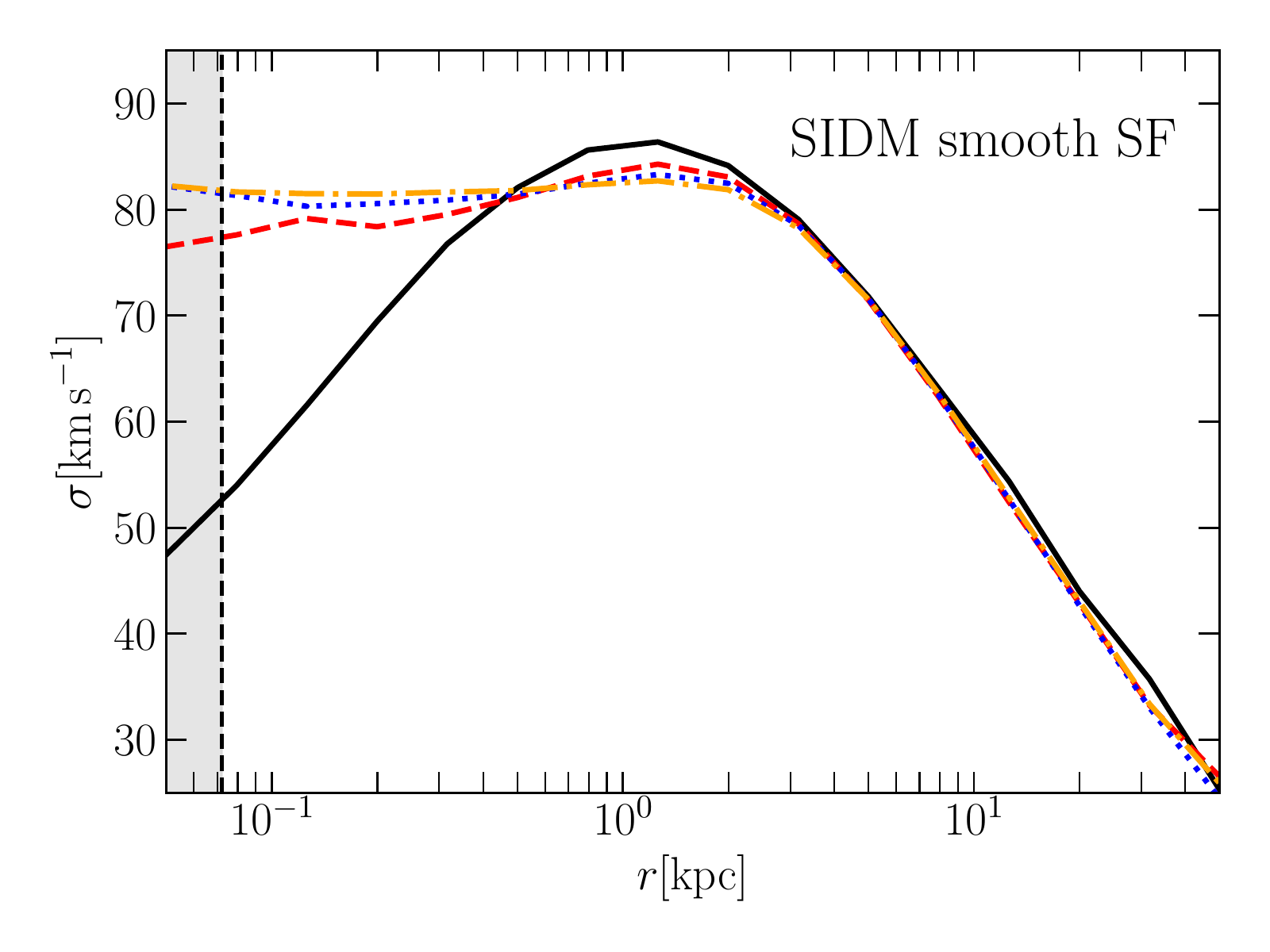}\\
    \includegraphics[width=0.48\linewidth,trim={0.5cm 0.5cm 0.5cm 0.5cm},clip=true]{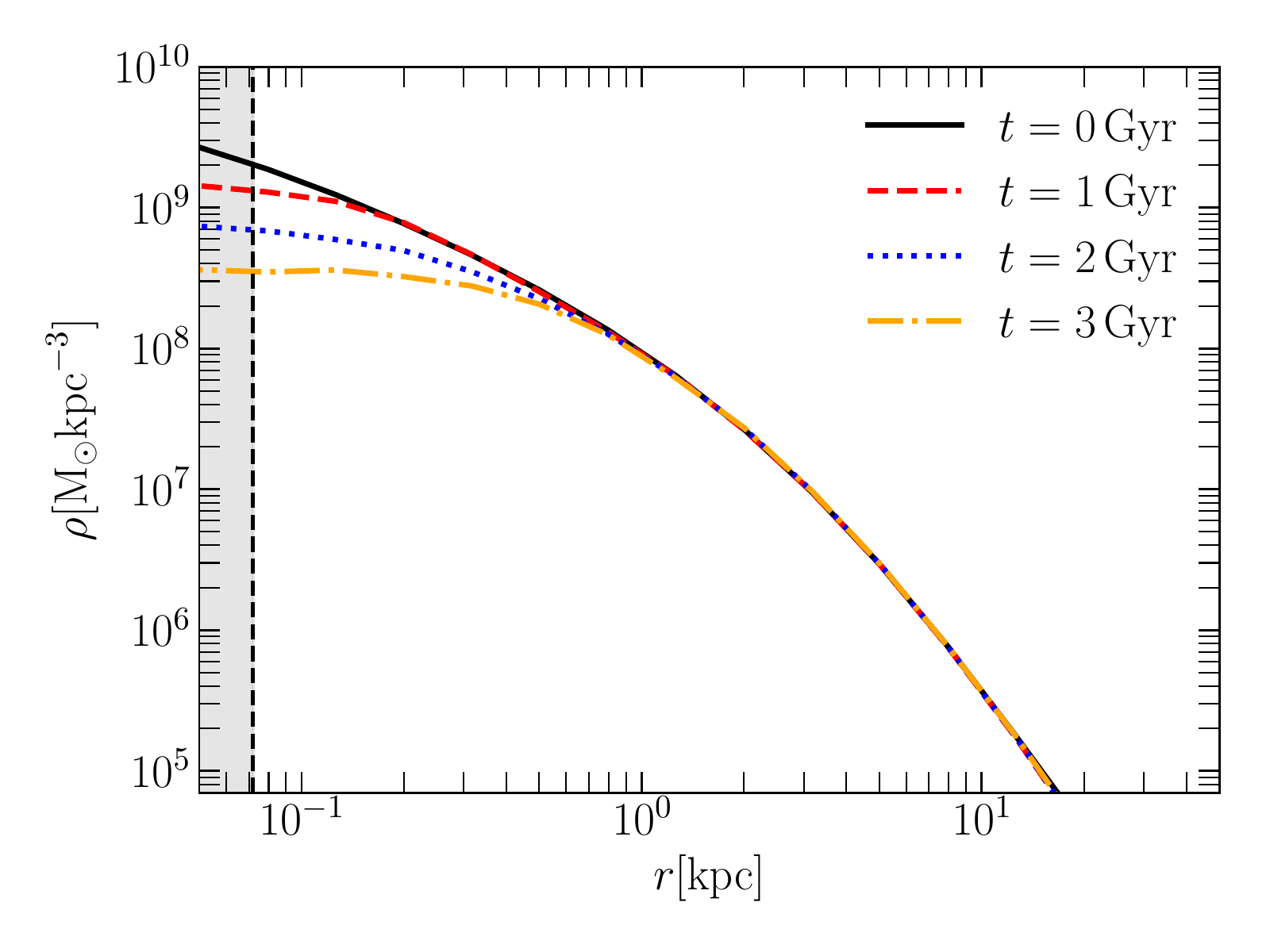}
    \includegraphics[width=0.48\linewidth,trim={0.5cm 0.5cm 0.5cm 0.5cm},clip=true]{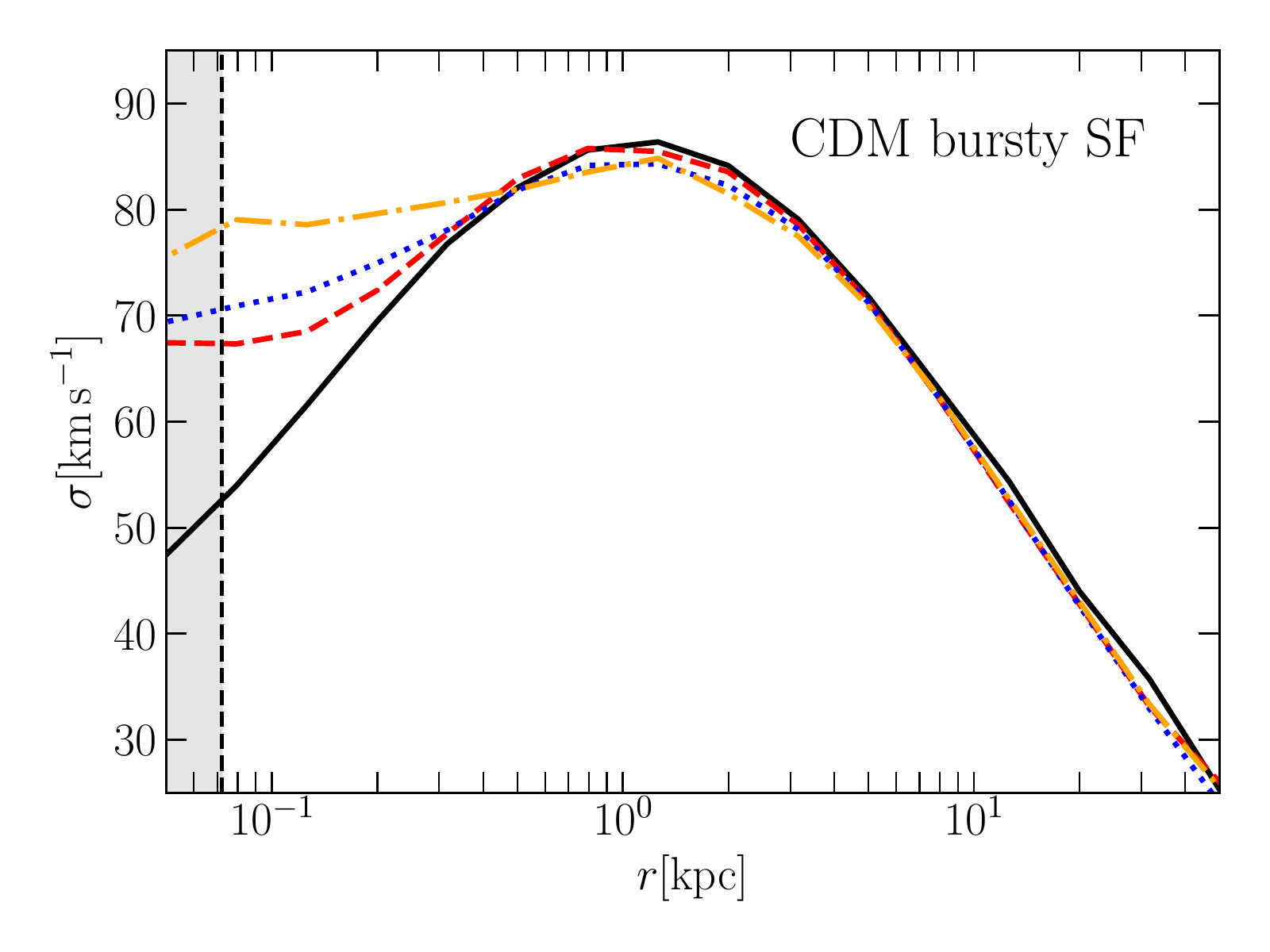}\\
    \caption{Evolution of density profile (left panels) and velocity dispersion profile (right panels) of the DM halo in three different simulations. In the top row we show results of a CDM run with $n_{\rm th} = 0.1\,{\rm cm^{-3}}$, in the middle row $\sigma_T/m_\chi = 1\,{\rm cm^2g^{-1}}$ and $n_{\rm th} = 0.1\,{\rm cm^{-3}}$ and in the bottom row we show results of a CDM run with $n_{\rm th} = 100\,{\rm cm^{-3}}$. We show spherically averaged profiles measured at the times indicated in the legends. The grey area corresponds to the radial range in which the profile is not converged according to the \citet{Power:2002sw} criterion.}
    \label{fig:dm_profiles}
\end{figure*}

We find striking differences in the evolution of the density and velocity dispersion profiles of the DM halo between simulations with different momentum transfer cross sections and star formation thresholds. Fig. \ref{fig:dm_profiles} compares their evolution (density to the left, velocity dispersion to the right) for three different simulations. The top panels correspond to the CDM run with $n_{\rm th} = 0.1\,{\rm cm^{-3}}$, in the middle panels $\sigma_T/m_\chi = 1\,{\rm cm^2g^{-1}}$ and $n_{\rm th} = 0.1\,{\rm cm^{-3}}$, and in the bottom panels we show results of the CDM run with $n_{\rm th} = 100\,{\rm cm^{-3}}$. Several profiles are shown in each panel, calculated from snapshots that are spaced apart by $1\,{\rm Gyr}$ of simulation time as labelled in the legend. 

The DM density and velocity dispersion profiles show almost no evolution in the case in which both the star formation threshold and momentum transfer cross section are small (upper panels). In fact, the DM halo remains cuspy down to the smallest resolved radius. In the other two cases, however, a constant density core forms in the inner halo. For $\sigma_T/m_\chi = 1\,{\rm cm^2g^{-1}}$ and $n_{\rm th} = 0.1\,{\rm cm^{-3}}$ (middle panels), a $\sim 1\,{\rm kpc}$ core forms quickly and is fully formed after 
$\sim 2\,{\rm Gyr}$. The corresponding velocity dispersion profile is flat out to approximately the scale radius of the initial halo. Density and velocity dispersion profiles of the CDM simulation with $n_{\rm th} = 100\,{\rm cm^{-3}}$ are displayed in the bottom panels. After $\sim 3\,{\rm Gyr}$, the density profile closely resembles the SIDM density profile shown in the middle panel. However, the 
cusp-core transformation proceeds slower and we see that the corresponding velocity dispersion profile is not yet fully isothermal at the end of the simulation. Thus, while the timescales for impulsive (SNF driven) and adiabatic (SIDM related) core formation are slightly different, the resulting cored density profiles look remarkably similar. As a consequence, we cannot differentiate between those two core formation scenarios by means of their final DM density profiles.

\begin{figure}
    \centering
    \includegraphics[width=0.98\linewidth,trim={0.5cm 0.5cm 0.5cm 0.5cm},clip=true]{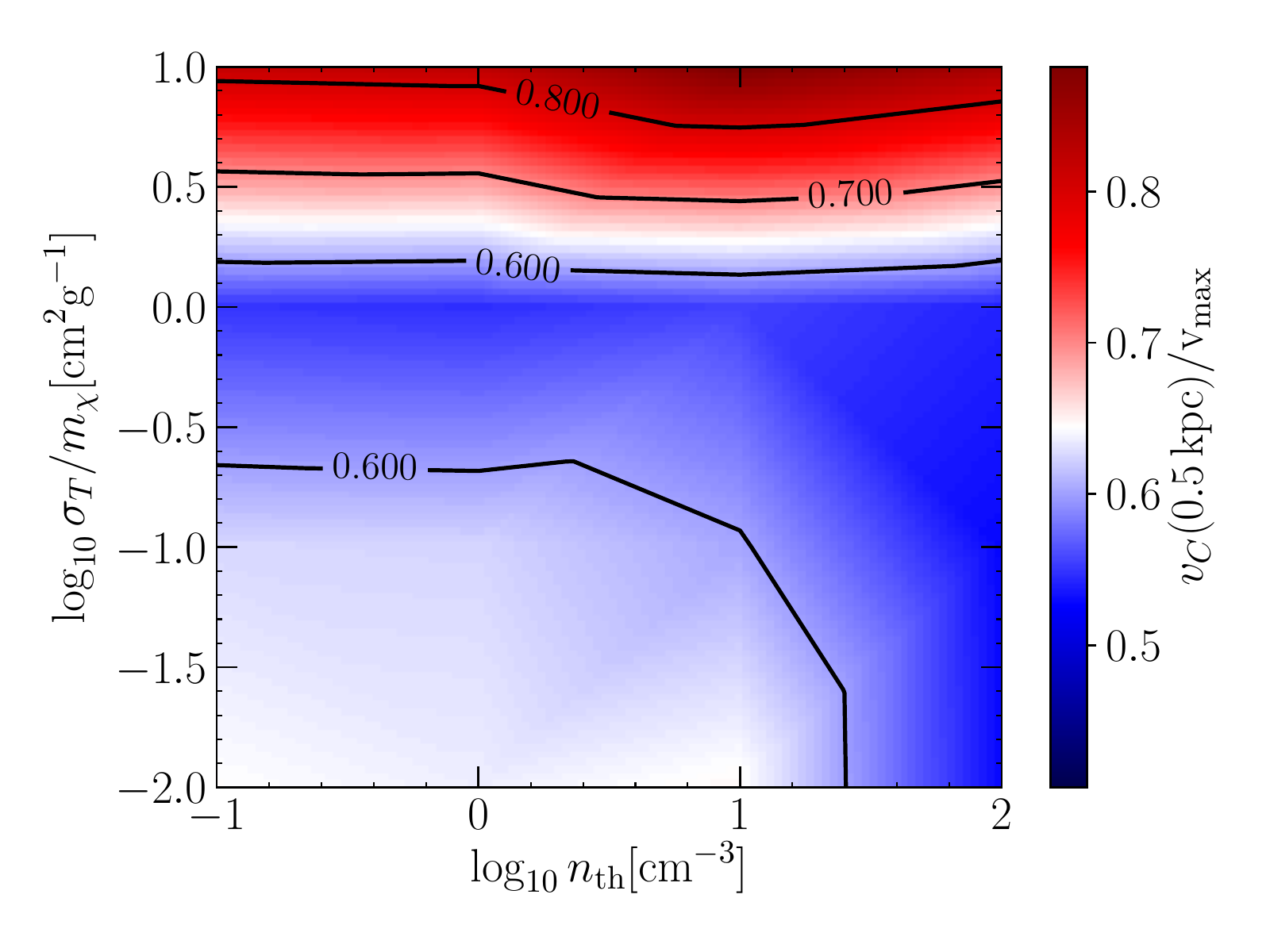}\\
    \includegraphics[width=0.98\linewidth,trim={0.5cm 0.5cm 0.5cm 0.5cm},clip=true]{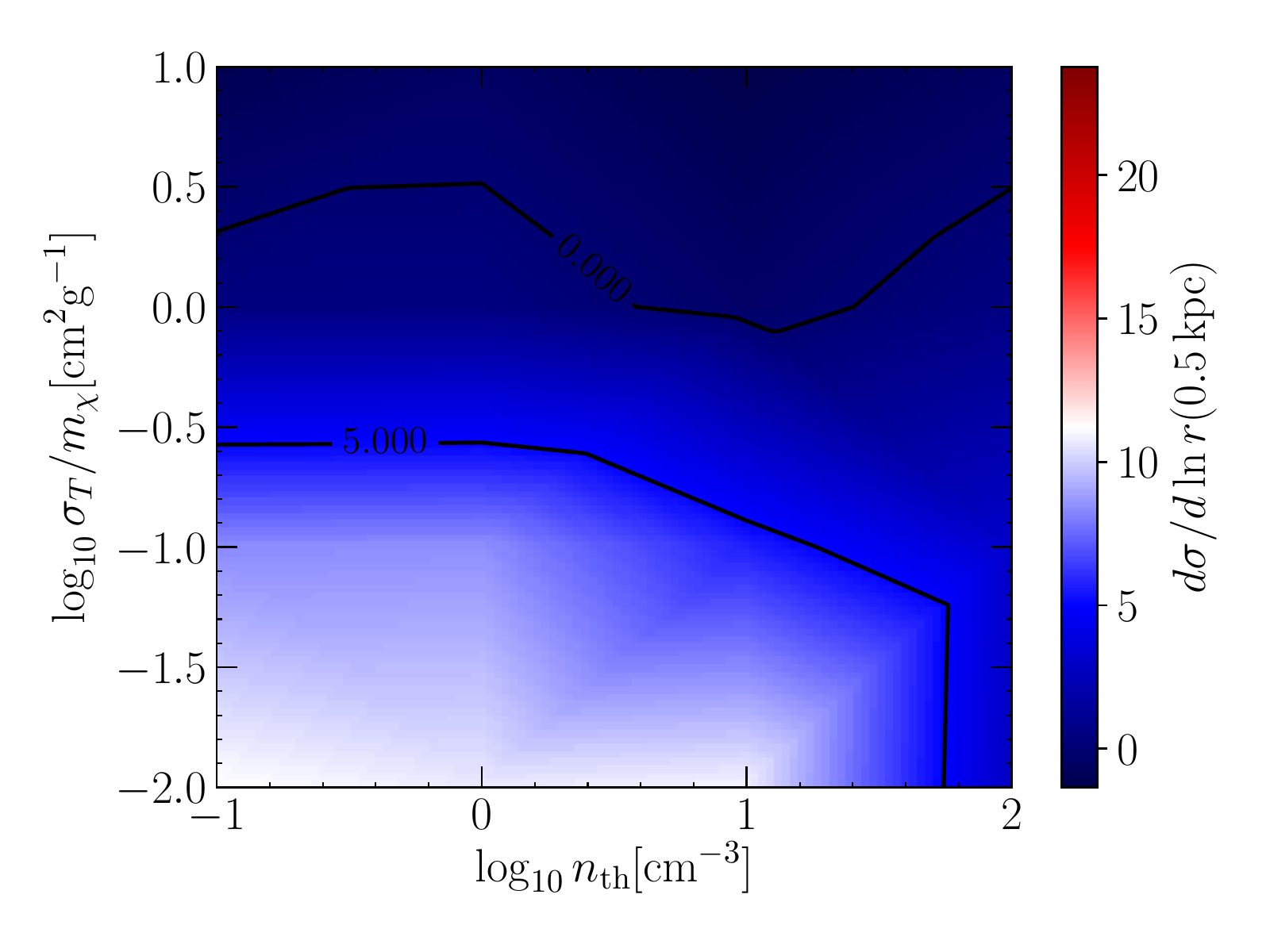}
    \caption{Two quantities that characterize cored and cuspy DM haloes at the characteristic radius of $0.5\,{\rm kpc}$ in our simulations: the circular velocity profile relative to its maximum value (upper panel), and the logarithmic slope of the velocity dispersion profile. These are shown 
    as a function of (logarithmic) star formation threshold and (logarithmic) transfer cross section per unit mass. 
    We adopt a value of $\sigma_T/m_\chi=0.01\,{\rm cm^2g^{-1}}$ to represent CDM in this plot.
    The colour maps show bilinear interpolations over all 16 simulations. The contour lines 
    show degenerate curves in parameter space along which $v_C(0.5\,{\rm kpc})/v_{\rm max}$ (upper panel) or $d\sigma/d\ln r (0.5\,{\rm kpc})$ (lower panel) assume the indicated values. The upper panel quantifies the (total) mass deficit in the inner part of the halo, whereas the lower panel focuses directly on whether the DM distribution in the inner halo is isothermal or not. Both panels correspond to results after $3\,{\rm Gyr}$. The colour map is chosen such that white colour corresponds to the CDM simulation with $n_{\rm th} = 0.1\,{\rm cm^{-3}}$ in which the DM halo remains cuspy (lower left corner). Deviations from this benchmark value in either direction are then coloured in red or blue as indicated by the colour bar.
    }
    \label{fig:dm_phasespace}
\end{figure}

\begin{figure}
    \centering
    \includegraphics[width=0.98\linewidth,trim={0.5cm 0.5cm 0.5cm 0.5cm},clip=true]{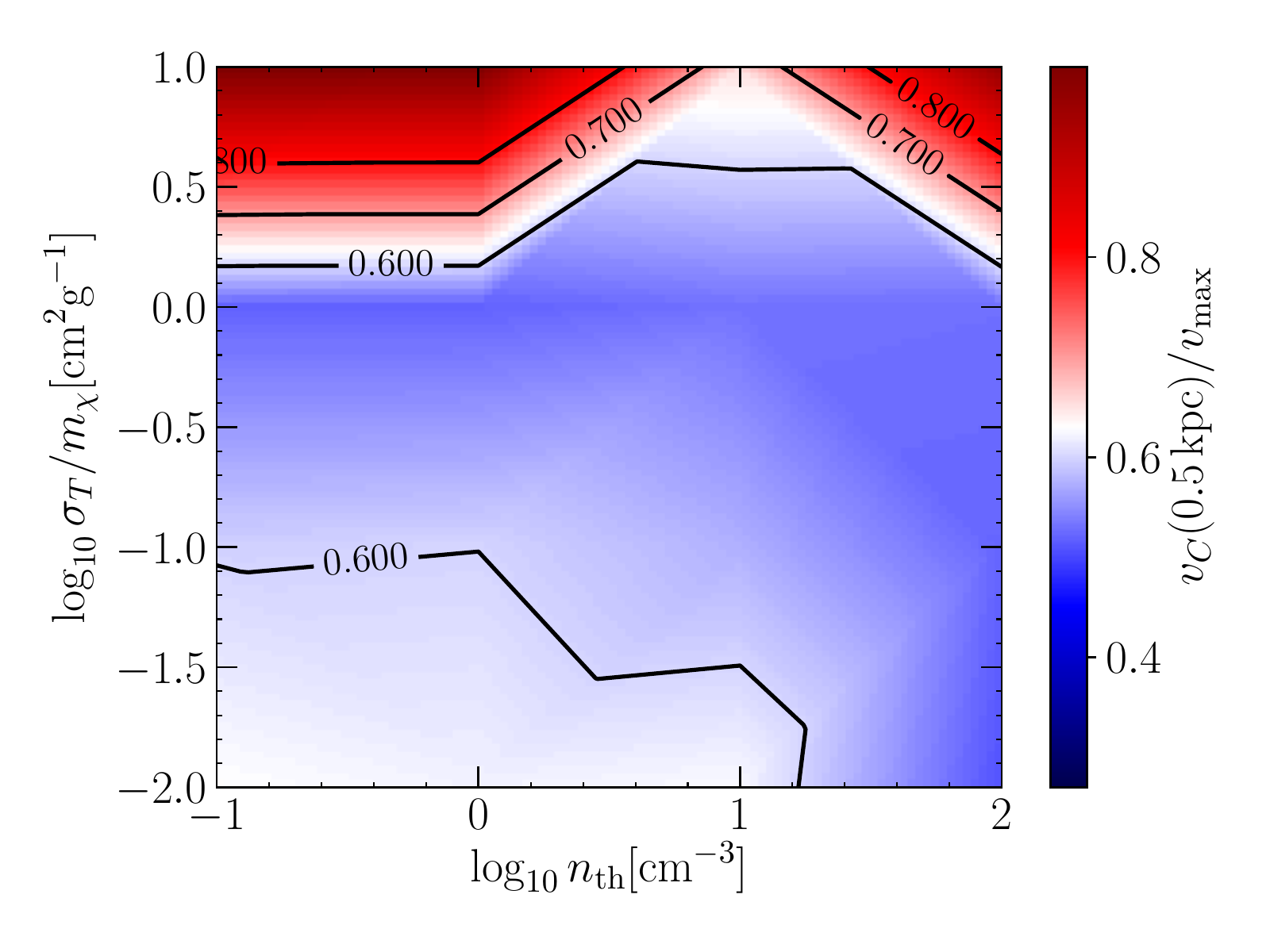}
    \caption{Same as upper panel of figure \ref{fig:dm_phasespace}, but after $4\,{\rm Gyr}$ of simulation time.}
    \label{fig:vc_4gy}
\end{figure}

The aim of our study is to compare the kinematic properties of baryons between simulations whose inferred final DM density profiles look nearly identical. More specifically, we look to identify structural differences between simulations in which one would observe a cored final DM density profile, and to relate those differences to the dominant core formation mechanism. Crucially, DM density profiles are not observed directly, but instead reconstructed from the measured rotation curves of observed galaxies. \citet{2020MNRAS.495...58S} introduced a method to categorize rotation curves by comparing the maximal circular velocity $v_{\rm max}$ with the circular velocity $v_C$ at a fiducial radius $r_{\rm fid} = 2 (v_{\rm max}/70\,{\rm km/s})\,{\rm kpc}$ (see also \citealt{2015MNRAS.452.3650O}). The authors state that a value of $v_C(r_{\rm fid})/v_{\rm max}\sim 0.7$ is typical for cuspy NFW haloes and that larger values correspond, on average, to adiabatically contracted haloes, whereas smaller values correspond to cored haloes. However, \citet{2020MNRAS.495...58S} also mention that while this ratio is a useful statistical measure to characterize rotation curves, it cannot be used to decide whether individual DM density profiles are cored or cuspy.
\citet{2021arXiv210301231B} showed that if one considers a set of (dwarf) galaxy-halo systems of a similar size and composition, the ratio of the circular velocity at a characteristic radius to the maximal circular velocity can be used to compare the final rotation curves of those systems, and can provide a relative measure for how cuspy/cored the final haloes are. 
Here, we chose 
$r=0.5\,{\rm kpc}$ as this characteristic radius, since Fig. \ref{fig:dm_profiles} suggests that the difference in enclosed mass between cuspy and cored final halo profiles is maximal at that radius. 
Hence, we adopt $v_C(0.5\,{\rm kpc})/v_{\rm max}$ as a measure for how cored or cuspy our simulated DM halo is at a given time. We stress again that by adopting this measure, we aim to focus on observable differences between different simulation suites. In Appendix \ref{app_addfigs}, we show that the general trends observed in Figs.~\ref{fig:dm_phasespace} and \ref{fig:vc_4gy} are preserved when adopting a more conventional measure for how cuspy/cored DM density profiles are, namely the logarithmic slope of the DM density profile in the central halo (see Fig.~\ref{fig:app_fig_one}). 

Fig. \ref{fig:dm_phasespace} shows $v_C(0.5\,{\rm kpc})/v_{\rm max}$ (upper panel) and $d\sigma/d\ln r (0.5\,{\rm kpc})$ (lower panel) as a function of (logarithmic) star formation threshold and (logarithmic) transfer cross section per unit mass after $3\,{\rm Gyr}$ of simulation time. The colour map is a bilinear interpolation in the (logarithmic) parameter space between all 16 simulations (see Section \ref{subsec:simparams}) and we have assigned a "self-interaction cross section" of $0.01\,{\rm cm^2g^{-1}}$ to the CDM runs to be able to include them in the Fig.\footnote{As mentioned above, CDM and SIDM are virtually indistinguishable for $\sigma_T/m_\chi \ll 0.1\,{\rm cm^2g^{-1}}$. We thus expect no important differences between CDM runs and potential runs with $\sigma_T/m_\chi = 0.01\,{\rm cm^2g^{-1}}$ and thus set them equal for the sake of presentation. 
In the remainder of the article, CDM runs will be interpreted as SIDM runs with $\sigma_T/m_\chi = 0.01\,{\rm cm^2g^{-1}}$ whenever interpolations over parameter space are presented.}. 
Fig. \ref{fig:dm_phasespace} demonstrates
that there are curves in the $\sigma_T/m_\chi$ - $n_{\rm th}$ parameter space along which the measured values of $v_C(0.5\,{\rm kpc})/v_{\rm max}$  (or $d\sigma/d\ln r (0.5\,{\rm kpc})$) are degenerate. Some of these curves are highlighted by the contour lines. We have constructed all colour maps here such that they refer to deviations from the CDM simulation with $n_{\rm th} = 0.1\,{\rm cm^{-3}}$, in which the halo remains cuspy (see Fig. \ref{fig:dm_profiles}). Quantities measured for this benchmark simulation are assigned white colour, while deviations into either direction are coloured blue or red.

A couple of interesting trends emerge in the upper panel of Fig. \ref{fig:dm_phasespace}. For star formation thresholds $n_{\rm th} \le 10\,{\rm cm^{-3}}$, the final mass distribution is solely determined by the self-interaction cross section. Cross sections up to $\sim 0.1\,{\rm cm^2g^{-1}}$ are rather ineffective at forming a core within a simulation time of $\sim 3\,{\rm Gyr}$. The most prominent cores are formed at values $\approx 1\,{\rm cm^2g^{-1}}$. However, for (much) larger cross sections an inversion of this effect occurs\footnote{Recall that we simulated cross sections of $\sigma_T/m_\chi = 1\,{\rm cm^2g^{-1}}$ and $\sigma_T/m_\chi = 10\,{\rm cm^2g^{-1}}$ and then interpolated between those simulations for the sake of presentation. Therefore, we cannot determine the exact cross section at which the inversion occurs from Fig.~\ref{fig:dm_phasespace}.}. 
In fact, for 
$\sigma_T/m_\chi\sim 10\,{\rm cm^2g^{-1}}$ we find that the final enclosed mass within $0.5\,{\rm kpc}$ is larger than in the baseline cuspy CDM case. This is due to the onset of the gravothermal collapse phase \citep{2002ApJ...568..475B, 2002ApJ...581..777C, 2011MNRAS.415.1125K, 2015ApJ...804..131P, 2020PhRvD.101f3009N}. Note that in a cosmological halo, the onset and progression of gravothermal collapse is expected to depend on the galaxy's mass aggregation history, the central DM density, and on the local environment (see Appendix \ref{appc} for a discussion). 

When increasing the star formation threshold, we find that at some value between $10\,{\rm cm^{-3}}$ and $100\,{\rm cm^{-3}}$, SNF becomes sufficiently impulsive to form a core in the DM profile that is of roughly the same size as the largest cores formed by SIDM. Combining a large star formation threshold with SIDM cross sections that would by themselves lead to the formation of cores does not change the value of $v_C(0.5\,{\rm kpc})/v_{\rm max}$ by much. Hence, the measured rotation curves are truly degenerate at this characteristic radius, indicating that DM cores of similar size form
for cross sections around $1\,{\rm cm^2g^{-1}}$ (regardless of the star formation threshold) and for smaller cross sections as long as the star formation thresholds is large enough. 
On the other hand, if the SIDM transfer cross section per unit mass is sufficiently large, the effect of gravothermal collapse 
always outweighs the effect of SNF, meaning that even at large star formation thresholds the circular velocity measured at the end of the simulation is always larger than in the benchmark simulation. 

The lower panel of Fig. \ref{fig:dm_phasespace} aims to 
provide a measure of
the dynamical differences between the 16 DM haloes after 
$3\,{\rm Gyr}$. 
We show the derivative $d\sigma/d\ln r$ at $r=0.5\,{\rm kpc}$, interpolated across the parameter space shown in the figure. For fully isothermal cores we expect values of $d\sigma/d\ln r \sim 0$. If we focus on the contour line in parameter space along which we found cored halo profiles in the upper panel of Fig. \ref{fig:dm_phasespace}, 
we see that the behaviour of the velocity dispersion profiles to some degree breaks this degeneracy between SIDM cores and SNF cores. SIDM cores are in general more isothermal than their SNF counterparts. This is a generalization of the statement that SIDM and SNF can lead to similar core sizes, but their DM components have a different dynamical structure (see \citealt{2019MNRAS.485.1008B}), at least over the simulated time interval. While the bottom panels of Fig. \ref{fig:dm_profiles} suggests that the core in CDM run with impulsive star formation becomes increasingly isothermal, we do not know whether a steady state similar to the quasi-equilibrium state of cored SIDM haloes will eventually be reached\footnote{No such steady state is found after 4 Gyr of simulated time either.}. Throughout the simulations presented here, the dynamical structure of the DM haloes is different for different core formation scenarios. 
Finally, we note that larger SIDM cross sections can lead to slightly negative gradients in $\sigma(r)$ at 
$0.5$~kpc, 
indicating that the core size has already decreased due to gravothermal collapse (see discussion below and Fig.~\ref{fig:app_fig_one}).

In some of our simulations, the DM haloes continue to evolve after $3\,{\rm Gyr}$, and thus the picture presented in Fig. \ref{fig:dm_phasespace} changes slightly. For illustration, we show $v_C(0.5\,{\rm kpc})/v_{\rm max}$ measured after $4\,{\rm Gyr}$ as a function of $\sigma_T/m_\chi$ and $n_{\rm th}$ in Fig. \ref{fig:vc_4gy}. Two trends are apparent when comparing Fig. \ref{fig:vc_4gy} to the upper panel of Fig. \ref{fig:dm_phasespace}. Firstly, the SIDM haloes with $\sigma_T/m_\chi = 0.1\,{\rm cm^2g^{-1}}$ are more cored after
an extra $1$~Gyr of evolution. This result is not surprising, since SIDM haloes with relatively weak self-interaction cross sections will still develop cores, albeit on longer timescales. Secondly, gravothermal collapse has progressed, heavily altering the dynamical structure of SIDM haloes with $\sigma_T/m_\chi = 10\,{\rm cm^2g^{-1}}$, making them ``cuspier'' on average. It is worth noting that for all of those four runs, gravothermal collapse has progressed to the point that the radius at which the circular velocity is maximal is now smaller than $0.5$ kpc. 

This is best appreciated in comparison with Fig.~\ref{fig:app_fig_one}, where we use $d\log\rho/d\log r$, the logarithmic slope of the DM density profile, as an alternative way to quantify how cored/cuspy the final DM density profiles are. For the runs with $\sigma_T/m_\chi = 10\,{\rm cm^2g^{-1}}$, we find that at $r = 500\,{\rm pc}$, the final profiles are significantly steeper than in the baseline CDM case. This indicates that in these haloes, gravothermal collapse has progressed to the point that they have formed extremely dense central cores with core radii $r_c \ll 500\,$pc (see e.g. \citealt{2015ApJ...804..131P}). This evolution is the furthest along in the run with $n_{\rm th} = 10\,{\rm cm^{-3}}$, leading again to a reduction in $v_C(0.5\,{\rm kpc})/v_{\rm max}$. Peculiar properties of this particular simulation are discussed in appendix \ref{appendix_3}. The 
CDM runs do not change appreciably in the additional $1\,{\rm Gyr}$, 
implying that residual evolution due to SNF occurs on longer timescales. 
Importantly, the degeneracy contours in parameter space along which haloes with a flat constant density core are located very similarly in the upper panel of Fig. \ref{fig:dm_phasespace} and Fig. \ref{fig:vc_4gy}.

\subsection{Galaxy sizes}\label{subsec:galsize}
The orbits of stars change in response to an evolving gravitational potential. However, their response may differ depending on whether the change in the potential is adiabatic or impulsive. In addition, different star formation histories will also lead to a distribution of newly formed stars that varies between simulations. As a result, the stellar mass distribution may differ between galaxies in our simulation suite, in particular also between those galaxies whose host haloes have formed a core either through SNF or through self-interactions between the DM particles.
\begin{figure}
    \centering
    \includegraphics[width=0.98\linewidth]{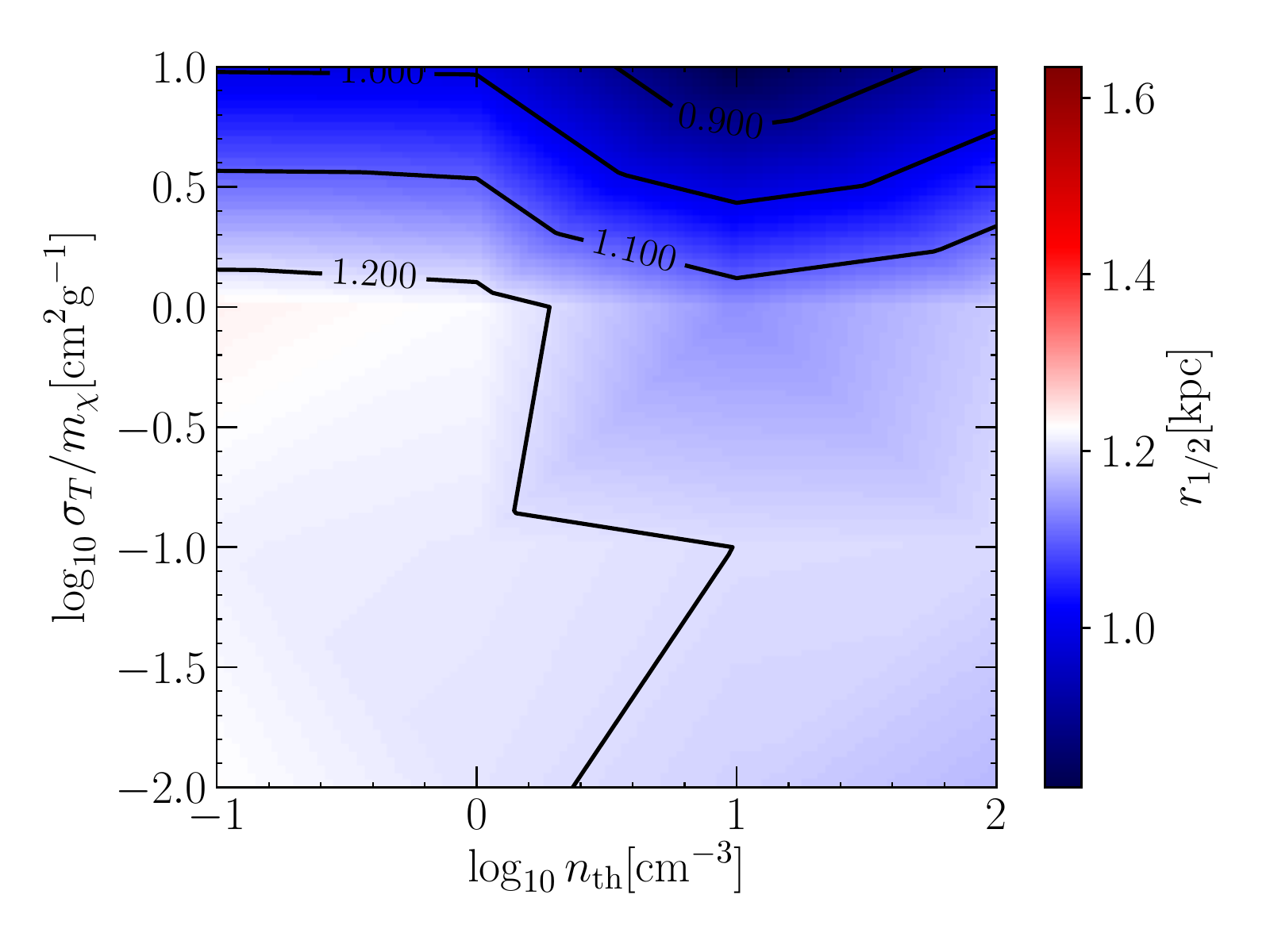}\\
    \includegraphics[width=0.98\linewidth]{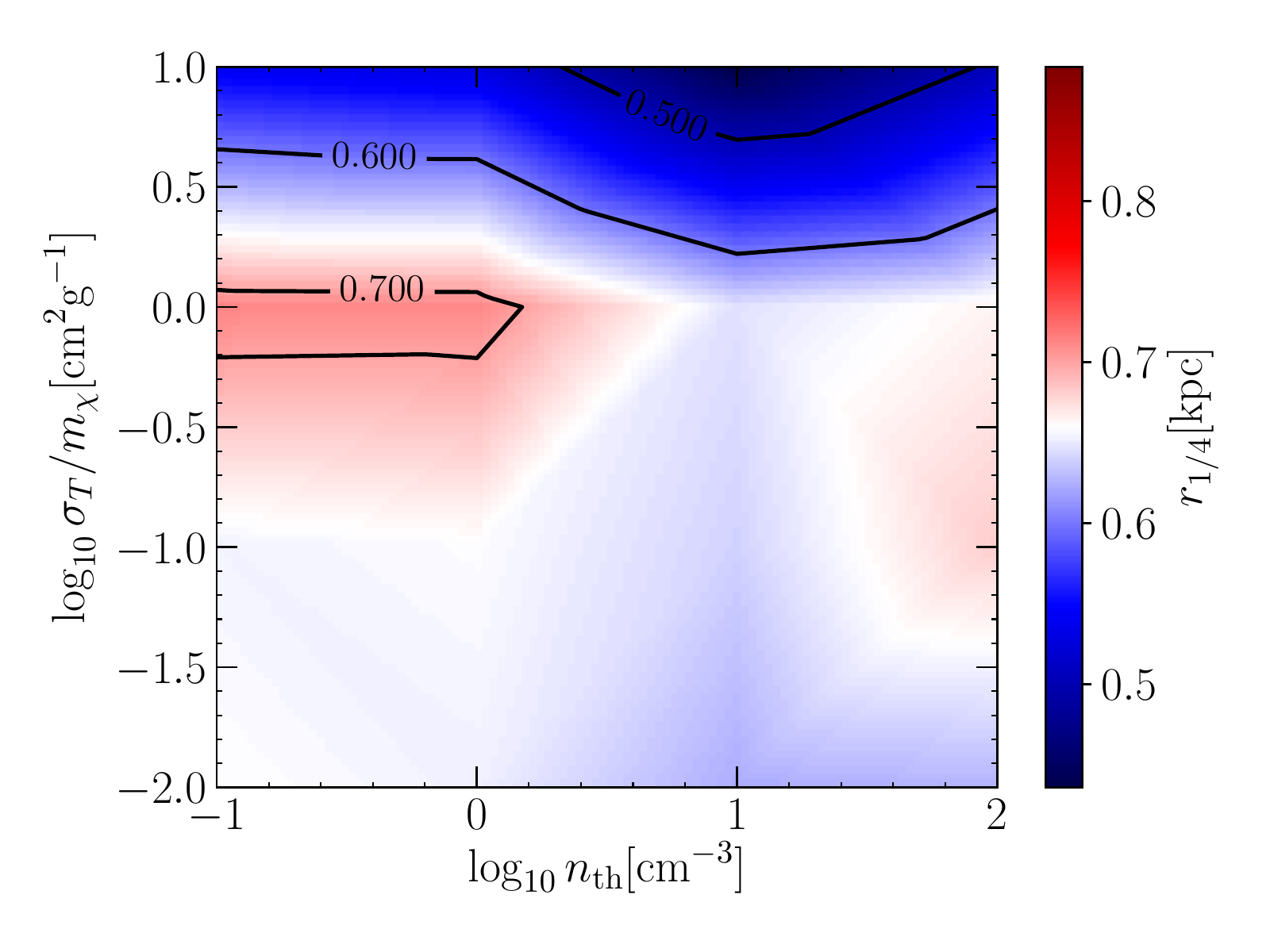}
    \caption{The galaxy half (stellar) mass radius $r_{1/2}$ (top panel) and quarter (stellar) mass radius $r_{1/4}$ (bottom panel) 
    as a function of $\sigma_T/m_\chi$ and $n_{\rm th}$, measured after $3\,{\rm Gyr}$. As in Fig. \ref{fig:dm_phasespace}, the
    parameter space shown is filled by interpolation across the 16 simulations in our suite.
    Contour lines indicate degeneracies (i.e. equal values of either $r_{1/2}$ or $r_{1/4}$) in parameter space. Notice that the half mass radius in all cases is larger than the typical core radius of $\sim 0.7\,{\rm kpc}$ found in Fig. \ref{fig:dm_profiles}. Thus, the impact of different core formation mechanisms is much more apparent in the quarter mass radius.}
    \label{fig:stellar_size}
\end{figure}

Fig. \ref{fig:stellar_size} shows different measures of the size of the simulated galaxy (after $3\,{\rm Gyr}$) 
as a function of star formation threshold and transfer cross section per unit mass. In the upper panel we show the half mass radius, whereas in the lower panel we show the quarter mass radius of the simulated galaxies. These radii are determined by calculating the enclosed stellar mass in spherical shells around the halo's centre of potential and then determining (using nested intervals and gsl ``akima" interpolation) the radius of the spherical shell that contains half (a quarter of) the total stellar mass. When calculating the stellar mass profile, we take into account all collisionless disc and bulge particles (see Section \ref{subsec:ICs}, as well as newly formed ``star" particles. 
The half mass (upper panel) radius is usually taken as a characteristic scale of galaxies. 
Given the simulation setup we have, in particular the values of the scale lengths for the gaseous and stellar discs chosen for the initial conditions in our simulations (see Section \ref{subsec:ICs}), the half mass radius ends up being
larger than the typical DM core radius by a factor of $\sim 1.5$ and is therefore not ideal to analyse how SNF and SIDM affect the stellar distribution.
Hence, we also look at the quarter mass radius (lower panel of Fig. \ref{fig:stellar_size}), which probes exactly the radial range of interest. 

A few trends are similar across both panels of Fig. \ref{fig:stellar_size}. At small SIDM 
cross sections ($\sigma_T/m_\chi \le 0.1\,{\rm cm^2g^{-1}}$), 
the star formation threshold hardly has any impact on the final galaxy size. 
We note that this result appears counter-intuitive at first. Among others, \citet{2005MNRAS.356..107R} and \citet{2012ApJ...755L..35M} have demonstrated that stellar particles react to a rapidly fluctuating gravitational potential in the same way that DM particles do, and that the formation of a shallow DM core should be accompanied by an expansion of the orbits of old stars (see also \citealt{2019MNRAS.485.1008B}). Moreover, \citet{2013MNRAS.429.3068T} showed that stronger feedback leads to more extended galaxies in idealized hydrodynamic simulations of an isolated dwarf galaxy, while \citet{2015MNRAS.448..792G} and \citet{2016ApJ...819..101G} derived similar results using cosmological zoom simulations of a single system. Contrary to our simulation suite, all of those latter works compared the results of simulations with a fixed star formation threshold, but different SNF efficiencies (realized in different ways in their simulations). What sets our simulation suite apart is that we regulate SNF through the star formation threshold. Since the runs with more impulsive feedback have larger star formation thresholds, the stars that newly form in these runs are more concentrated towards the centre of the galaxy, effectively creating a more compact galaxy. In our simulations, this effect competes with the expansion of the orbits of old stars due to feedback. The above mentioned results are qualitatively recovered when limiting our analysis to the stellar (disk) particles that were present at the beginning of the simulation. 

A key trend that can be observed across both panels of Fig.~\ref{fig:stellar_size} is that the galaxy size contracts significantly in those simulations in which the self-interaction cross section is large enough to trigger the gravothermal catastrophe. The most significant contraction is observed for the case in which $\sigma_T/m_\chi = 10\,{\rm cm^2g^{-1}}$ and $n_{\rm th} = 10\,{\rm cm^{-3}}$, where the gravothermal collapse proceeds somewhat faster than in simulations with the same self-interaction cross section and different star formation thresholds (see Appendix \ref{appendix_3} for a discussion of this run). 

The most interesting feature of Fig. \ref{fig:stellar_size} appears only in the bottom panel. For $n_{\rm th} \le 1{\rm cm^2g^{-1}}$ the galaxy becomes more extended if $\sigma_T/m_\chi \sim 1\,{\rm cm^2g^{-1}}$, precisely the case in which {\it i}) the DM density profile forms a core in an adiabatic way due to SIDM and {\it ii}) SNF does not cause impulsive changes in the gravitational potential. In simulations in which the DM halo adiabatically forms a core, the stellar tracers follow the adiabatic evolution of the gravitational potential, resulting in a less bright and more extended galaxy. \citet{Vogelsberger2014} performed SIDM simulations with $\sigma_T/m_\chi \sim 1\,{\rm cm^2g^{-1}}$ with a baryonic physic implementation having effectively a low star formation threshold. The authors find that the stellar distribution of their simulation traces the evolution of the DM, forming a core that is related to the DM core (see Fig. 8 in \citet{Vogelsberger2014}), which is in very good agreement with the results we find here. 

Of key importance is the observation that the above described expansion of the galaxy in simulations with SIDM-induced core formation is only observed in simulated galaxies with a smooth star formation history -- and not when impulsive feedback is also present. This highlights that even if the evolution of the DM is governed by the effect of SIDM when the cross section is $\gtrsim 1\,{\rm cm^2g^{-1}}$, SNF remains an important perturber to the dynamics of the stars. The galaxy sizes measured after $4\,{\rm Gyr}$ are very similar to the ones presented in Fig. \ref{fig:stellar_size}. The only difference is a strong additional spatial contraction of the galaxies in the simulations with $\sigma_T/m_\chi = 10\,{\rm cm^2g^{-1}}$. An important conclusion from Fig. \ref{fig:stellar_size} is thus that the spatial distribution of stars can, potentially, indicate the presence (or absence) of impulsive feedback in an SIDM universe. However, we note that decreasing the initial ratio of DM to baryonic matter in the centre of the galaxy, which leads to (much) burstier star formation on average, results in a significant expansion of the simulated galaxy in runs with particularly impulsive feedback. We discuss this in Appendix \ref{app_bar_sig} and explore how the combination of two baryonic signatures may allow for a firmer conclusion about the pre-dominant core formation mechanism. 

Inspired by the results of \citet{2019MNRAS.485.1008B}, we now focus on the dynamical properties of the stars and the gas to search for further observable signatures of either the adiabatic or impulsive cusp-core transformation scenarios.

\subsection{Line-of-sight gas dynamics} \label{subsec:gas_dyn}
\begin{figure}
    \centering
    \includegraphics[width=\linewidth,trim={0.5cm 0.5cm 0.5cm 0.5cm},clip=true]{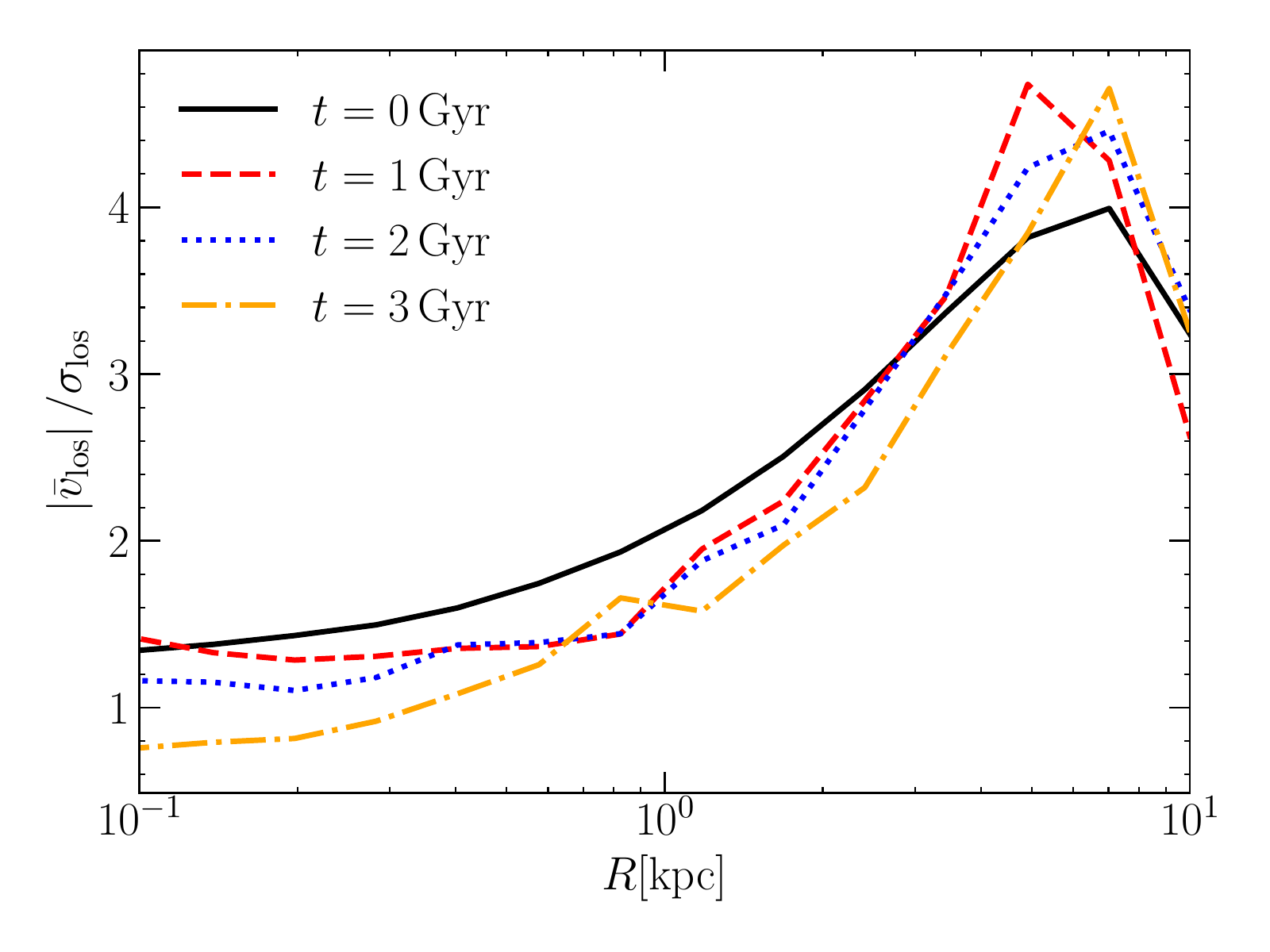}
    \caption{Evolution of the ratio of the average line-of-sight speed to the line-of sight velocity dispersion as a function of cylindrical radius for the 
    CDM simulation with $n_{\rm th} = 100\,{\rm cm^{-3}}$ (bursty star formation). 
    Different lines correspond to different times as indicated in the legend.}
    \label{fig:vbs_100_0}
\end{figure}

Impulsive injection of energy, momentum, and ejecta mass from supernovae into the surrounding interstellar medium (ISM) causes random motion in the gas, perturbing its circular streaming motion. 
Fig. \ref{fig:vbs_100_0} 
shows the ratio between the line-of-sight speed and the line-of-sight rms (root mean squared) velocity of the gas in the CDM  simulation in which $n_{\rm th} = 100\,{\rm cm^{-3}}$ (bursty star formation),
calculated at different
times as indicated in the legend. 
To calculate this ratio, we first determine the centre of potential from all simulation particles with a shrinking spheres method and subsequently calculate the total angular momentum vector from all particles that are part of the rotating disc (gas cells, ``disc" particles, and ``star" particles). 
We then use the normalized total angular momentum vector to rotate the galaxy into a coordinate system whose origin is the centre of potential and whose vertical axis is aligned with the normalized total angular momentum vector. 

Afterwards, we calculate $|\bar{v}_{\rm los}|/\sigma_{\rm los}$ for an edge-on galaxy configuration. 
Since our initial conditions were set up with an axisymmetric baryonic disc, we here assume that our simulated discs remain axisymmetric by the end of the simulations. 
Without loss of generality, we can then assume that the line of sight is aligned with the $y$-axis in our new coordinate system.\footnote{We have verified that the results do not change significantly when choosing the $x$-axis instead.} Averaged line-of-sight speed and velocity dispersion are then calculated in logarithmic bins of cylindrical radius, with $\bar{v}_{\rm los} = \bar{v}_y$ and $\sigma_{\rm los} = \sigma_{v_y}$. Initially, the gas streams without any random motion, and thus the observed dispersion in Fig. \ref{fig:vbs_100_0} is simply a consequence of the projection of the circular motion performed by the gas into the line-of-sight direction. 
As the simulation progresses, successive injection of momentum, energy and mass 
into the ISM causes additional random motion in the gas, in particular near the centre of the galaxy. After $3\,{\rm Gyr}$, we see that the inner value of $|\bar{v}_{\rm los}|/\sigma_{\rm los}$ has dropped below one. The effect of the reduction is particularly significant at very small radii
$\lesssim 200\,{\rm pc}$. 

\begin{figure}
    \centering
    \includegraphics[width=\linewidth,trim={0.5cm 0.5cm 0.5cm 0.5cm},clip=true]{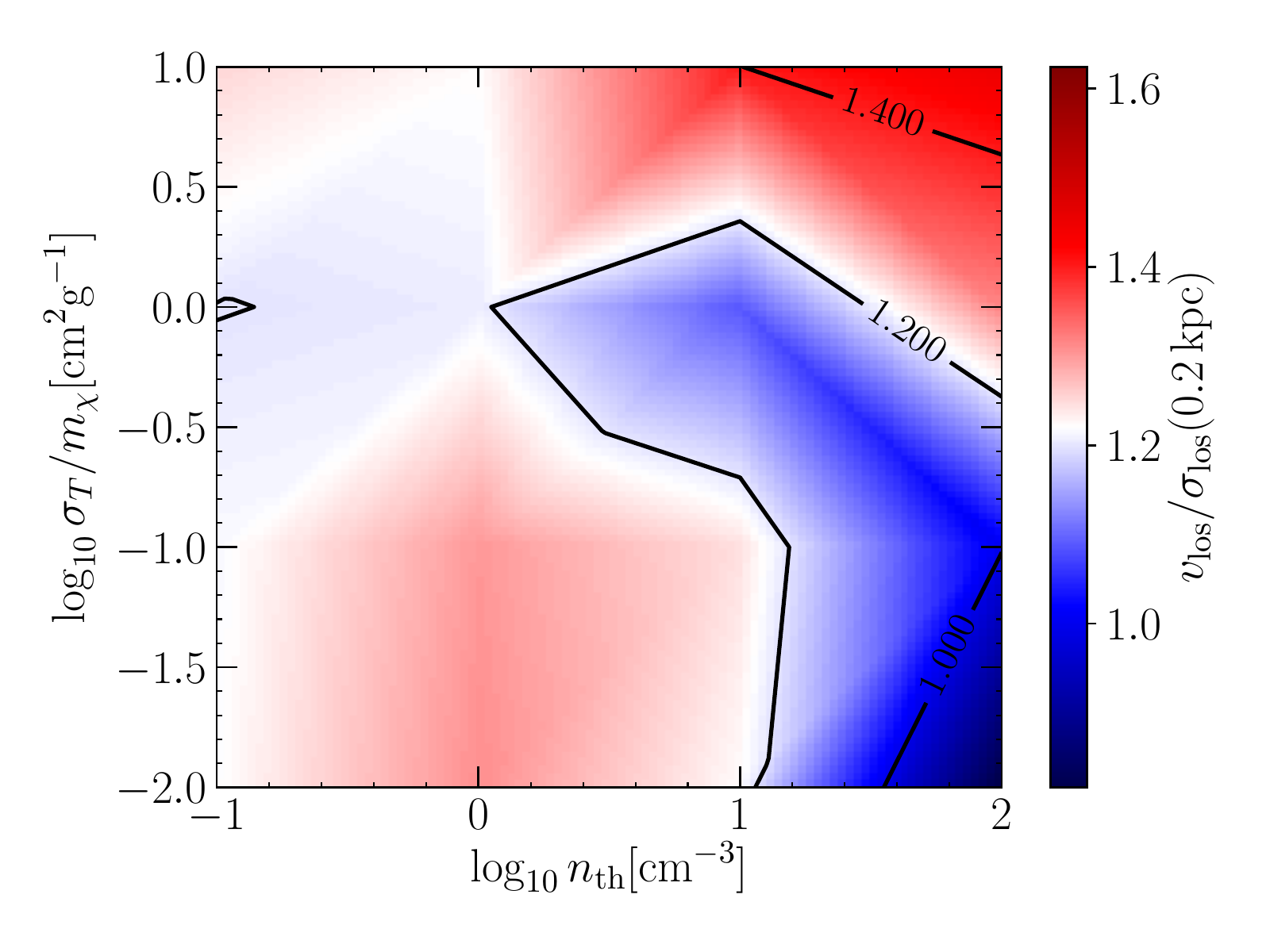}\\
    \includegraphics[width=\linewidth,trim={0.5cm 0.5cm 0.5cm 0.5cm},clip=true]{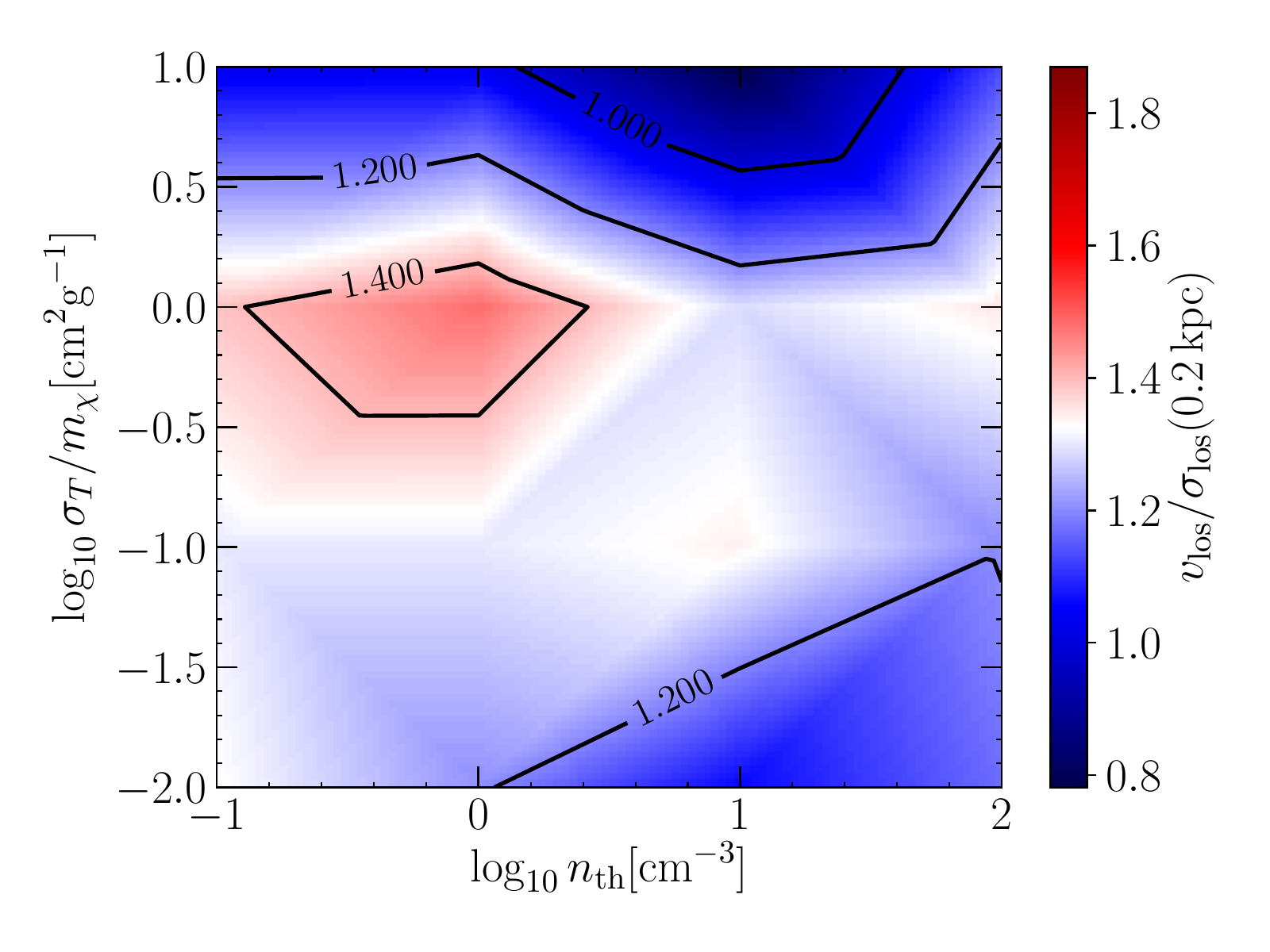}
    \caption{A measure of the relative impact of random motions over rotation: the ratio $|\bar{v}_{\rm los}|/\sigma_{\rm los}$, after $3\,{\rm Gyr}$ (top panel) and $4\,{\rm Gyr}$ (bottom panel). 
    We show this ratio at a cylindrical distance of $200\,{\rm pc}$ from the centre of the galaxy 
    as a function of the parameters $n_{\rm th}$ and $\sigma_T/m_\chi$. 
    A tendency towards smaller ratios (i.e. larger random motion) arises in simulations with cored haloes and bursty star formation
    ($n_{\rm th} \ge 10\,{\rm cm^{-3}}$), i.e., roughly within the contour line of value 1.2, which is roughly the baseline cuspy CDM value with smooth star formation. The strength of this trend is however transitory, being
    significantly stronger at $3\,{\rm Gyr}$ (top panel) than at $4\,{\rm Gyr}$ (bottom panel). 
    At $t = 4$ Gyr, we find that random gas motion increases significantly in the central region of galaxies whose host haloes undergo gravothermal collapse.}
    \label{fig:vbs_phase}
\end{figure}

\begin{figure*}
    \centering
    \includegraphics[width=0.48\linewidth,trim={0.5cm 0.5cm 0.5cm 0.5cm},clip=true]{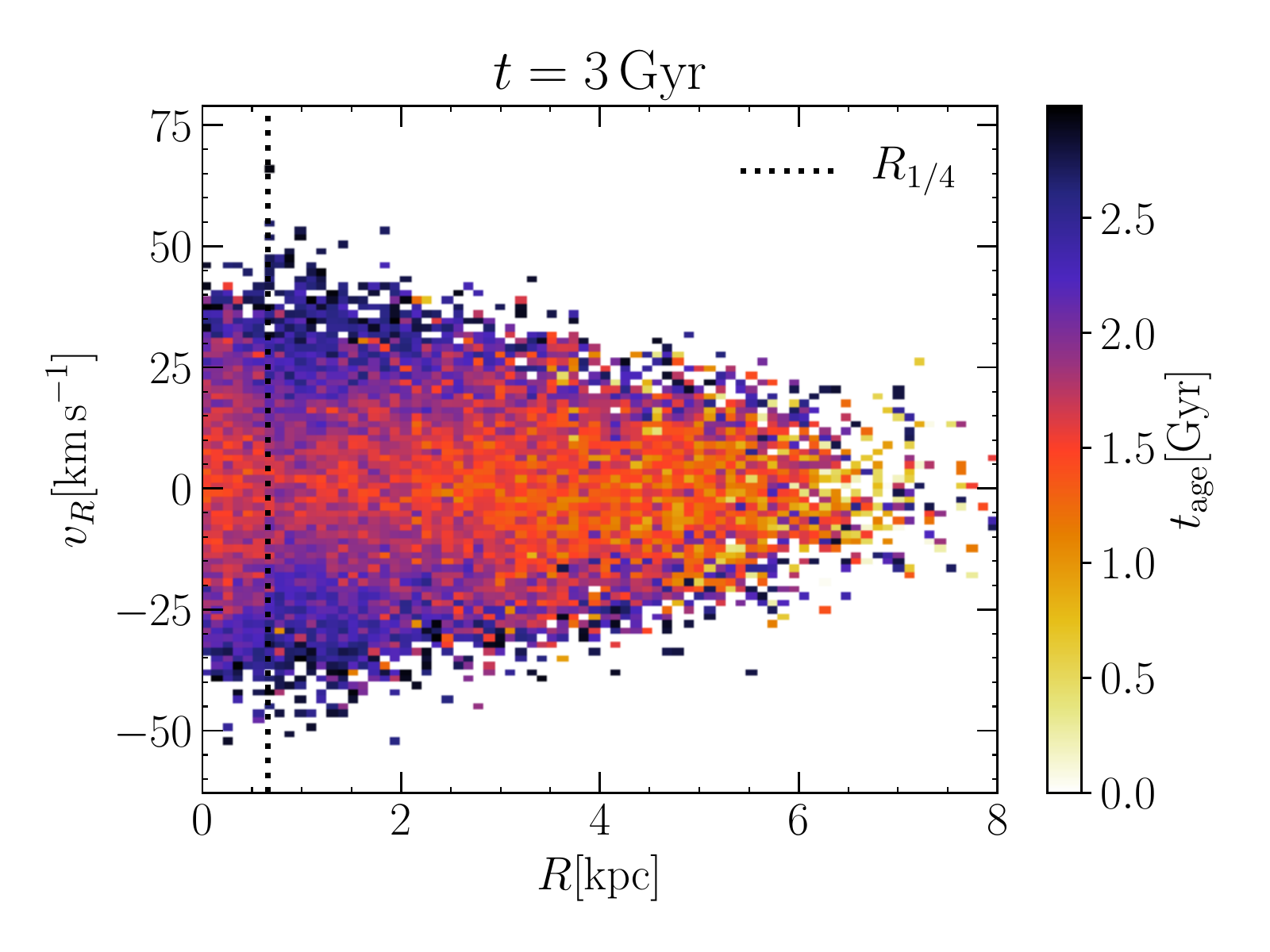}
    \includegraphics[width=0.48\linewidth,trim={0.5cm 0.5cm 0.5cm 0.5cm},clip=true]{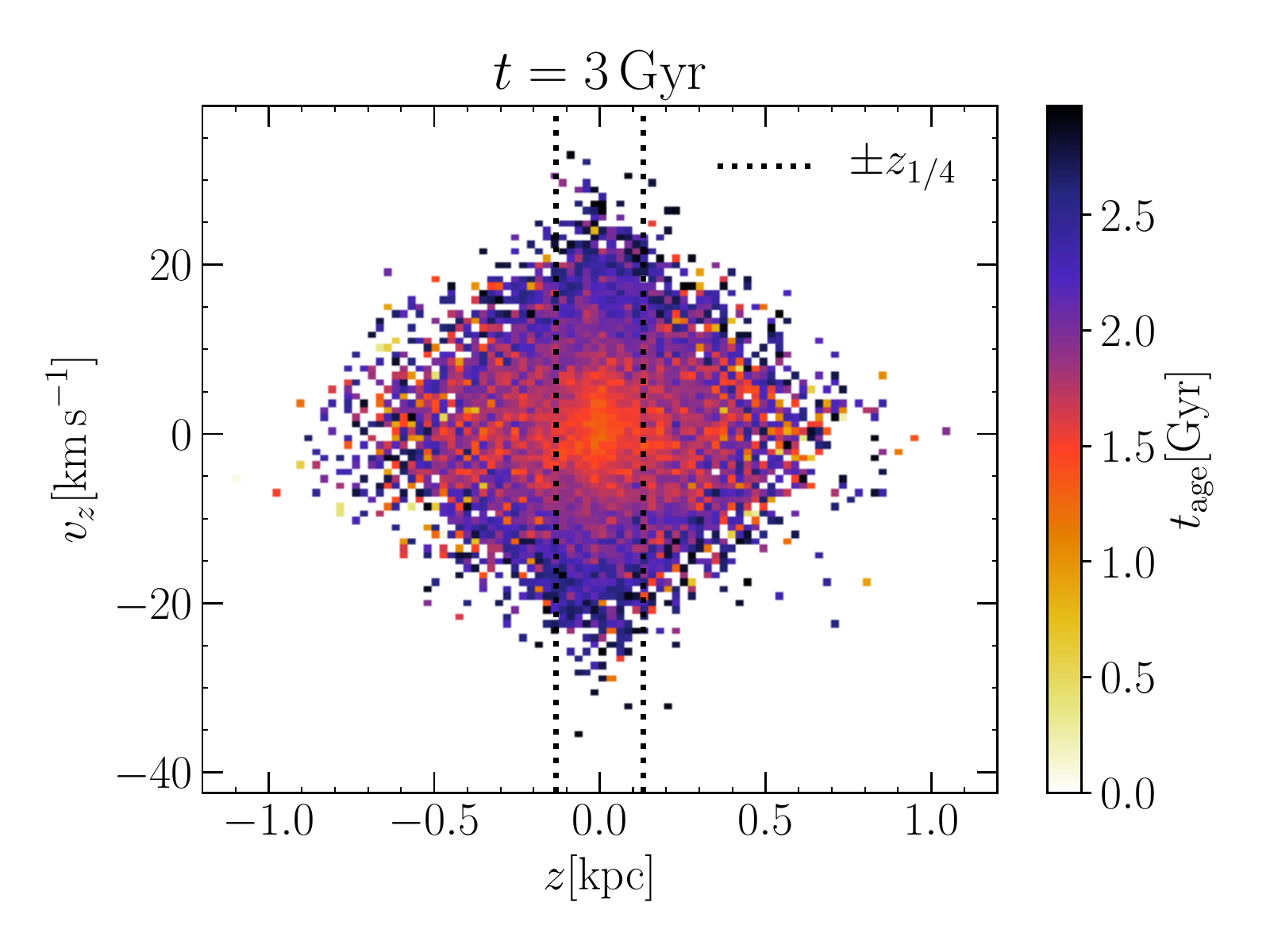}\\
    \includegraphics[width=0.48\linewidth,trim={0.5cm 0.5cm 0.5cm 0.5cm},clip=true]{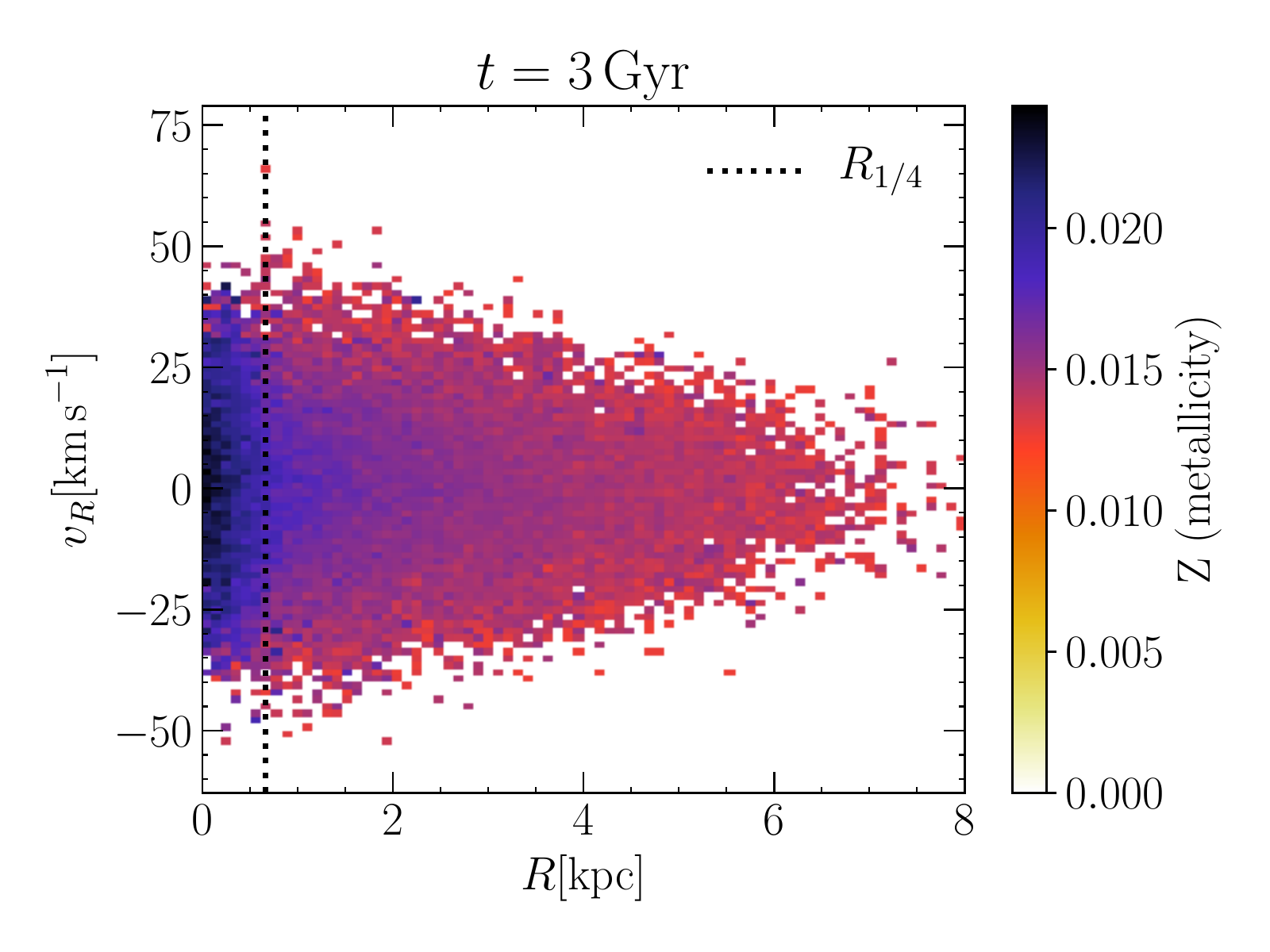}
    \includegraphics[width=0.48\linewidth,trim={0.5cm 0.5cm 0.5cm 0.5cm},clip=true]{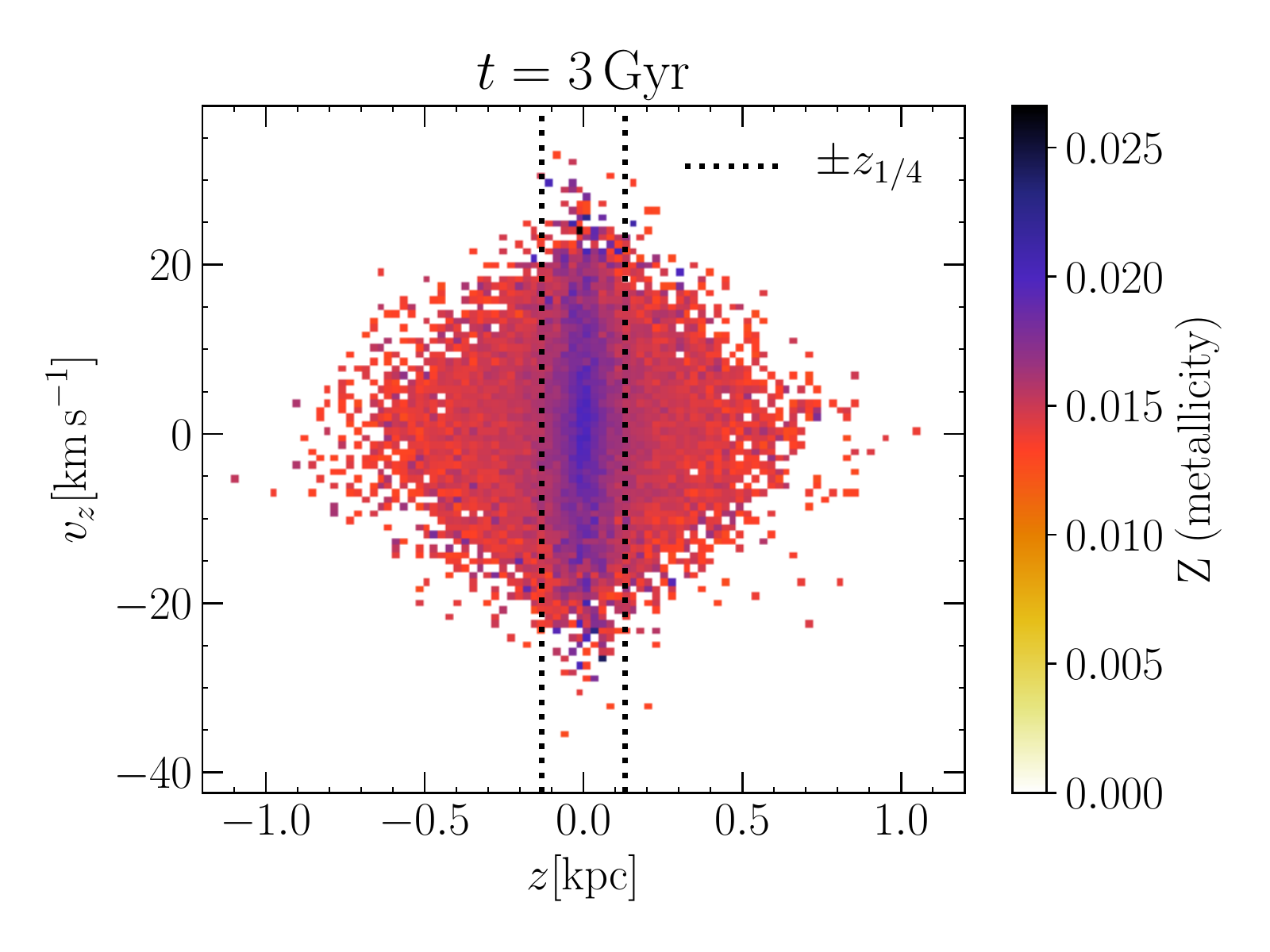}
    \caption{Average age (upper panels) and metallicity (lower panels) of stars formed after $3\,{\rm Gyr}$ in the CDM simulation with $n_{\rm th} = 0.1\,{\rm cm^{-3}}$ (smooth star formation) as a function of phase space coordinates:
    cylindrical radius $R$ and velocity $v_R$ (left panels), 
    vertical coordinate $z$ and the vertical velocity $v_z$ (right panels). The data is averaged in phase space bins. Black dotted vertical lines denote either the quarter mass radius ($R_{1/4}$, left column) or $z_{1/4}$ (right column, see text). }
  \label{fig:age_met_grad}
\end{figure*}

If this increase in random gas motion is in fact a direct consequence of SNF, then the different values of $n_{\rm th}$ used in our simulation suite should lead to a systematic 
difference in $|\bar{v}_{\rm los}|/\sigma_{\rm los}$ across simulations. In the upper (lower) panel of Fig. \ref{fig:vbs_phase}, we show $|\bar{v}_{\rm los}|/\sigma_{\rm los}$ measured at a cylindrical radius of $0.2\,{\rm kpc}$ from the centre of the galaxy after $3\,{\rm Gyr}$ ($4\,{\rm Gyr}$) 
as a function of star formation threshold and SIDM transfer cross section. Over large parts of the parameter space, the degree of random motion in the gas close to the centre of the galaxy is nearly constant across simulations. However, in some cases with large star formation thresholds (impulsive SNF), the central value of $|\bar{v}_{\rm los}|/\sigma_{\rm los}$ is reduced
after $3\,{\rm Gyr}$ (relative to the baseline cuspy CDM case with smooth star formation), 
indicating an increase in random motion within the gas. 

This increase in random motion is particularly strong for $n_{\rm th} = 100\,{\rm cm^{-3}}$ and $\sigma_T/m_\chi \le 0.1\,{\rm cm^2g^{-1}}$. 
The strength of this distinction between smooth and bursty star formation (in CDM and SIDM with $\sigma_T/m_\chi \le 1\,{\rm cm^2g^{-1}}$) is however quite dependent on the simulation time. For instance, in the lower panel of Fig. \ref{fig:vbs_phase},
which corresponds to $t=4\,{\rm Gyr}$, the difference across different star formation thresholds is much smaller than in the upper panel. This implies that impulsive SNF can lead to a significant increase in the random motion of the gas that is rather short lived. Thus, while hypothetical observations of very chaotic gas motion in dwarf galaxies with cored host haloes would hint at a recent impulsive starburst event, SNF cannot be ruled out as the cause of a DM core if no such increased random motion is observed.

We notice as well that due to the increasing impact of the gravothermal collapse phase on the central properties of the galactic system, there is a strong increase in random gas motion in the simulations with $\sigma_T/m_\chi = 10\,{\rm cm^2g^{-1}}$ by $t=4\,{\rm Gyr}$. Therefore, we predict that hypothetical observations of compact dwarf galaxies with very fast-rising rotation curves and a large amount of random gas motion could indicate the presence of haloes that have gravothermally collapsed due to very large SIDM cross sections. 

Leaving gravothermally collapsed haloes aside, we find that a significant increase in random motion of the gas in the centre of a galaxy with a cored host halo can only be observed if SNF is impulsive (right corner in both panels of Fig. \ref{fig:vbs_phase} relative to the CDM case with smooth star formation baseline in the lower left corner). 
  
\subsection{Age and metallicity gradients}\label{subsec:age_met_grad}

The stellar evolution module of \texttt{SMUGGLE} keeps track of several properties of individual star particles, among them their formation time and metallicity. Fig. \ref{fig:age_met_grad} shows projections of the phase space distribution of the stellar age and metallicity of newly formed stars after $3\,{\rm Gyr}$ in the benchmark CDM simulation with $n_{\rm th} = 0.1\,{\rm cm^{-3}}$ (smooth star formation). In the upper (lower) panels, we show the average stellar age (metallicity) as a function of the phase space coordinates $R$ and $v_R$ (left panels) and $z$ and $v_z$ (right panels). 
For presentation purposes, the data is averaged in $100\times 100$ equally spaced bins in phase space. 
The average stellar age appears to be approximately independent of the cylindrical radius $R$ (upper left panel of Fig. \ref{fig:age_met_grad}). However, older stars seem to be on orbits with a relatively large vertical extent (upper right panel of Fig. \ref{fig:age_met_grad}). This hints at a slow migration of stars out of the disc plane over the course of the simulation. In contrast, 
we observe a clear radial and vertical gradient in metallicity, 
which has a straightforward physical explanation. 
As outlined in Section \ref{subsec:sev}, metals that are ejected by supernovae are distributed among the neighbouring gas cells. Since the gas is initially densest in the centre of the galaxy, this is where most stars form and hence where most supernovae occur. This larger supernova rate leads to a more metal-rich ISM in the inner galaxy and in turn to second generation stars with larger metallicities than in the the outskirts of the galaxy, explaining the observed metallicity gradient.

The stellar age and metallicity distributions
shown in Fig. \ref{fig:age_met_grad} are calculated after $3\,{\rm Gyr}$ in the CDM simulation with smooth star formation 
in which the DM halo does not form a core (see Fig. \ref{fig:dm_profiles}), i.e., the gravitational potential remains approximately constant. On the other hand, in a halo with an evolving gravitational potential, (adiabatic or impulsive) cusp-core transformation 
can alter the phase space distribution of the stars in dwarf galaxies \citep{2019MNRAS.485.1008B}. Thus, we surmise that the age and metallicity distributions will look distinctly different in our simulations in which the DM haloes' final density profiles are cored.

\begin{figure*}
    \centering
    \includegraphics[width=0.48\linewidth,trim={0.5cm 0.5cm 0.5cm 0.5cm},clip=true]{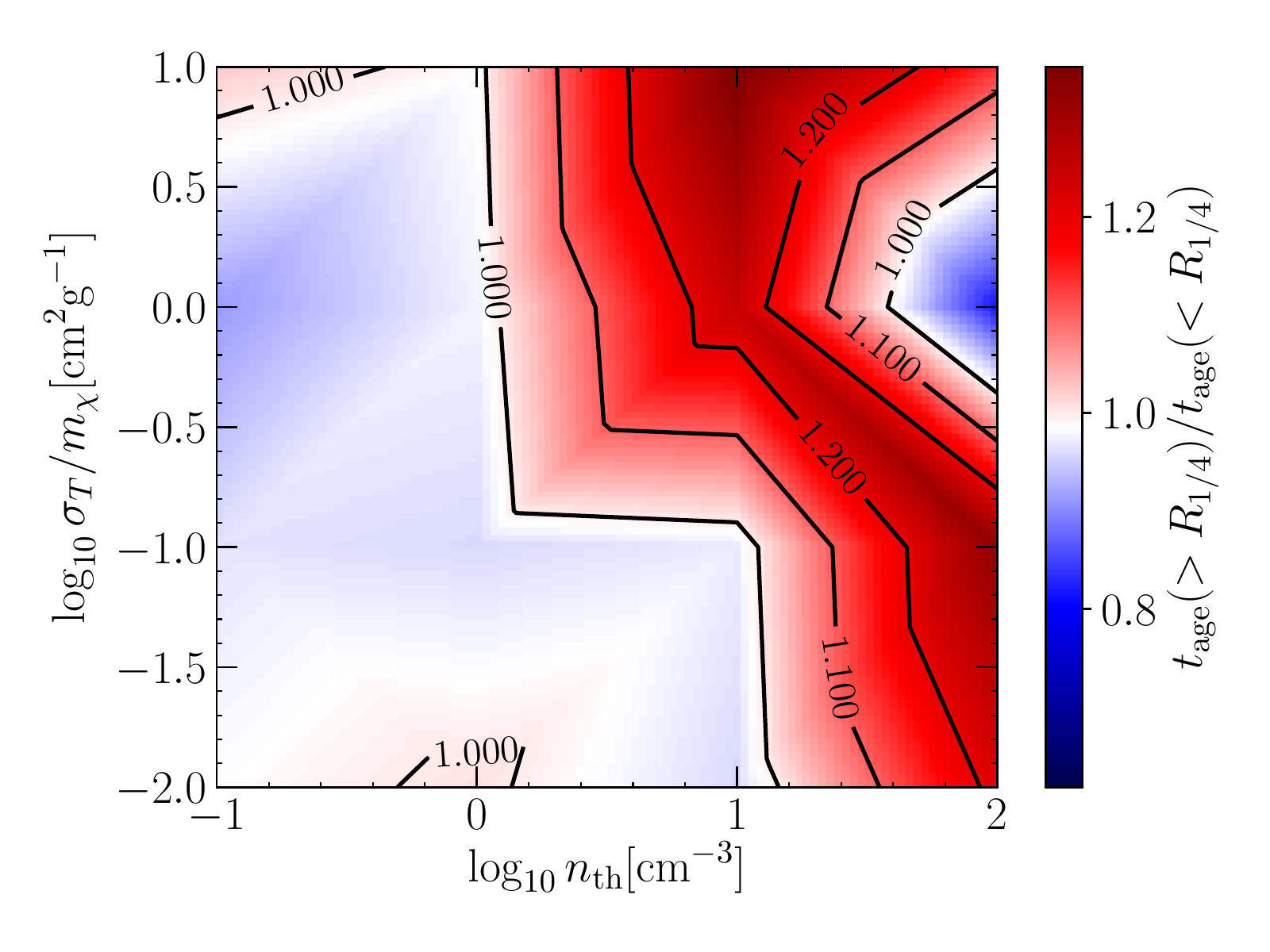}
    \includegraphics[width=0.48\linewidth,trim={0.5cm 0.5cm 0.5cm 0.5cm},clip=true]{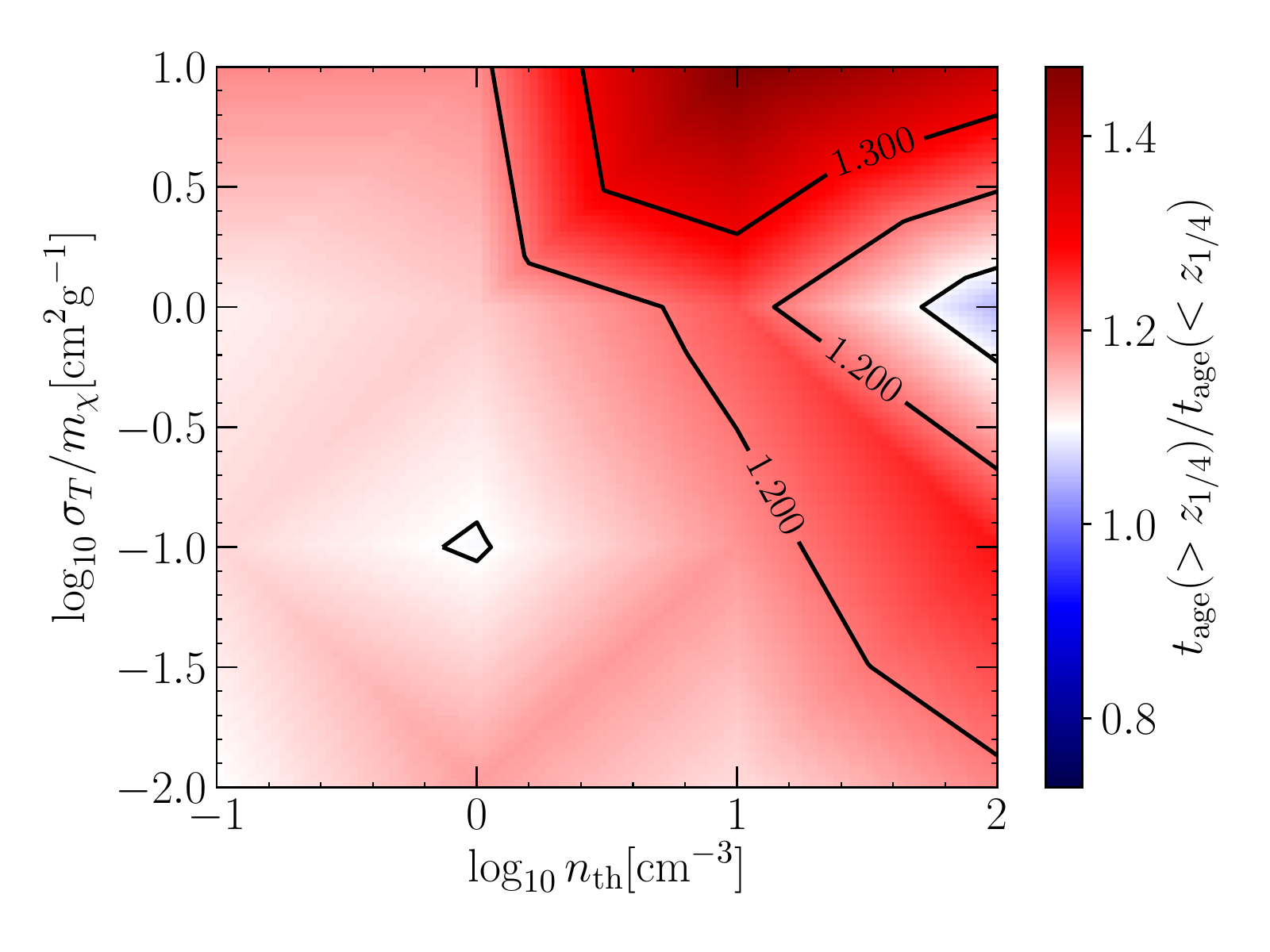}\\
    \includegraphics[width=0.48\linewidth,trim={0.5cm 0.5cm 0.5cm 0.5cm},clip=true]{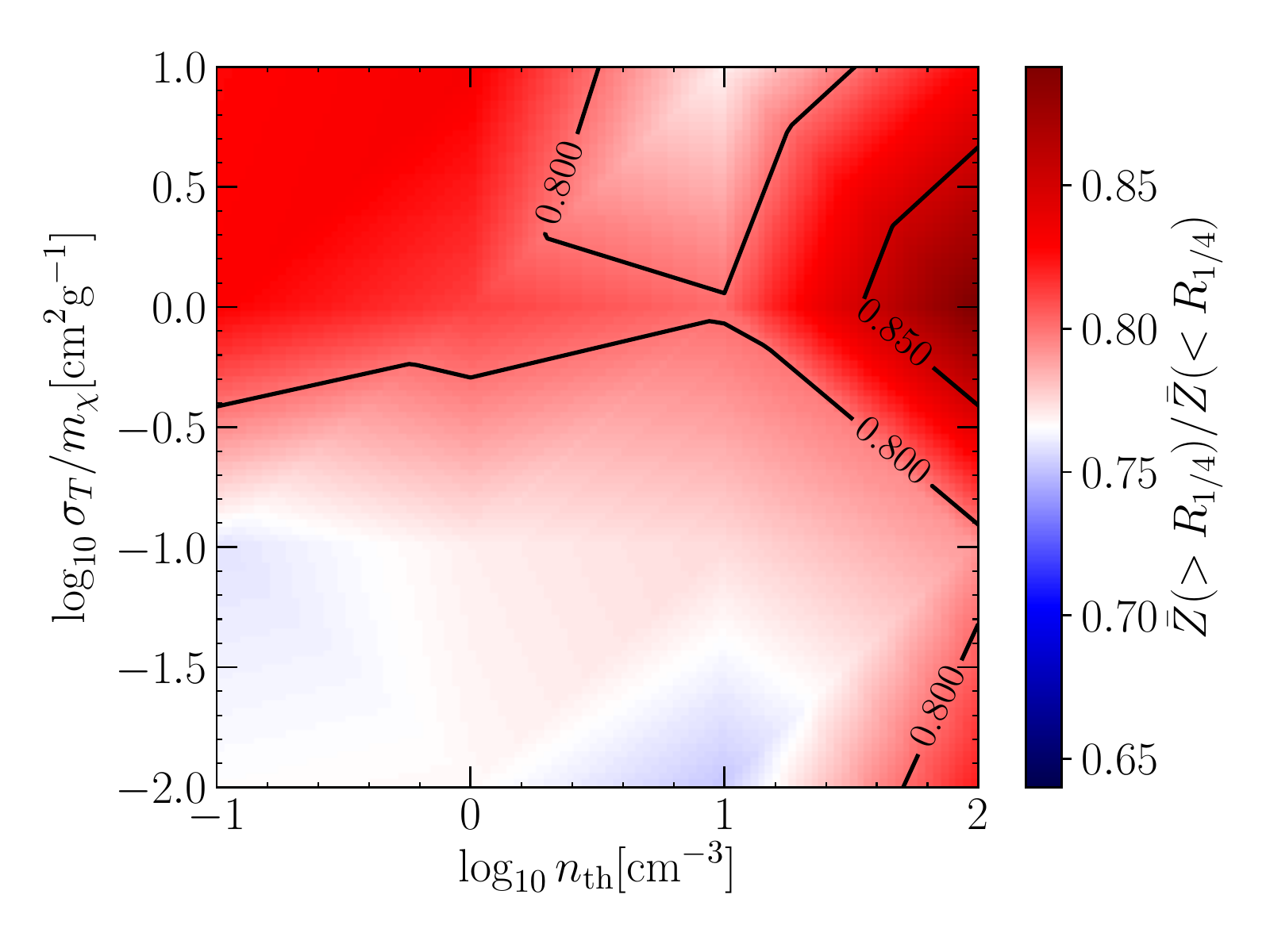}
    \includegraphics[width=0.48\linewidth,trim={0.5cm 0.5cm 0.5cm 0.5cm},clip=true]{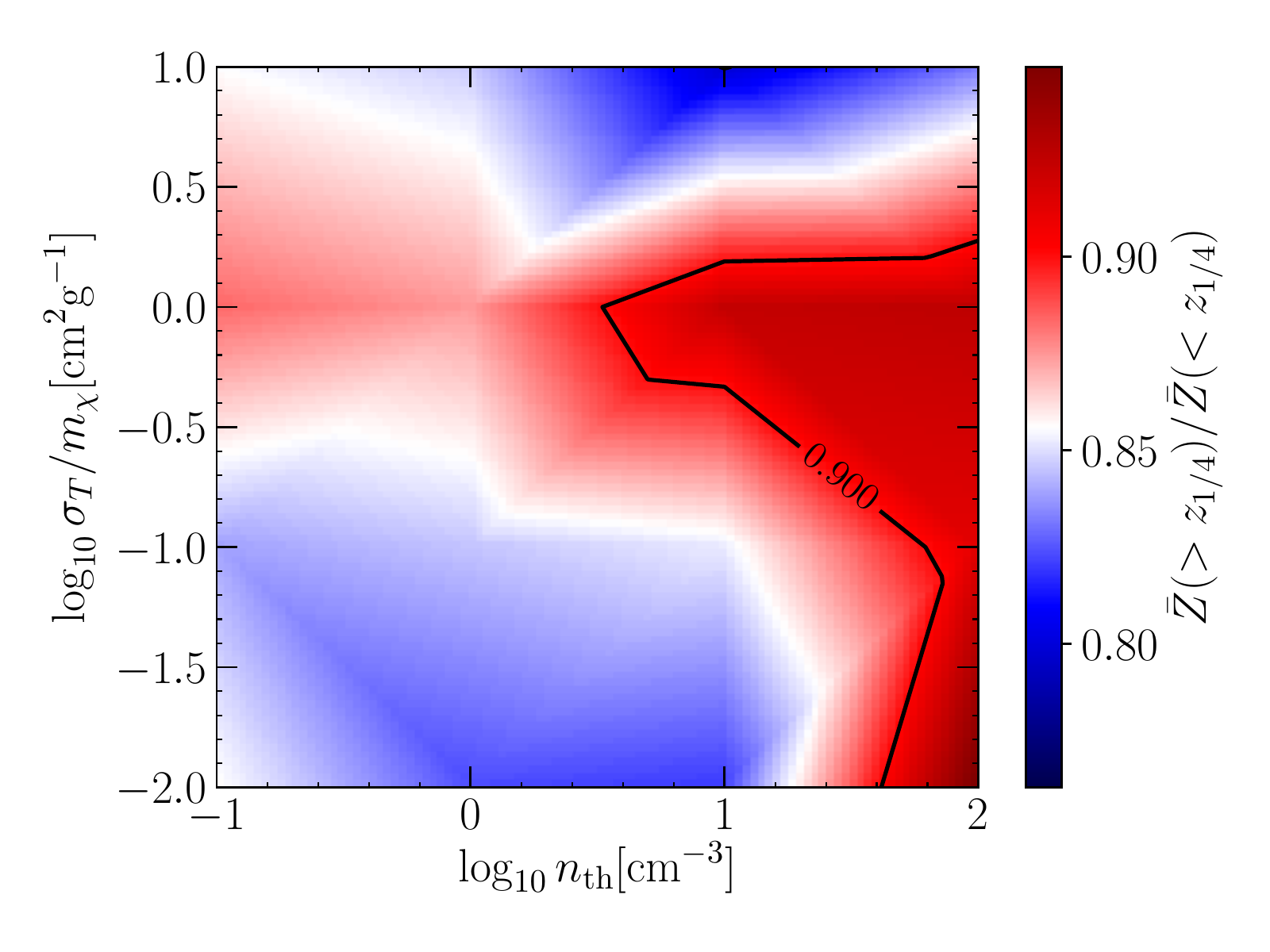}
    \caption{Different ratios characterizing age gradients and metallicity gradients in our simulations as a function of star formation threshold and SIDM transfer cross section, calculated after $3\,{\rm Gyr}$. 
    In the upper left panel, we show the ratio between average stellar ages in the outer and inner regions of the simulated galaxy across the disc plane, where the quarter mass (cylindrical) radius (of the stellar distribution) is used as the boundary. The upper right panel shows a similar plot, but vertically, perpendicular to the plane of the disc, with a boundary for inner and outer regions given by $z_{1/4}=0.2 R_{1/4}$ (see also Fig. \ref{fig:age_met_grad}). 
    The lower panels are as the upper panels but 
    for stellar metallicities instead of ages. Since the quarter mass radius lies well within the radial range that is strongly affected by core formation (see Fig. \ref{fig:dm_profiles}), these ratios are a good 
    probe of how adiabatic or impulsive core formation mechanisms affect the stellar phase space distributions.}
    \label{fig:grad_paramspace}
\end{figure*}

\begin{figure*}
    \centering
    \includegraphics[width=0.48\linewidth,trim={0.5cm 0.5cm 0.5cm 0.5cm},clip=true]{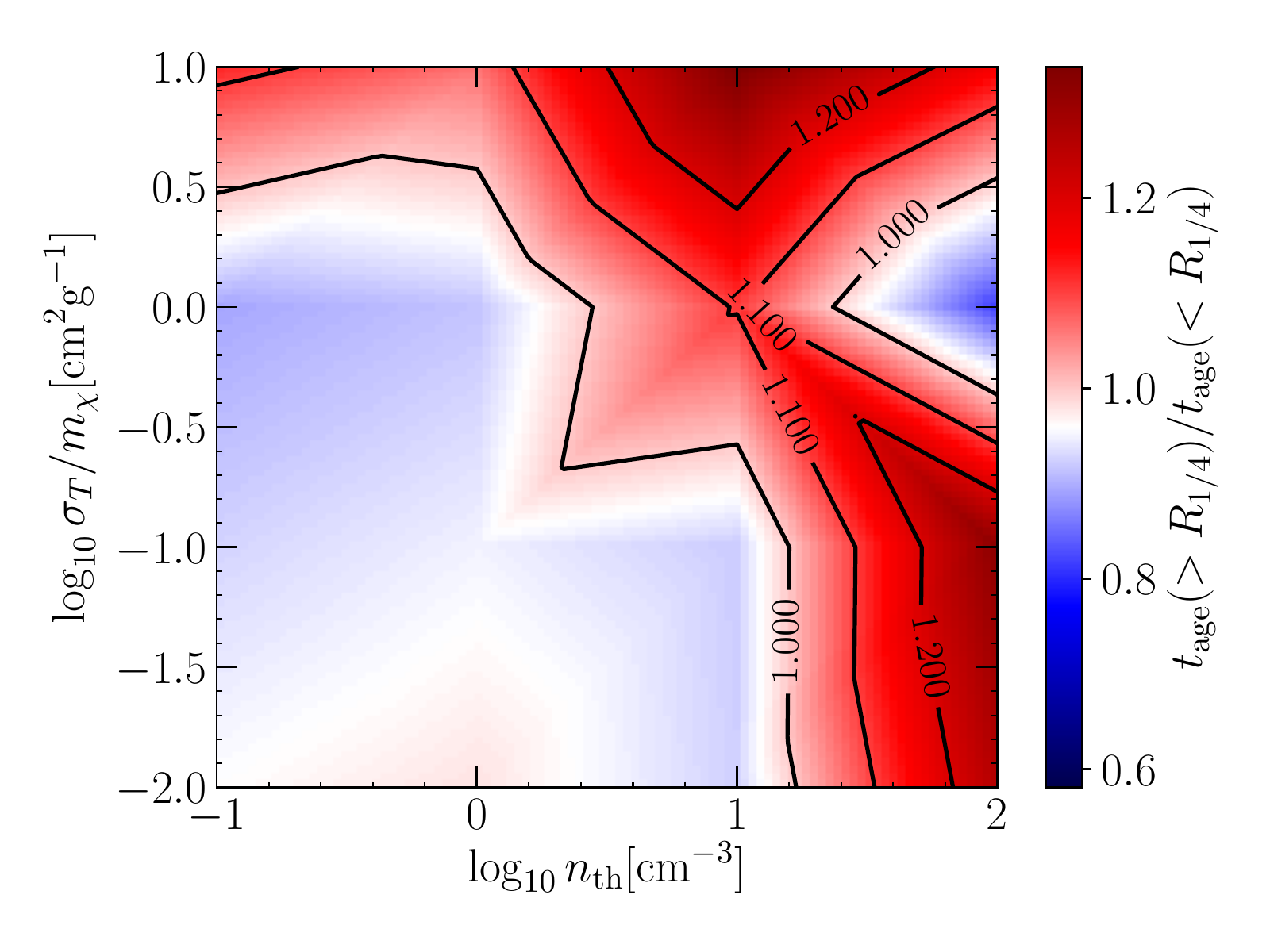}
    \includegraphics[width=0.48\linewidth,trim={0.5cm 0.5cm 0.5cm 0.5cm},clip=true]{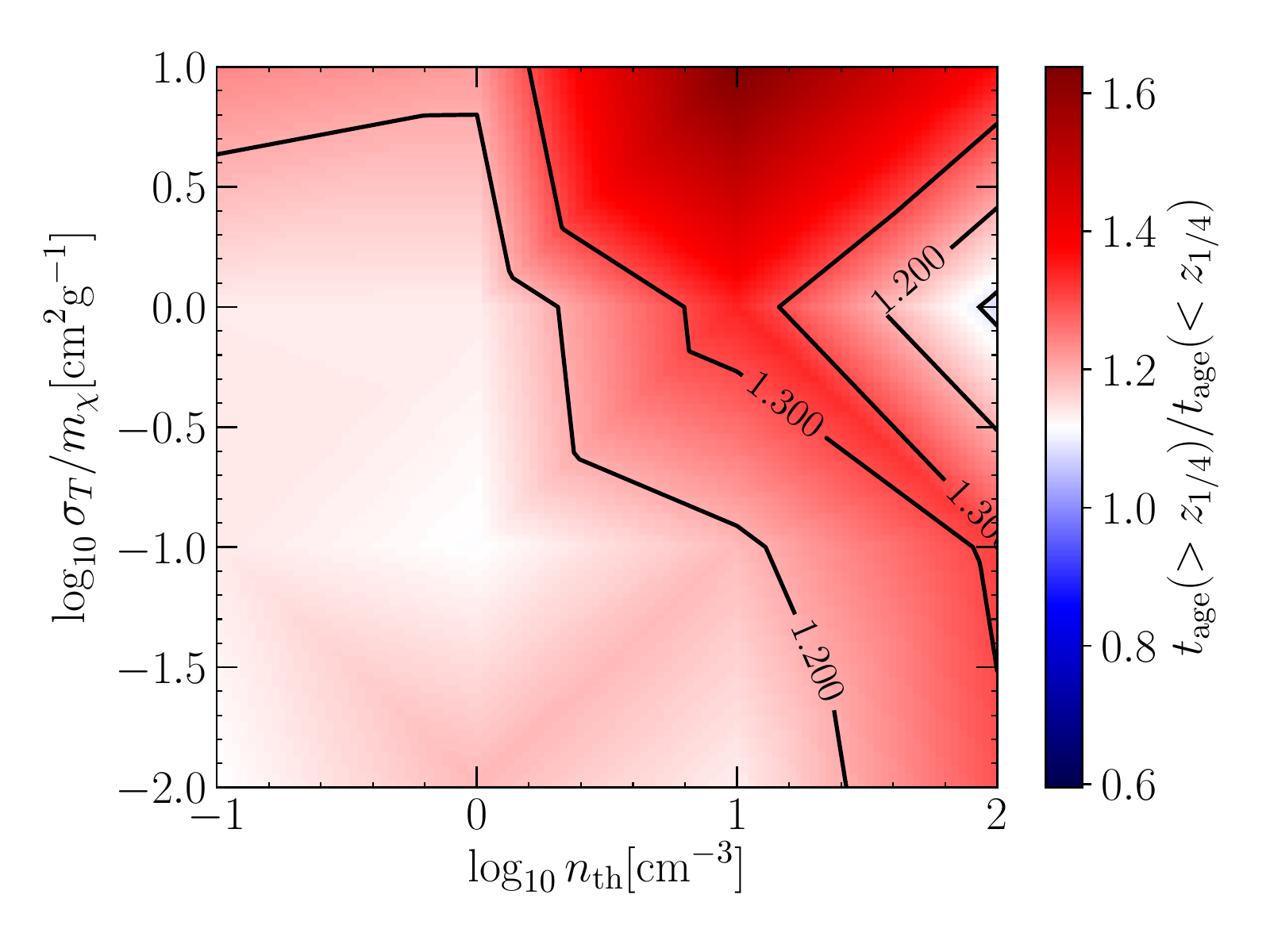}\\
    \includegraphics[width=0.48\linewidth,trim={0.5cm 0.5cm 0.5cm 0.5cm},clip=true]{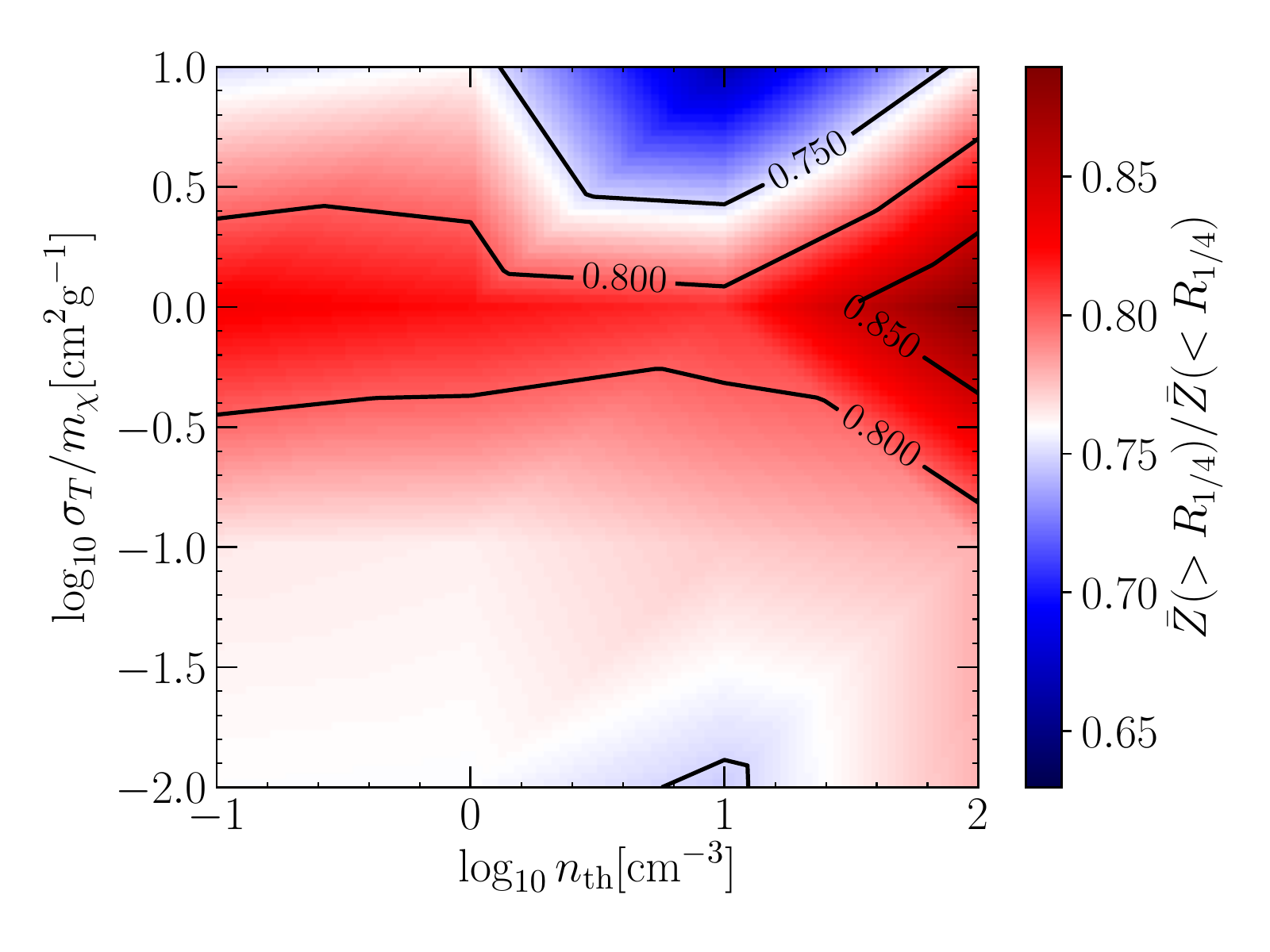}
    \includegraphics[width=0.48\linewidth,trim={0.5cm 0.5cm 0.5cm 0.5cm},clip=true]{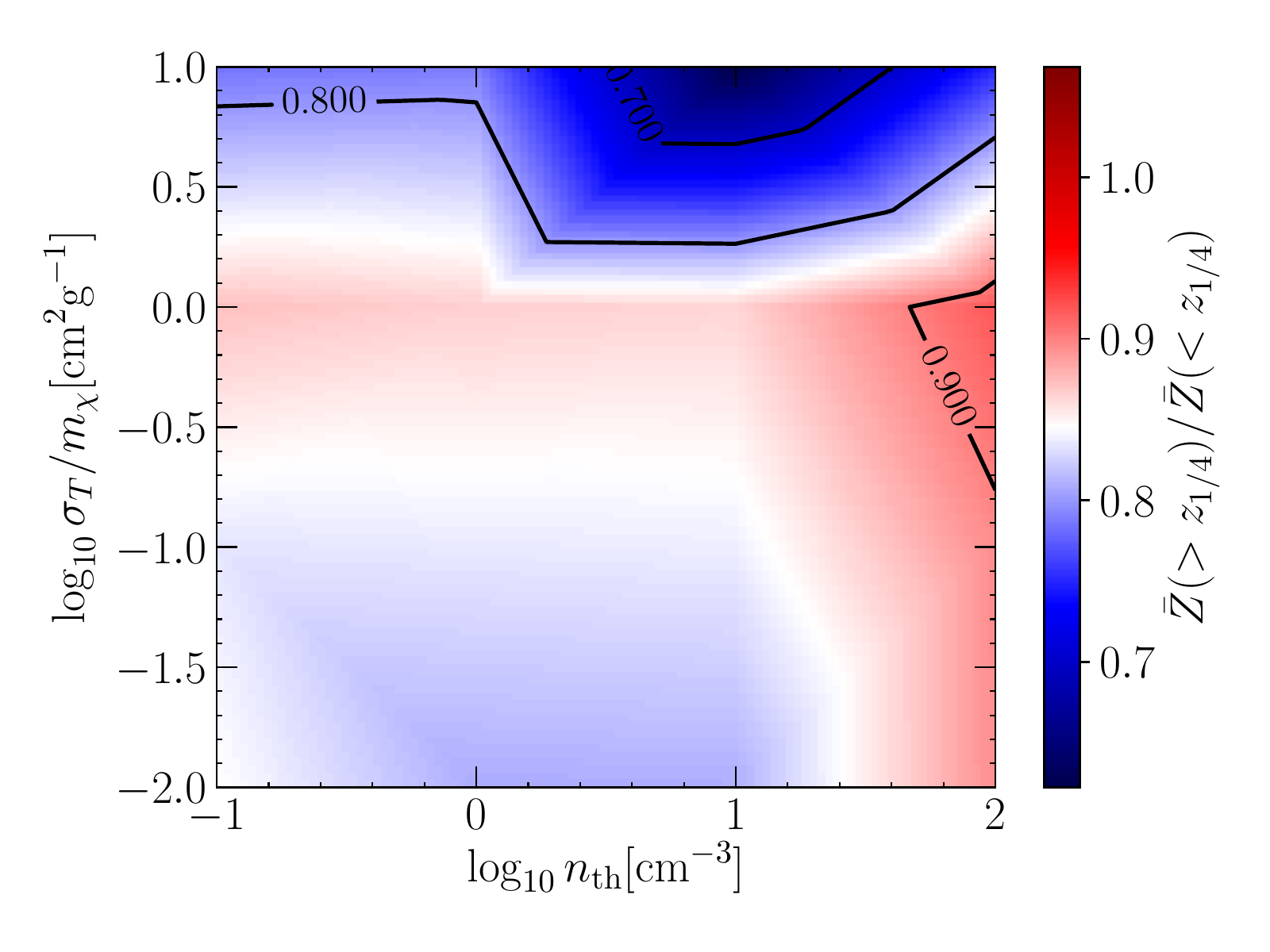}
    \caption{Same as Fig. \ref{fig:grad_paramspace}, but after $4\,{\rm Gyr}$ of simulation time.}
    \label{fig:new}
\end{figure*}

Figs. \ref{fig:grad_paramspace} and \ref{fig:new} compare the age and metallicity gradients measured after $3\,{\rm Gyr}$ and  $4\,{\rm Gyr}$, respectively, across the parameter space of the simulations.
In order to facilitate a comparison 
between 
different values of $\sigma_T/m_\chi$ and $n_{\rm th}$, we quantify the steepness of the gradients by taking the age/metallicity ratios between stars with radial/vertical distances from the centre of the galaxy that are larger than a certain characteristic scale to those that are smaller than that scale. 
Through gravity, the process that causes core formation affects the stellar distribution as well. This makes a comparison (across different simulations) of the ratio of the averaged properties of stars outside of the core to the averaged properties of stars inside the core particularly interesting. From Figs. \ref{fig:dm_phasespace} and \ref{fig:stellar_size}, we know that the enclosed mass within the stellar quarter mass radius ($r_{1/4}$) is a good proxy for whether the DM halo 
is cored or cuspy. In particular, in simulations with core formation, the stellar quarter mass radius approximately separates the central core from the rest of the halo. We therefore choose this radius as the characteristic scale to characterize the age/metallicity gradients in Figs. \ref{fig:grad_paramspace} and \ref{fig:new}. Since we are interested in the gradients along the disc plane and perpendicular to it, we define the characteristic cylindrical radius $R_{1/4}=r_{1/4}$, and the characteristic vertical scale $z_{1/4}=0.2R_{1/4}$ in line with the initial ratio between vertical and radial scale height of the disc (see also Fig. \ref{fig:age_met_grad}).

The upper left panel of Fig. \ref{fig:grad_paramspace} shows the ratio between the average age of stars with cylindrical radii larger than $R_{1/4}$ and stars with cylindrical radii smaller than $R_{1/4}$
at $t=3\,{\rm Gyr}$. 
For $n_{\rm th} \le 1\,{\rm cm^{-3}}$ this ratio is close to unity, indicating that there is no discernible age gradient around radii similar to $R_{1/4}$. In most simulations with larger star formation thresholds, we measure a significant, positive age gradient, 
particularly in those simulations with $n_{\rm th} \ge 10\,{\rm cm^{-3}}$ (bursty star formation) in which the DM halo forms a core. An exception is the run with $n_{\rm th} = 100\,{\rm cm^{-3}}$ and $\sigma_T/m_\chi = 1\,{\rm cm^2g^{-1}}$, where the age gradient is negative at the scale of the quarter mass radius: $t_{\rm age}(>R_{1/4})/t_{\rm age}(<R_{1/4}) < 1$ (see Appendix \ref{appendix_3} for further discussion). 
The general trend of simulations with larger star formation thresholds having (on average) older stars in the outer parts of the galaxy than in the inner parts can be explained by the mechanism of impulsive SNF. 

The rapid change in the gravitational potential caused by the supernova cycle triggered in early starbursts that causes the formation of the DM core also results in an outward migration of some of the stars that were present in the inner regions at the time. 
The exception seen for the simulation with $n_{\rm th} = 100\,{\rm cm^{-3}}$ and $\sigma_T/m_\chi = 1\,{\rm cm^2g^{-1}}$ is likely caused by a massive starburst occurring just before $t=1\,{\rm Gyr}$ (see Figs. \ref{fig:sfhs} and \ref{fig:sfr}, and discussion in Appendix \ref{sec:discussion}), which leads to a SN-driven gas outflow that effectively 
shuts off star formation for a long time. At later times, the supernova feedback mechanism is not energetic enough to cause older stars to migrate into the outskirts of the galaxy. The net result is that at the end of this simulation, the stellar population within $R=R_{1/4}$ is older (on average) than outside of it.   

The upper left panel of Fig. \ref{fig:new}, showing the age ratio after $4\,{\rm Gyr}$ of simulation time, essentially confirms the same picture. However, given the growing impact of the gravothermal collapse phase for large cross sections, the age gradient is noticeably different (larger than one) than for smaller cross sections, even for low star formation thresholds (non-impulsive supernova feedback).

The upper right panels of Figs. \ref{fig:grad_paramspace} and \ref{fig:new} show a similar ratio as in the upper left panel, but in this case perpendicular to the plane of the disc instead of along it. Displayed is the average age of stars with $|z| > z_{1/4}$ divided by the average age of stars with $|z| < z_{1/4}$. 
At both times shown in Figs. \ref{fig:grad_paramspace} and \ref{fig:new},
the age gradient in the vertical direction (upper right panels) shows a very similar pattern as in the (cylindrical) radial direction on the upper left panels.  

The lower left panels of Figs. \ref{fig:grad_paramspace} and \ref{fig:new} show the ratio of average stellar metallicities outside of $R_{1/4}$ to average stellar metallicities within $R_{1/4}$ after $3\,{\rm Gyr}$ and $4\,{\rm Gyr}$, respectively. After $3\,{\rm Gyr}$, this ratio is a mostly featureless constant across the parameter space of the simulations, 
except for a slightly reduced metallicity gradient in the simulation with $\sigma_T/m_\chi = 1\,{\rm cm^2g^{-1}}$ and $n_{\rm th} = 100\,{\rm cm^{-3}}$, where star formation is strongly reduced for a while and almost no stars with high metallicities are formed (see Appendix \ref{appendix_3}). The spatial metallicity distribution is fairly even in this case, compared to the other simulations. In general, metallicity gradients are 
negative, since the ISM is more metal-rich in the centre of galaxies. 

This is reflected in metallicity ratios (measured with respect to the quarter mass radius) which are smaller than 1.  
In simulations with $\sigma_T/m_\chi = 1\,{\rm cm^2g^{-1}}$, where the DM halo forms an isothermal core due to DM self-interactions, the observed metallicity gradients (along the cylindrical radial direction) are shallower than in other simulations after 4 Gyr of simulation time. In Fig.~\ref{fig:new}  
a clear difference emerges between simulations in which the DM haloes have formed cores adiabatically, and 
simulations in which they have not. A potential explanation for this arises from Fig. \ref{fig:stellar_size}. Galaxies that are hosted by haloes with SIDM-induced cores have 
stellar distributions in which the central density of stars is smaller (i.e. the galaxies are less compact) than in the other simulated galaxies. Thus, the ejected metals from SNF 
are more evenly distributed within a larger volume around the centre of the galaxy. 

Finally, the lower right panel of Fig. \ref{fig:grad_paramspace} (Fig. \ref{fig:new}) shows the ratio between the average metallicity of stars with $|z| > z_{1/4}$ and stars with $|z| < z_{1/4}$ after $3\,{\rm Gyr}$ ($4\,{\rm Gyr}$). 
At both times we find that this ratio is closer to unity in simulations in which a shallow core has formed than in other simulations, and slightly more so in simulations with impulsive SNF. However, the difference between cored galaxies with smooth star formation and cored galaxies with bursty star formation is not very pronounced. Thus, while shallow vertical metallicity gradients indicate the presence of a core, they are not as useful to identify the predominant core formation mechanism as the radial age gradients discussed above. 

Our main observations can be summarized as follows. Moderate, but significant, positive radial stellar age gradients 
appear in simulations in which supernova feedback is impulsive. Observing them in galaxies whose DM haloes have cored density profiles does not rule out SIDM as the mechanism responsible for the formation of the DM cores, but it suggests that supernova feedback is the dominant 
mechanism of cusp-core transformation.  
Shallow vertical metallicity gradients (in comparison to other simulations) are a characteristic feature of galaxies with shallow DM cores. However, this is true regardless of the core formation mechanism. The effect is larger if SNF is impulsive, but vertical metallicity gradients can likely not be used to differentiate between SIDM and SNF. 
Vertical age gradients and radial metallicity gradients exhibit less obvious features than radial age gradients and vertical metallicity gradients.

Finally, we stress that our choice to characterize the age/metallicity gradients through the age/metallicity ratios as outlined above was motivated by the connection of $R_{1/4}$ and $z_{1/4}$ to the size of the cores that form in our simulations. However, it is also possible to quantify the gradients in a way that is independent of any particular scale. One alternative way to quantify the grandients is explored in Appendix \ref{app_addfigs}. Fig.~\ref{fig:newgradients} is similar to Fig.~\ref{fig:grad_paramspace}, but we show the slope of a linear fit to the ages of stars as a function of cylindrical radius ($\Delta t_{\rm age}/\Delta R$, which we calculate using the stellar particle masses as weights) instead of the age ratio, and make equivalent replacements for the other three panels. We find that the key features of Fig.~\ref{fig:grad_paramspace} are unchanged, in particular when considering the quantities with the most pronounced features, i.e., radial age gradients and vertical metallicity gradients.

\section{Summary}\label{sec:sum}

We explored the differences between two core formation mechanisms that can, under certain conditions, lead to constant density cores of near identical size in the DM host haloes of dwarf galaxies. 
Particularly, we focused on 
how the dynamical properties of gas and stars in dwarf galaxies might be affected by either impulsive (SNF driven) or adiabatic (SIDM driven) core formation. 

To that end, we performed a suite of 16 high-resolution hydrodynamical simulations, evolved for $4$~Gyr, of an idealized 
SMC-size galaxy embedded within a live DM halo with an initially cuspy Hernquist density profile. 
Our simulations included both a stellar evolution and feedback prescriptions using the \texttt{SMUGGLE} model \citep{2019MNRAS.489.4233M} and self-interactions between the DM particles \citep{Vogelsberger2012}, all within the framework of the \texttt{AREPO} code \citep{Springel:2009aa}. We present a detailed description of core formation in \texttt{SMUGGLE}  for CDM haloes in a companion paper (\citealt{2021arXiv211000142J}) and focus here on the comparison between dark matter cores formed due to SNF versus due to dark matter self-interactions, all evolved with the same \texttt{SMUGGLE} baryonic treatment. 

Starting from identical initial conditions, each simulation was performed with a different combination of SIDM momentum transfer cross section ($\sigma_T/m_\chi$) and star formation threshold ($n_{\rm th}$). Through these two parameters we controlled the efficiency of the SIDM-driven and SNF-driven mechanisms of cusp-core transformation, respectively. The values of these parameters 
were chosen in order to probe star formation regimes from smooth (low $n_{\rm th}$) to bursty (high $n_{\rm th}$, see Fig. \ref{fig:sfhs}), as well as to probe the regimes from collisionless DM (CDM; $\sigma_T/m_\chi=0$) to strong self interactions as large as $10\,{\rm cm^2g^{-1}}$. 
We showed for which combinations of self-interaction cross section and star formation threshold, the initially cuspy halo develops a $\mathcal{O}(1)$~kpc size constant density core (see Figs. \ref{fig:dm_phasespace}, \ref{fig:vc_4gy}, and \ref{fig:app_fig_one}). In particular, we identified a degenerate line in the $\sigma_T/m_\chi-n_{\rm th}$ parameter space plane along which the final simulated DM haloes are cored. 

Moreover, we found that adiabatically formed cores (SIDM cores) tend to be fully isothermal, while those formed 
through impulsive SNF are not, at least within the timescales of our simulations. To be more precise, our results indicate that SIDM cores fully thermalize significantly faster than those formed through SNF (see bottom panel of Fig. \ref{fig:dm_phasespace}).  
SIDM haloes with $\sigma_T/m_\chi = 10\,{\rm cm^2g^{-1}}$ undergo gravothermal collapse after $\sim 2.5$ Gyr in our simulations. 
Their density profiles are cored for a short while before they collapse and form very steep central density cusps (see Figs. \ref{fig:dm_phasespace} and \ref{fig:vc_4gy}). 

To differentiate between SIDM and SNF as core formation mechanisms, we compare several observable quantities between simulations. A few clear trends emerge. Galaxies within cored host haloes form 
extended stellar distributions
that follow the gravitational potential of the host halo if {\it i}) the core was formed adiabatically through SIDM and {\it ii}) star formation is smooth 
instead of bursty, i.e., SNF is not impulsive (Fig. \ref{fig:stellar_size}). Impulsive SNF can cause positive stellar age gradients (Figs. \ref{fig:grad_paramspace}, \ref{fig:new}, and \ref{fig:newgradients}) and increased random motion in the gas (Fig. \ref{fig:vbs_phase}). Ubiquitous observations of turbulent gas or positive stellar age gradients within cored DM haloes would therefore suggest that impulsive SNF has caused the cusp-core transformation. 

The vertical metallicity gradients of stars 
in cored haloes are systematically shallower than the vertical metallicity gradients of stars in haloes that remain cuspy (Figs. \ref{fig:grad_paramspace}, \ref{fig:new}, and \ref{fig:newgradients}). This feature is slightly more pronounced in haloes with SNF-induced cores, but the difference to haloes with SIDM-induced cores is likely insufficient to use vertical metallicity gradients to differentiate between the two. The dynamical properties of galaxies embedded in haloes which have undergone gravothermal collapse are systematically different from the dynamics of all other simulated galaxies. 

The parameter space plots in our results are to be indicative of general trends, not precise predictions, due to the fact that our simulations are idealized. In particular, our specific choice of initial conditions enabled us to compare the structural properties of the baryons between galaxies whose host haloes formed a core either through SNF or SIDM (or not at all) -- simply by varying $n_{\rm th}$ and $\sigma_T/m_\chi$. Assuming a substantially lower initial DM to baryon ratio in the centre of the galaxy will lead to universally burstier star formation histories (Fig. \ref{fig:c12_sfhs}), which in turn results in the formation of constant density cores for all 16 combinations of $n_{\rm th}$ and $\sigma_T/m_\chi$ (Fig. \ref{fig:rv_log_slopes}). None the less, the largest (positive) radial age gradients still correspond to the simulations with the burstiest star formation histories, and combining observations of age gradients and galaxy sizes may still allow us to differentiate between galaxies with cores that have formed predominantly because of SNF -- or SIDM (see Fig. \ref{fig:rv_bar} and related discussion). When moving beyond our idealized setup, we expect that the final structural and dynamical properties of dwarf galaxies will not only depend on the strength of SIDM and SNF, but also on their initial DM to baryon ratios -- as well as their dynamical histories.  

To obtain an accurate quantitative understanding of the degeneracies/interplay between SNF and SIDM cross section in the inner structure of dwarf-size haloes, as well as the role of mergers, a comprehensive exploration of the $(n_{\rm th},\sigma_T/m_\chi)$ parameter space in a full cosmological setting is required. Although such an undertaking is computationally expensive, requiring large suites of cosmological simulations with sufficiently high resolution, our idealized runs strongly suggest it would be fruitful, leading to detailed predictions regarding the properties of the visible components (gas and stars) that are truly distinct between these mechanisms of cusp-core transformation based on either baryonic physics or new DM physics.

On the observational front, searching for the trends we have found in this work could prove to be quite significant to 
understand how dwarf-size DM haloes develop cores. For instance, if positive age gradients were observed in (the central region of) most 
dwarf galaxies with cored host haloes, SNF would likely be impulsive. This would strongly suggest that SNF is the main
mechanism that drives the cusp-core transformation in these galaxies.
Finally, studies of the dynamical properties of kinematic tracers (e.g. \citealt{2019MNRAS.485.1008B}) may reveal whether SNF is impulsive enough to be a feasible mechanism of cusp-core transformation, provided we have a way of identifying orbital families of stars in observational data.

\section*{Data Availability}
The data underlying this article were accessed from the Garpur supercomputer. The derived data generated in this research will be shared on reasonable request to the corresponding author.

\section*{Acknowledgments}

JB and JZ acknowledge support by a Grant of Excellence from the Icelandic Research Fund (grant number
173929). The simulations in this paper were carried out
on the Garpur supercomputer, a joint project between
the University of Iceland and University of Reykjav\'ik
with funding from the Icelandic Research Fund. LVS acknowledges support from NSF and NASA through grants AST-1817233, CAREER 1945310 and NASA ATP 80NSSC20K0566.
PT acknowledges support from National Science Foundation (NSF) grants AST-1909933, AST-2008490 and National Aeronautics and Space Administration (NASA) Astrophysics Theory Program (ATP) grant 80NSSC20K0502.

\bibliographystyle{mnras}
\bibliography{manuscript}

\appendix

\section{Caveats}\label{sec:discussion}
In this article, we have presented a suite of 16 high-resolution hydrodynamical simulations of an isolated SMC-size dwarf galaxy in a live DM halo. Across simulations, we have changed the star formation threshold $n_{\rm th}$, regulating the ``burstiness'' of star formation and hence the ability of SNF to drive large-scale gas outflows that rapidly change the gravitational potential. Furthermore, we have tested the impact of 
self-interactions between the DM particles 
by changing the momentum transfer cross section per unit mass across simulations. By exploring this two-dimensional parameter space, we can study the two most viable mechanisms of core formation in DM haloes: adiabatic due to the impact of SIDM, and impulsive due to SNF. We have then searched for differences 
in the distribution function of gas and stars across the simulations. In the following, we briefly discuss some caveats of our analysis. In particular, we outline how a different choice of initial conditions, and of the values of the gravitational softening may impact our results. Moreover, we discuss how to interpret the results presented in Section \ref{sec:results} in light of the inherent stochasticity of the ISM and stellar feedback model, and the fact that a single snapshot (time output) of each simulation in our simulations suite corresponds to only a single possible realization of the evolved distribution function.

\subsection{Initial conditions} \label{app_ics}

Since our simulations are not cosmological, the simulated halo has no cosmological assembly 
history. In cosmological simulations, haloes and galaxies form from initial conditions that are not arbitrary, but (statistically) fixed by the assumed cosmological model and constrained on large scales by the observed perturbations in the cosmic microwave background, which set the cosmic fractions of dark matter and baryons. The relative amount of DM or baryons in a galaxy is thus a prediction of cosmological simulations which is obtained from the full structure formation and evolution process, coupled to the baryonic physics model, and not an initial condition. With our choice of initial conditions, we aim to mimic an isolated virialized dynamical system that is similar (in scale) to the Small Magellanic Cloud, with initial structural parameters as in \citet{2012MNRAS.421.3488H}. 

With our choice of initial conditions, the structure of the halo and the galaxy, as well as the relative amounts of DM, gas, and stars are fixed. Since the efficiency of DM self-interactions as a mechanism of cusp-core transformation depends on the density of DM in the central halo, whereas the efficiency of SNF depends on the relative amount of baryonic matter in the central galaxy, our choice of initial conditions can thus have a large impact on the efficiency of SIDM or SNF as mechanisms of cusp-core transformation
For that reason, our isolated simulations can make no definitive quantitative statements about 
the exact properties (including formation timescales) of DM cores formed through these mechanisms in realistic dwarf-size haloes formed in a cosmological setting. 

For SIDM, the observed core formation for $\sigma_T/m_\chi \sim \,1\,{\rm cm^2g^{-1}}$ (see middle panels of Fig. \ref{fig:dm_profiles}) is approximately  
in agreement with cosmological simulations of SIDM haloes (see e.g. \citealt{Vogelsberger2012} and \citealt{Vogelsberger2014}). We therefore conclude that the predictions of our SIDM simulations, including the timescales for core formation and the adiabatic nature od the cusp-core transformation, are reasonable.  

For SNF, the situation is more complex. Most hydrodynamical cosmological simulations find that SNF can form cores only in the mass range of bright dwarfs (see e.g. \citealt{2020MNRAS.497.2393L}). However, \citet{Read2016} find, based on high-resolution simulations of isolated haloes, that SNF can form cores event at the scale of ultra faint galaxies. Crucially, it is unclear whether the results of cosmological simulations are more correct in this regime or not. The baryon fractions assumed by \citet{Read2016} are rather large, and it is possible that they cannot be realized in haloes with a cosmological formation history. However, due to the high resolution of \cite{Read2016}'s simulations, the effects of SNF can be modeled much more accurately than in large cosmological simulations. 
using modern ISM and stellar feedback models. 
As far as our initial conditions are concerned, it is worth mentioning that the assumed stellar mass of our SMC-size system 
is at the upper end of what is allowed by abundance-matching results for the stellar-to-halo mass relation (see e.g. \citealt{2010ApJ...710..903M,Behroozi:2019}). In principle, this may imply increased efficiency of SNF. However, our galaxy is also very extended initially, and hence star formation is not very concentrated towards the centre of the halo. Moreover, the halo is initially very concentrated, which increases the central DM to gas ratio. Both of these features result in a smoother star formation history and have an adverse effect on the core formation efficiency of SNF (\citealt{2021arXiv210301231B}). In Appendix \ref{appc}, we investigate the impact of decreasing the initial DM to gas ratio in the dwarf's centre by repeating our simulation suite, starting from initial conditions in which we have assumed a lower initial halo concentration ($c_{200} = 12$). 

\subsection{Gravitational softening and concentration of baryons}

The choice of the gravitational softening length can significantly affect how DM haloes respond to SNF with a large star formation threshold. \citet{2020MNRAS.499.2648D} suggest that for large star formation thresholds, small softening lengths need to be adopted in order for SNF to efficiently transform cusps into cores. For collisionless simulations of gravitationally self-bound haloes, \citet{Power:2002sw} have conducted a convergence study and derived an optimal force softening length 
\begin{equation}
       \epsilon_{\rm opt} = 4\frac{r_{\rm 200}}{\sqrt{N_{\rm 200}}},
\end{equation}
where $r_{200}$ is the halo's virial radius and $N_{200}$ is the number of DM simulation particles contained within $r_{200}$. 
For our simulated halo, this corresponds to
$\epsilon_{\rm opt} \sim 50\,{\rm pc}$. 

In hydrodynamical simulations, however, the choice of the force softening length is less clear. Before running our final simulation suite, we tested different simulation settings. In particular, we ran the CDM simulation with $n_{\rm th}  =100\,{\rm cm^{-3}}$ with different choices for the gravitational softening length and found that the cusp-core transformation does not occur for $\epsilon = \epsilon_{\rm opt}$. Our final choice of $\epsilon = 24\,{\rm pc} \sim 0.5\,\epsilon_{\rm opt}$ is on the higher end of the force softening lengths for which core formation does occur in the CDM run with $n_{\rm th}  =100\,{\rm cm^{-3}}$. The reason for this dependence on the force softening is simple. In runs with larger star formation thresholds, gas needs to be denser for stars to form. Hence, larger gravitational forces need to be resolved on small scales, for which smaller softening lengths are required. 

Apart from the gravitational softening length, how baryonic matter is initially distributed within the inner DM halo can also significantly change the impact of SNF on the inner DM distribution. 
We performed a CDM test run with 
$n_{\rm th} = 100\,{\rm cm^{-3}}$ in which we omitted the stellar bulge when setting up the initial conditions (see Section \ref{subsec:ICs}). In this simulation, significantly less stars formed compared to the simulation in which the stellar bulge is included in the initial conditions. Moreover, star formation was less concentrated towards the centre of the halo. As a consequence, the DM density profile remained cuspy. In part, this is explained simply by the reduction in star formation which inevitably means reduced SN activity. Additionally, SNF which is less concentrated is less effective at forming cores (\citealt{2021arXiv210301231B}). 

The question then arises why the inclusion of the stellar bulge changes the picture this much, despite it accounting for only half a per cent of the total baryonic mass in the simulation. The answer must be that without it, the gradient of the gravitational potential in the inner halo is too shallow, since both the stellar and the gaseous disk are very extended and have no appreciable density gradient towards the centre. Including the very concentrated bulge generates a steeper gradient in the central potential, and causes cooling gas to fall into the centre and reach the large densities required for star formation.        

We therefore stress that 
our results should not be understood as absolute predictions. Instead, our goal is to study in controlled/idealized simulations the (key) parameter space of the two cusp-core transformation mechanisms: SNF and DM self-interaction, and explore the similarities and differences between DM cores formed in these scenarios. 
 
\begin{figure*}
    \centering
    \includegraphics[width=0.49\linewidth,trim={0.5cm 0.5cm 0.5cm 0.5cm},clip=true]{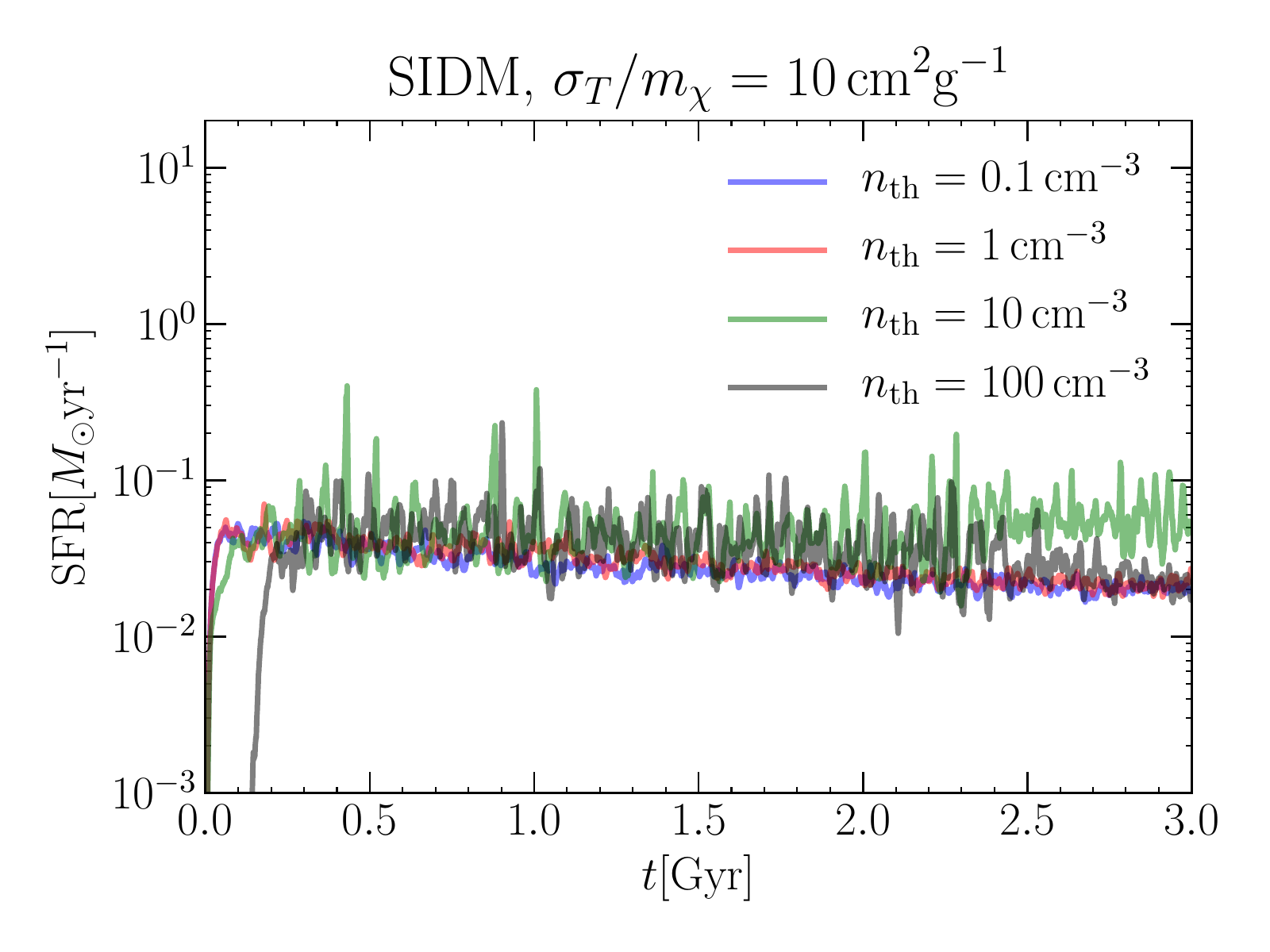}
    \includegraphics[width=0.49\linewidth,trim={0.5cm 0.5cm 0.5cm 0.5cm},clip=true]{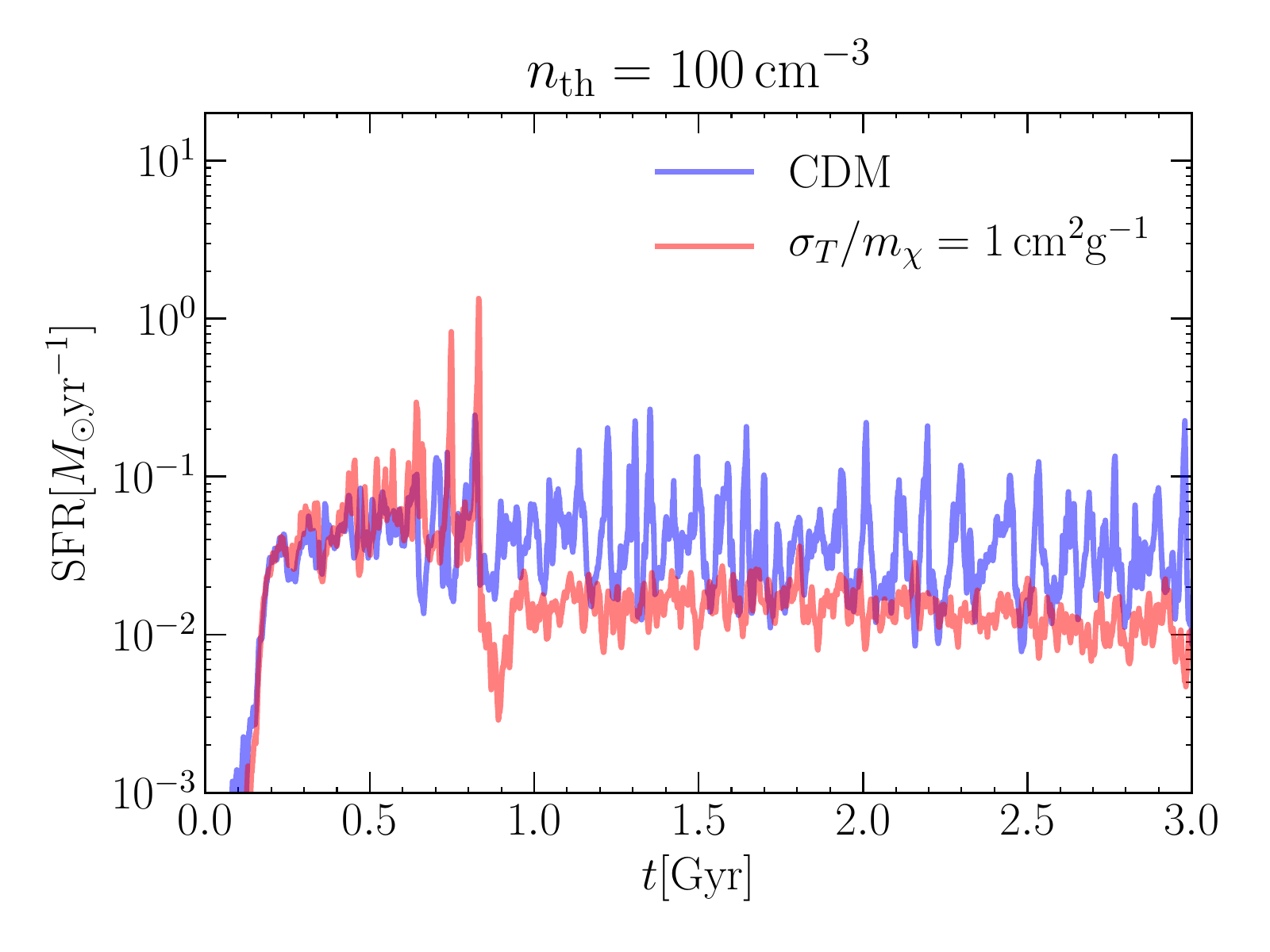}
    \caption{Star formation rate as function of time (star formation history) for different simulations. On the left panel we show the star formation histories for all SIDM simulations with $\sigma_T/m_\chi = 10\,{\rm cm^2g^{-1}}$. The star formation thresholds are as indicated in the legend. Notice the bursty star formation history of the simulation with $n_{\rm th} = 10\,{\rm cm^{-3}}$, with four star burst events during the first gigayear. 
    On the right panel, we compare the star formation histories of two simulations with $n_{\rm th} = 100\,{\rm cm^{-3}}$. The blue line corresponds to the CDM run, whereas the red line is the run with $\sigma_T/m_\chi = 1\,{\rm cm^2g^{-1}}$. Star formation in the latter is strongly suppressed at $\sim 0.9\,{\rm Gyr}$ after a few strong starbursts.} 
    \label{fig:sfr}
\end{figure*}

\subsection{Interpreting our results} \label{appendix_3}

Most of our results are presented 
in the $n_{\rm th}-\sigma_T/m_\chi$ parameter space, bilinearly interpolated from the 16 simulations in our suite with each interpolation point corresponding to an estimate of the outcome of an actual simulation in that point in parameter space.
There is one significant caveat to this way of presentation which affects how our results should be interpreted. We use hydrodynamical simulations 
to evolve the distribution function of a self-gravitating system in time, starting from well-defined initial conditions. Each snapshot that is taken at a later time corresponds to a single realization 
of the distribution function at that time, when in reality an ensemble average of different realizations would be required to determine the most likely evolved state of the system. 

In DM-only simulations, this is not a big issue since gravity is fully deterministic in the sense that different DM-only simulations will produce essentially the same results, above the (coarse-grained) scale resolved in the simulations. 
However, the stellar evolution model used in the suite of hydrodynamical simulations presented in this article is stochastic. In particular, star formation and supernovae are implemented as probabilistic random processes (see Sections \ref{subsec:sev}). Therefore, two simulations defined by identical parameters and starting from the same initial condition are not guaranteed to produce the same results. Instead, the evolved state of a system at a fixed time may differ between such simulations. It is thus possible that some of the signatures presented in Section \ref{sec:results} may be statistical outliers whose occurrence is related to the stochastic implementation of star formation and supernova feedback.

Within our suite of 16 simulations, two of them yield results that are ``peculiar'' when compared to the other simulations. The first one is the simulation in which $\sigma_T/m_\chi = 10\,{\rm cm^2g^{-1}}$ and $n_{\rm th} = 10\,{\rm cm^{-3}}$. Here, the onset of the gravothermal collapse phase is triggered earlier 
than in the other simulations with the same self-interaction cross section. As a result, essentially all observables that we consider in this article are very different between this simulation and all other simulations, in particular after $4\,{\rm Gyr}$ of simulation time (see Fig.~\ref{fig:vc_4gy} for the rotation curve (see also Fig.~\ref{fig:app_fig_one}), the lower panel of Fig. \ref{fig:vbs_phase} for the gas random motion, and Fig. \ref{fig:new} for age and metallicity gradients). Particularly striking are the steep age and metallicity gradients of the stars in this simulation.

The second simulation whose results are strikingly different from that of the other runs is the one with $\sigma_T/m_\chi = 1\,{\rm cm^2g^{-1}}$ and $n_{\rm th} = 100\,{\rm cm^{-3}}$. In particular, it is the only simulation in which SNF is impulsive and the age gradient of stars in the simulated galaxy is negative: on average, stars at larger radii are younger than at smaller radii. This is a rather unexpected feature, since stars should form at all radii all the time and we expect that impulsive SNF causes a migration of older stars from the centre into the outskirts of the galaxy, an expectation that is generally confirmed by the results of all other simulations in which SNF is impulsive (see Figs. \ref{fig:grad_paramspace} and \ref{fig:new}). 

In Fig. \ref{fig:sfr}, we attempt to explain the behaviour observed in those two simulations by looking at their star formation histories. On the left panel, we compare the star formation histories of all the simulations with very large SIDM momentum transfer cross section ($\sigma_T/m_\chi = 10\,{\rm cm^2g^{-1}}$. Interestingly, we find that star formation is very bursty early on in the simulation with $n_{\rm th} = 10\,{\rm cm^{-3}}$. In particular, we can identify four strong star burst events within the first Gyr of the simulation. Overall, star formation in this run is even burstier than in the simulation with $n_{\rm th} = 100\,{\rm cm^{-3}}$. While we do not know the exact reason for this behaviour, we argue that it implies that baryons are very concentrated towards the centre of the galaxy in the beginning of the simulation. The early gravothermal collapse observed in this simulation, along with all the ``odd'' signatures outlined above, is thus likely the result of a complex interplay between baryonic physics and DM self-interactions.

The right panel of Fig. \ref{fig:sfr} compares the star formation history of the CDM simulation with $n_{\rm th} = 100\,{\rm cm^{-3}}$ (bursty star formation) to the SIDM simulation with $\sigma_T/m_\chi = 1\,{\rm cm^2g^{-1}}$ and $n_{\rm th} = 100\,{\rm cm^{-3}}$. 
We see that initially stars in the two simulations form at a similar rate. After $\sim 700\,{\rm Myr}$ however, the star formation histories start to deviate significantly and at
$t\sim 900\,{\rm Myr}$, 
there is a very large spike in the star formation rate of the SIDM simulation, followed by a sharp drop down to a steady, smooth, and rather low rate. In the CDM case, on the other hand, star formation continues in bursty cycles, with spikes that occur every $\sim 100\,{\rm Myr}$ on average. The impulsive SNF episodes following those spikes in star formation cause the migration of older stars into the outskirts of the galaxy. Since they are absent in the SIDM simulation, no strong stellar age gradient forms. The large spike in SNF activity that followed the large peak in star formation in the SIDM simulation has driven most of the gas out of the centre of the galaxy. Subsequently, gas no longer accumulates in the centre of the galaxy and star formation proceeds more or less smoothly at random locations in the galaxy. 
Hence, the final stellar age gradients are significantly different from simulations with similarly strong SNF mainly due to the very strong initial star formation activity, a consequence of the stochastic implementation of the stellar evolution model. 

\section{Theory-motivated alternative figures}\label{app_addfigs}

The motive of our article is to highlight potential ways to differentiate between adiabatic and impulsive core formation mechanisms, using observations. For that reason, most of the figures in our main article aim to highlight the differences between observable quantities that are likely to be related to the nature of the dominant core formation mechanism. This applies to Figs.~\ref{fig:dm_phasespace}, \ref{fig:vc_4gy}, and \ref{fig:grad_paramspace} in particular.

\begin{figure*}
    \centering
    \includegraphics[width=0.49\linewidth,trim={0.5cm 0.5cm 0.5cm 0.5cm},clip=true]{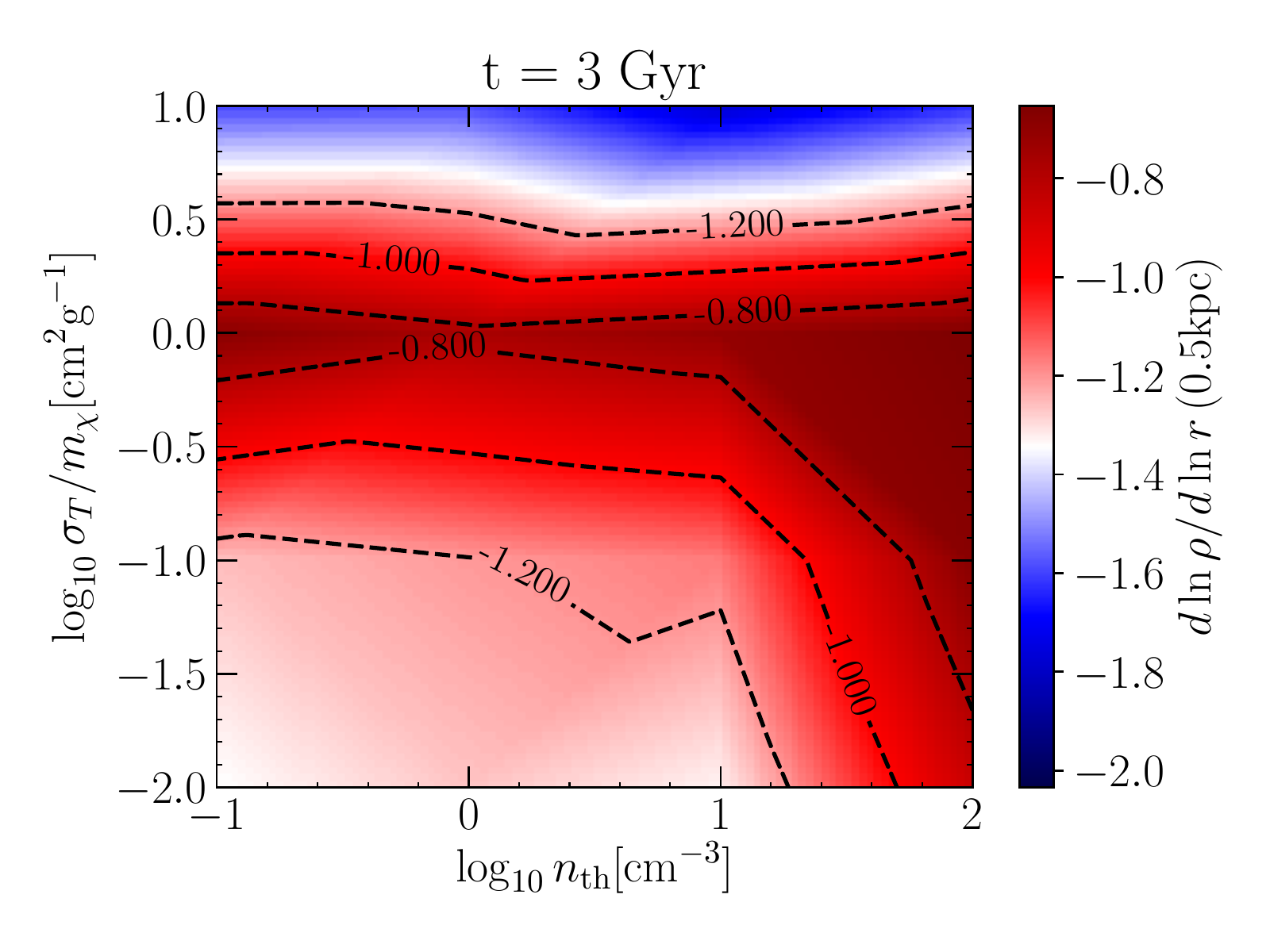}
    \includegraphics[width=0.49\linewidth,trim={0.5cm 0.5cm 0.5cm 0.5cm},clip=true]{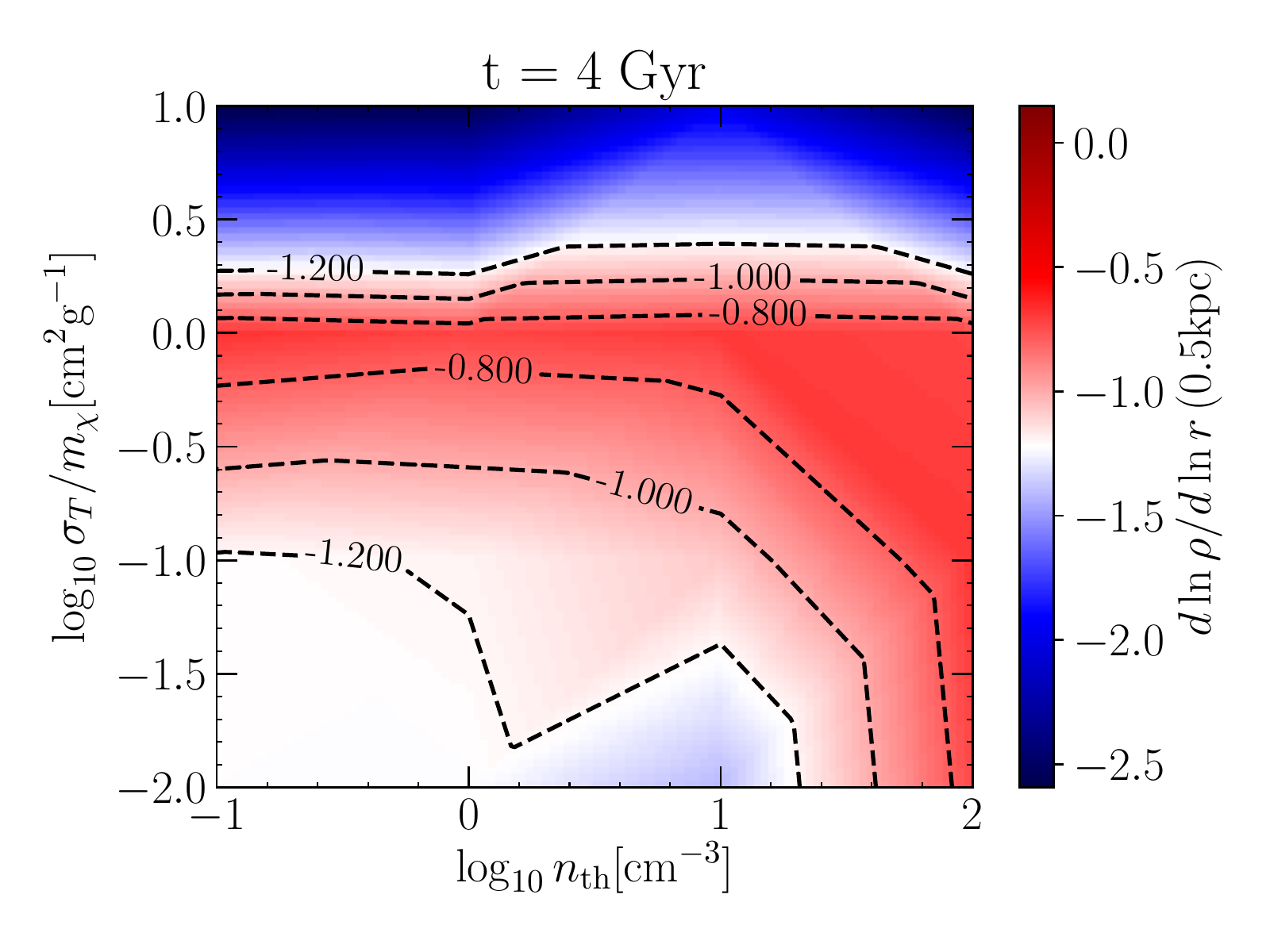}
    \caption{Logarithmic slope of the DM density profile at $r = 500\,{\rm pc}$ after 3 Gyr (left-hand panel) and 4 Gyr (right-hand panel) of simulation time, interpolated between the simulations in our simulation suite in the usual way. The divergent colour map is in reference to the CDM run with $n_{\rm th} = 0.1\,{\rm cm^{-3}}$ (lower left corner), and shows more cored profiles in red and more cuspy profiles in blue. Notice that the general trend is the same as can be observed from the measure based on the circular velocity curves shown in Figs.~\ref{fig:dm_phasespace} and \ref{fig:vc_4gy}. While the progression of the gravothermal collapse in simulations with $\sigma_T/m_\chi = 10\,{\rm cm^2g^{-1}}$ is captured better by comparing the logarithmic slope as shown here, we emphasize that DM density profiles cannot be observed directly.}
    \label{fig:app_fig_one}
\end{figure*}

For the upper panel of Fig.~\ref{fig:dm_phasespace}, as well as Fig.~\ref{fig:vc_4gy}, we chose to adopt the ratio between the (theoretically calculated) circular velocity at $r = 500$ pc and the maximal circular velocity as a measure for how cored or cuspy the final density profiles of simulated DM haloes are. While this quantity is closely related to an observable -- the galaxy rotation curve -- it is not commonly used to determine whether the central density profiles of DM haloes are cored or cuspy, both because it is derived from the integrated density profiles, and because baryons contribute to the rotation curves as well. Fig.~\ref{fig:app_fig_one} demonstrates that our findings hold when adopting a more established, theory-motivated measure to characterize the central DM density profile, namely the density profile's logarithmic slope in the inner kpc. Specifically, we show $d\ln\rho/d\ln r(0.5\,{\rm kpc})$ (as in \citealt{2015MNRAS.448..792G}), interpolated in simulation parameter space as outlined throughout Section \ref{sec:results}. Shallower slopes correspond to more cored profiles. Fig.~\ref{fig:app_fig_one} displays the same key features as Figs.~\ref{fig:dm_phasespace} and \ref{fig:vc_4gy}. In particular, cored final DM density profiles are found in the same area in parameter space. A slight difference is that the progression of gravothermal collapse is better captured in the right-hand panel of Fig.~\ref{fig:app_fig_one} than in Fig.~\ref{fig:vc_4gy}. This is because after 4 Gyrs, the circular velocity curve of simulated DM haloes with $\sigma_T/m_\chi = 10\,{\rm cm^2g^{-1}}$ is maximal for $r < 500\,{\rm pc}$ (see also Section \ref{subsec:profiles}). 

\begin{figure*}
    \centering
    \includegraphics[width=0.49\linewidth,trim={0.5cm 0.5cm 0.5cm 0.5cm},clip=true]{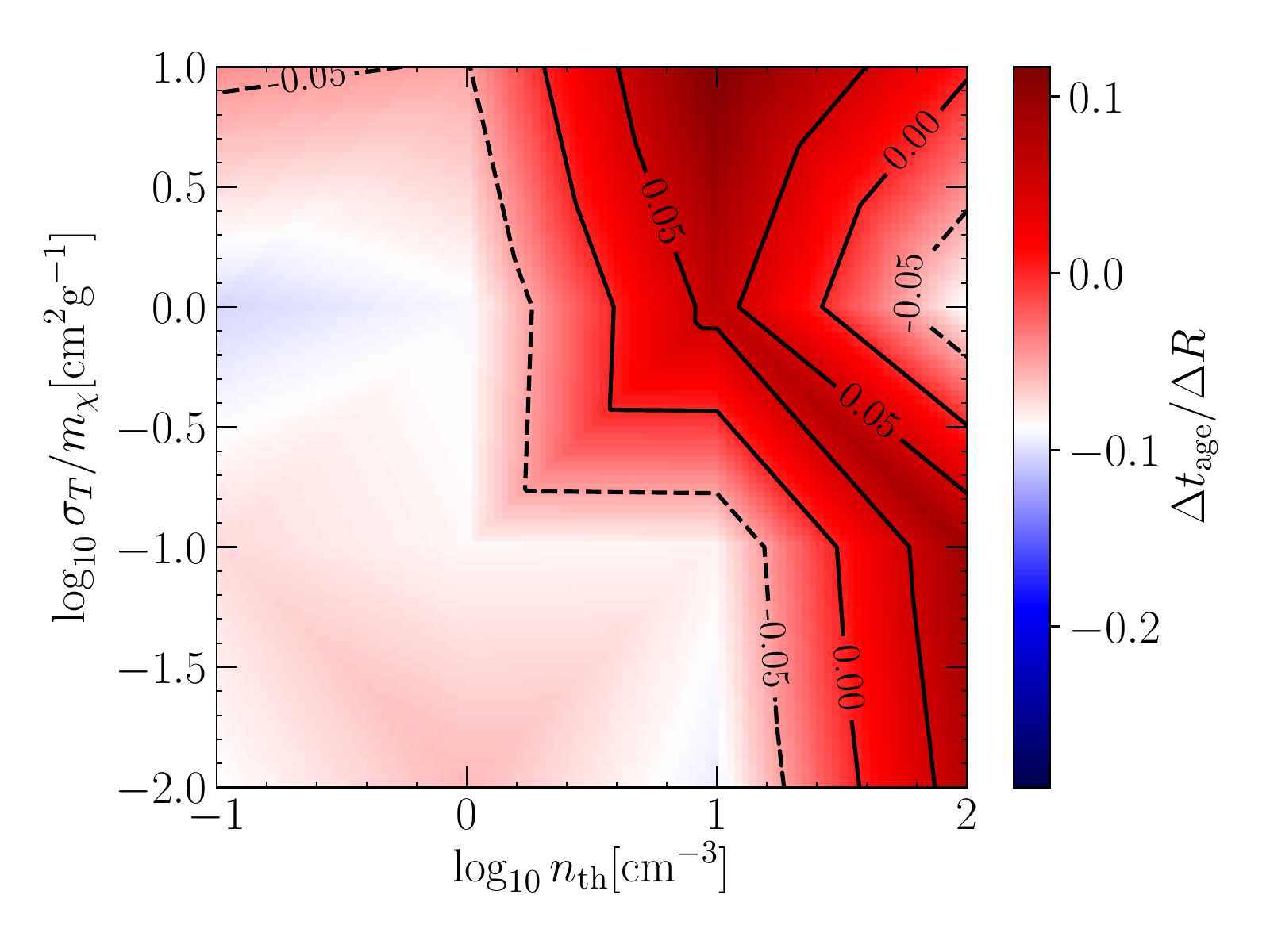}
    \includegraphics[width=0.49\linewidth,trim={0.5cm 0.5cm 0.5cm 0.5cm},clip=true]{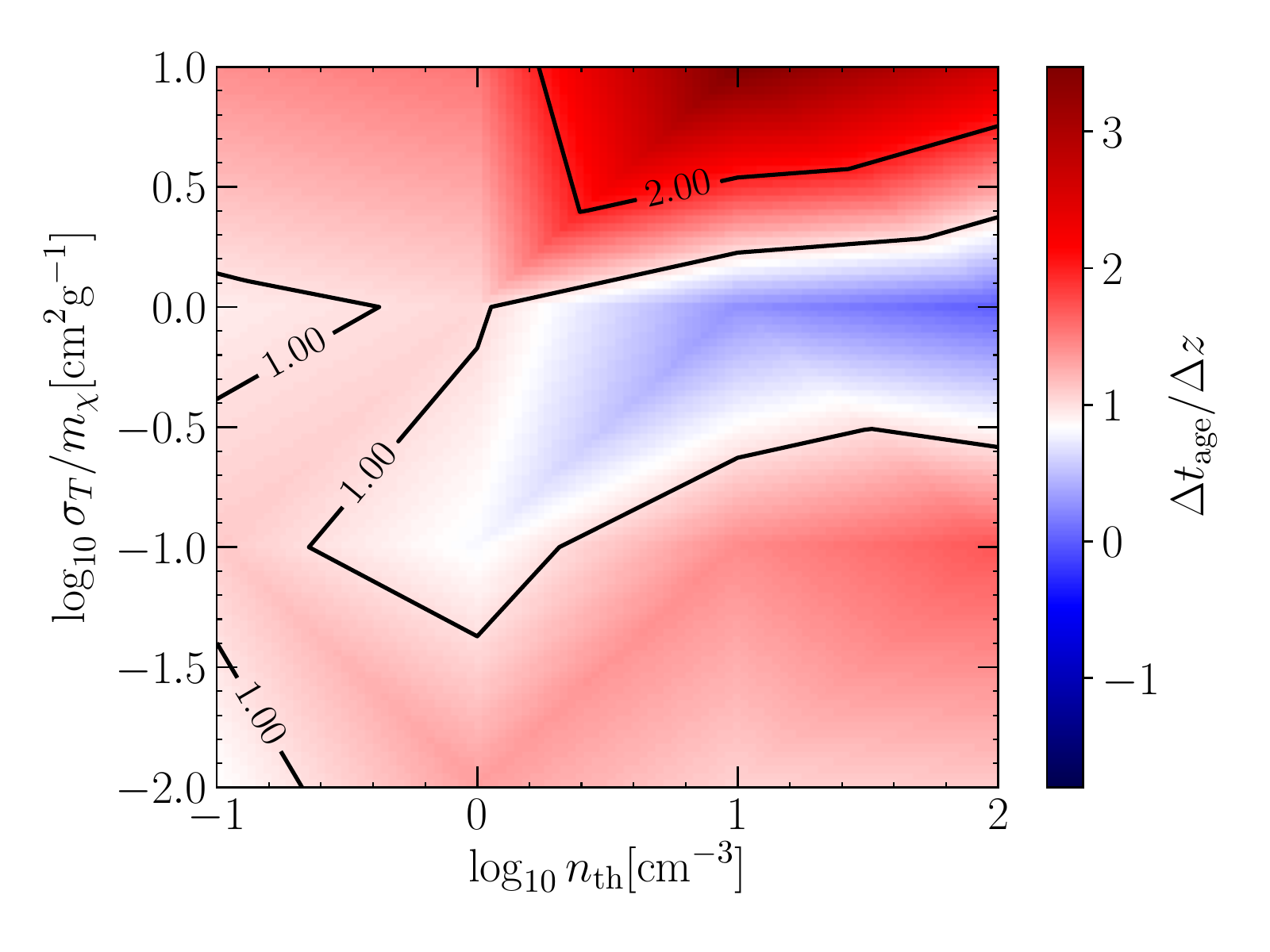}\\
    \includegraphics[width=0.49\linewidth,trim={0.5cm 0.5cm 0.5cm 0.5cm},clip=true]{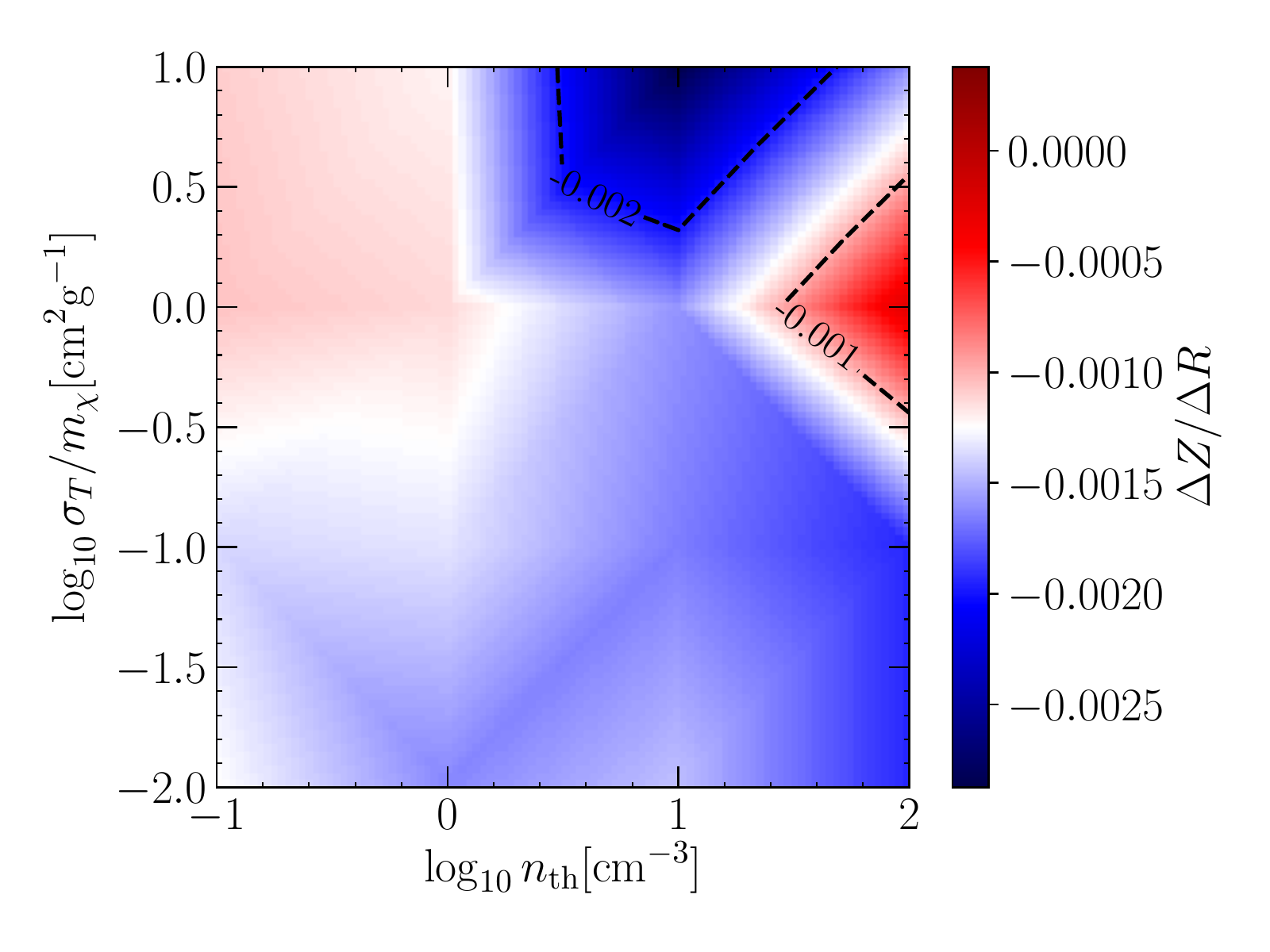}
    \includegraphics[width=0.49\linewidth,trim={0.5cm 0.5cm 0.5cm 0.5cm},clip=true]{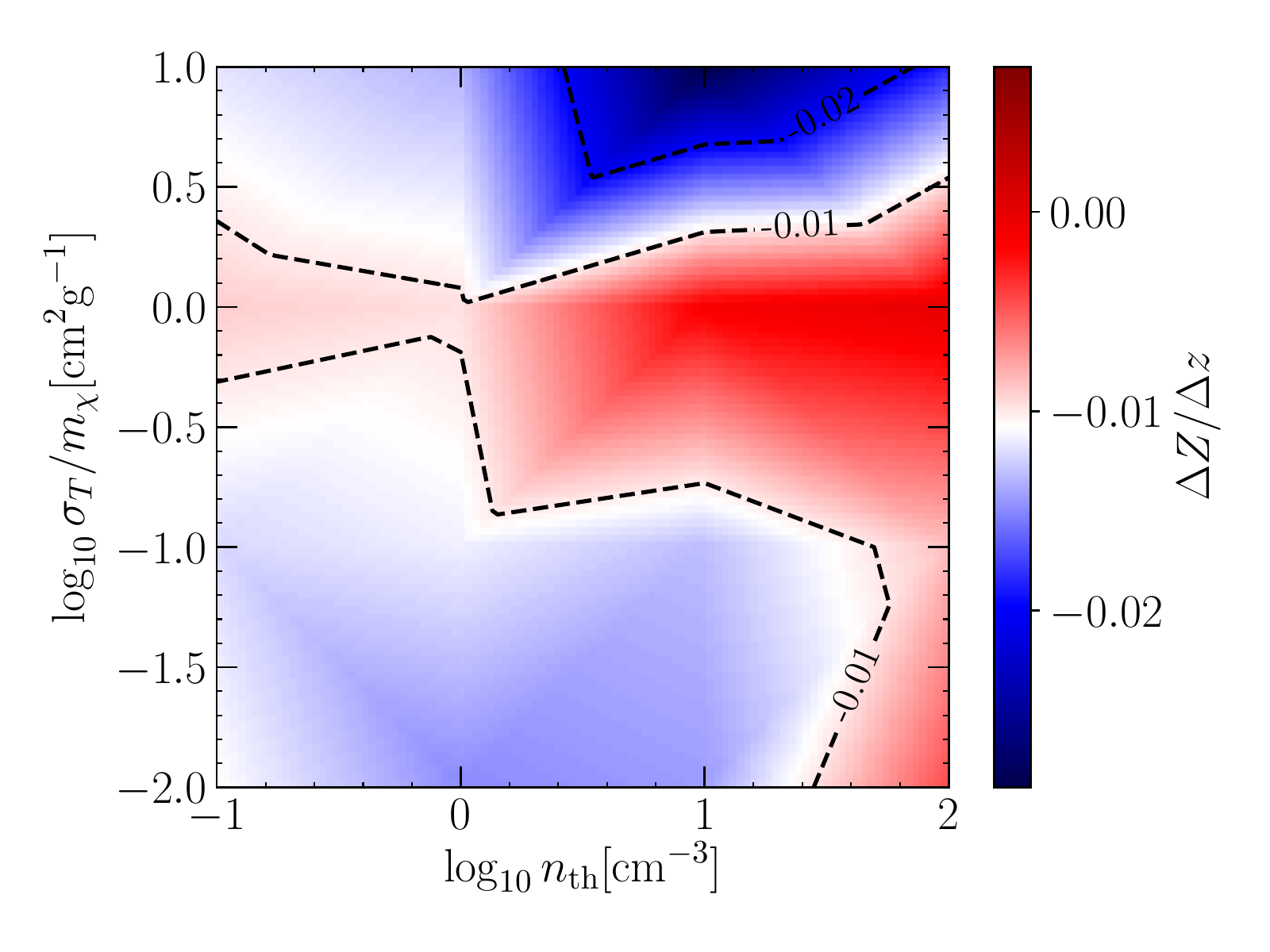}
    \caption{As in Fig.~\ref{fig:grad_paramspace}, we show bilinear interpolations between simulations in our suite over quantities that characterize the radial/vertical age/metallicity gradients. Instead of age/metallicity ratios, we characterize the gradients by the slopes of linear fits to the ages/metallicities of the stars formed after $3$ Gyr of simulated time as functions of the radial/vertical coordinate. Starting from the upper left panel, and in clockwise order, we show bilinear interpolations over the fitted slopes of the functions $t_{\rm age}(R)$, $t_{\rm age}(z)$, $Z(z)$, and $Z(R)$. Notice that the clear signal regions that, in Fig.~\ref{fig:grad_paramspace}, presented in the radial age gradient and the vertical metallicity gradient, appear nearly identical here. The upper right panel and the lower left panel differ somewhat from the corresponding panels of Fig.~\ref{fig:grad_paramspace}, but notice that the two outstanding features of those panels, namely the divergent behaviour of the two ``special'' simulations (see appendix \ref{appendix_3}), are still reproduced.}
    \label{fig:newgradients}
\end{figure*}

In Fig.~\ref{fig:grad_paramspace} we characterized the stellar age and metallicity gradients by age/metallicity ratios. For example, we evaluated radial age gradients using the ratio of the average age of stellar particles with $R> R_{1/4}$ to the average age of stellar particles with $R< R_{1/4}$. We chose the quarter mass radius as our scale of reference because it is an observable quantity that, in simulations in which cores had formed, turned out to (approximately) separate the central core from the remainder of the DM halo. However, we can also estimate age and metallicity gradients independently of any particular scale. In Fig.~\ref{fig:newgradients}, we show the slopes of linear fits to the functions $Z(R),\,Z(z),\,t_{\rm age}(R),\,t_{\rm age}(z)$, interpolated in the parameter space of our simulation suite as outlined in Section \ref{sec:results}. Linear fits are calculated using the masses of individual stellar particles as weights. For instance, we parameterize 
\begin{align}
    Z(R) = a + \frac{\Delta Z}{\Delta R}R,
\end{align}
with 
\begin{align}
    \frac{\Delta Z}{\Delta R} &= \frac{\sum_i m_iR_i(Z_i-\overline{Z})}{\sum_i m_iR_i(R_i-\overline{R})}\\
    a &= \overline{Z}-b\overline{R},
\end{align}
where $\overline{R}$ and $\overline{Z}$ are mass averaged radius and metallicity, and the sums go over all newly formed stellar particles. The remaining age and metallicity gradients are calculated analogously. We find only slight differences between Figs.~\ref{fig:grad_paramspace} and \ref{fig:newgradients}. All key features (in particular in the radial age gradient and the vertical metallicity gradient) are preserved.

\section{Changing the initial conditions} \label{appc}
The goal of our simulation suite was to study the divergent baryonic signatures of two distinct core formation mechanisms. To that end, we conducted a suite of 16 simulations of an isolated galaxy embedded in a DM halo, and we effectively varied the strength of both, SIDM and SNF, by changing the transfer cross section per unit mass to regulate the strength of the former and the density threshold for star formation to regulate the strength of the latter. Within our setup, we identified a region in $\sigma_T/m_\chi-n_{\rm th}$ parameter space in which cores of similar sizes formed (Fig.~\ref{fig:dm_phasespace}) and then demonstrated that this degeneracy is broken by the properties of the baryons in the simulations. 

While these results are solid predictions for the differences between adiabatic, SIDM-driven core formation and impulsive, SNF-driven core formation, they should not be taken as hard predictions for the properties of present day dwarf galaxies under the condition that $\sigma_T/m_\chi$ and $n_{\rm th}$ take a certain fixed set of values. The reason for that is that our choice of initial conditions is likely to have a key impact on the final results. SNF works as a core formation mechanism because the cold gas from which stars are formed temporarily dominates the gravitational potential in the galaxy's centre, before being rapidly driven out by supernovae, thus causing large fluctuations in the center of the gravitational potential (see \citealt{Pontzen:2011ty}). Evidently, this implies that the initial ratio of DM to baryonic matter in the centre of the dwarf galaxy can have a significant impact on the effectiveness of SNF for a given $n_{\rm th}$ (see e.g. \citealt{2015MNRAS.454.2092O}, \citealt{2016ApJ...819..101G}). Likewise, the scattering rate of SIDM depends on the local DM density as well as the cross section (e.g. \citealt{Vogelsberger2012}). Here, we fix this initial ratio a priori through the choice of initial conditions. In cosmological simulations, this initial ratio will be scattered around a mean that is set by the assumed cosmology, with the value for a specific galaxy also depending on its formation time, environment and assembly history. 

We emphasize again that our results, in particular regarding SNF, should be seen as predictions for the impact of SNF given a certain star formation history. However, to illustrate the role of our choice of initial conditions in combination with the numerical parameters chosen, we repeated our simulation suite (until $t=3\,{\rm Gyr}$), leaving all parameters as in Section \ref{subsec:ICs} except for the halo concentration, which we changed to $c_{200} = 12$ in order to investigate the impact of a significantly lower initial DM to baryon ratio in the centre of the simulated dwarf. We report key results below and comment on how our findings may change in a cosmological setup.  

\subsection{Star formation histories}
\begin{figure*}
    \centering
    \includegraphics[width=0.48\linewidth,trim={0.5cm 0.5cm 0.5cm 0.5cm},clip=true]{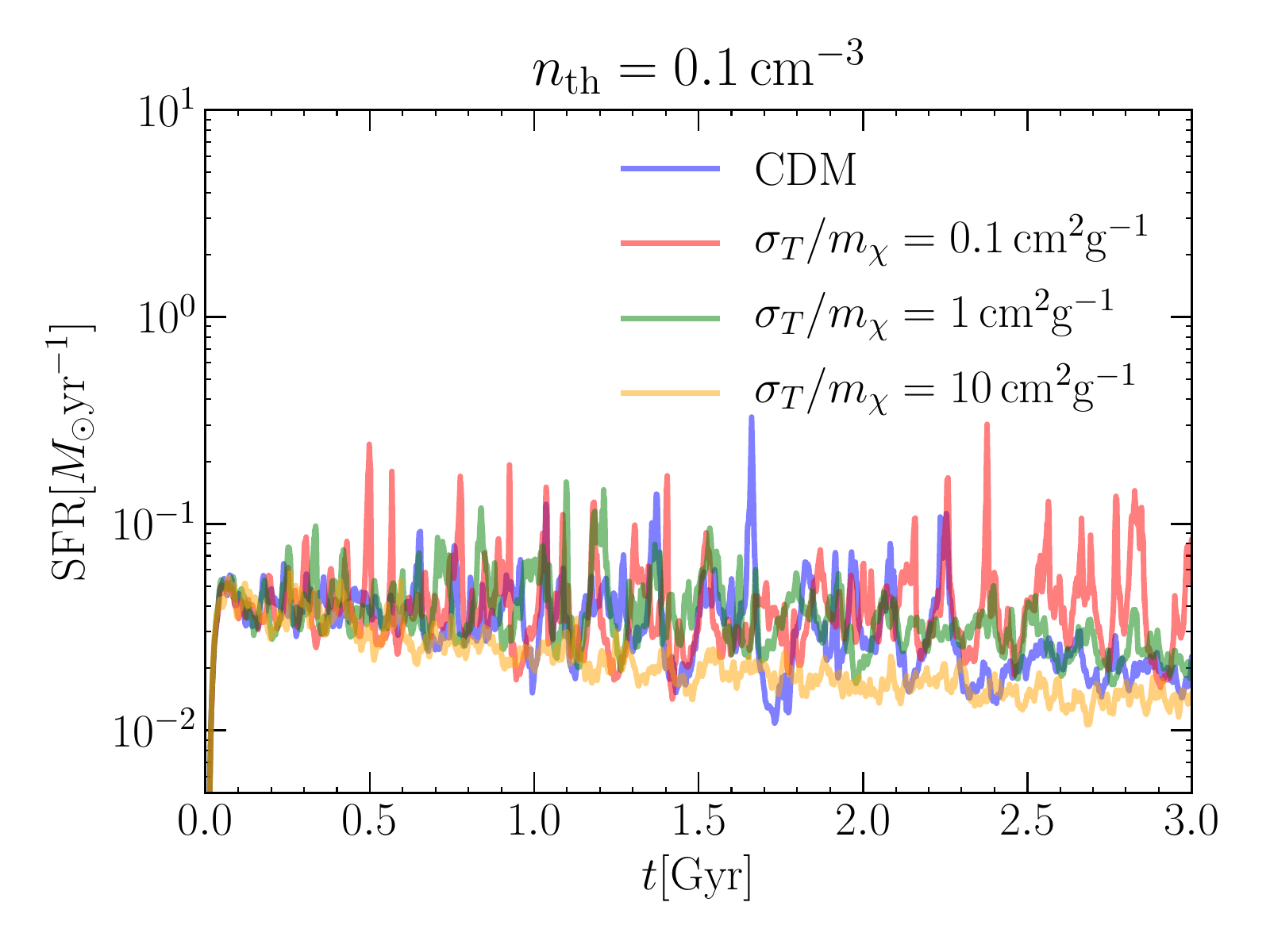}
    \includegraphics[width=0.48\linewidth,trim={0.5cm 0.5cm 0.5cm 0.5cm},clip=true]{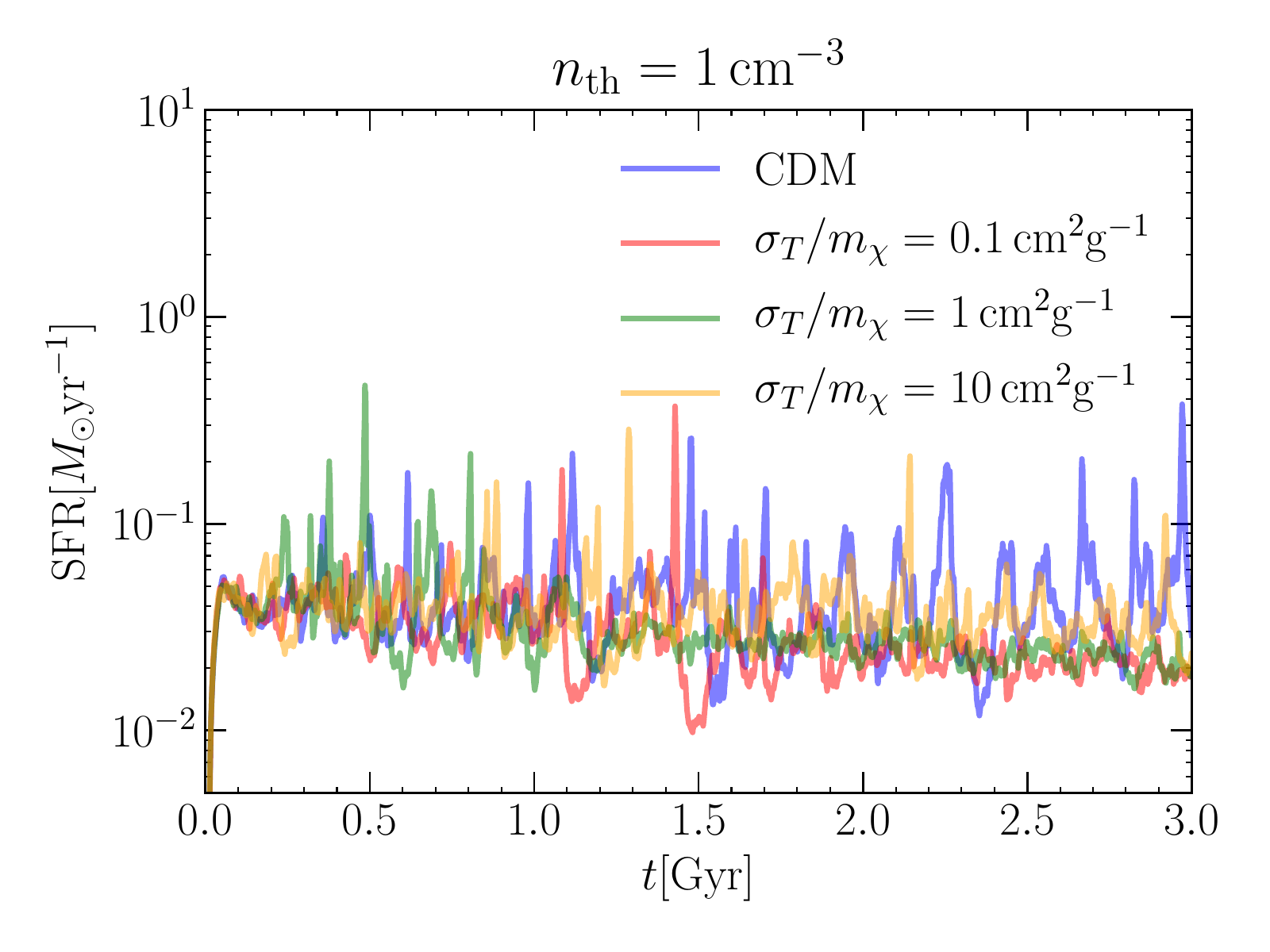}
    \includegraphics[width=0.48\linewidth,trim={0.5cm 0.5cm 0.5cm 0.5cm},clip=true]{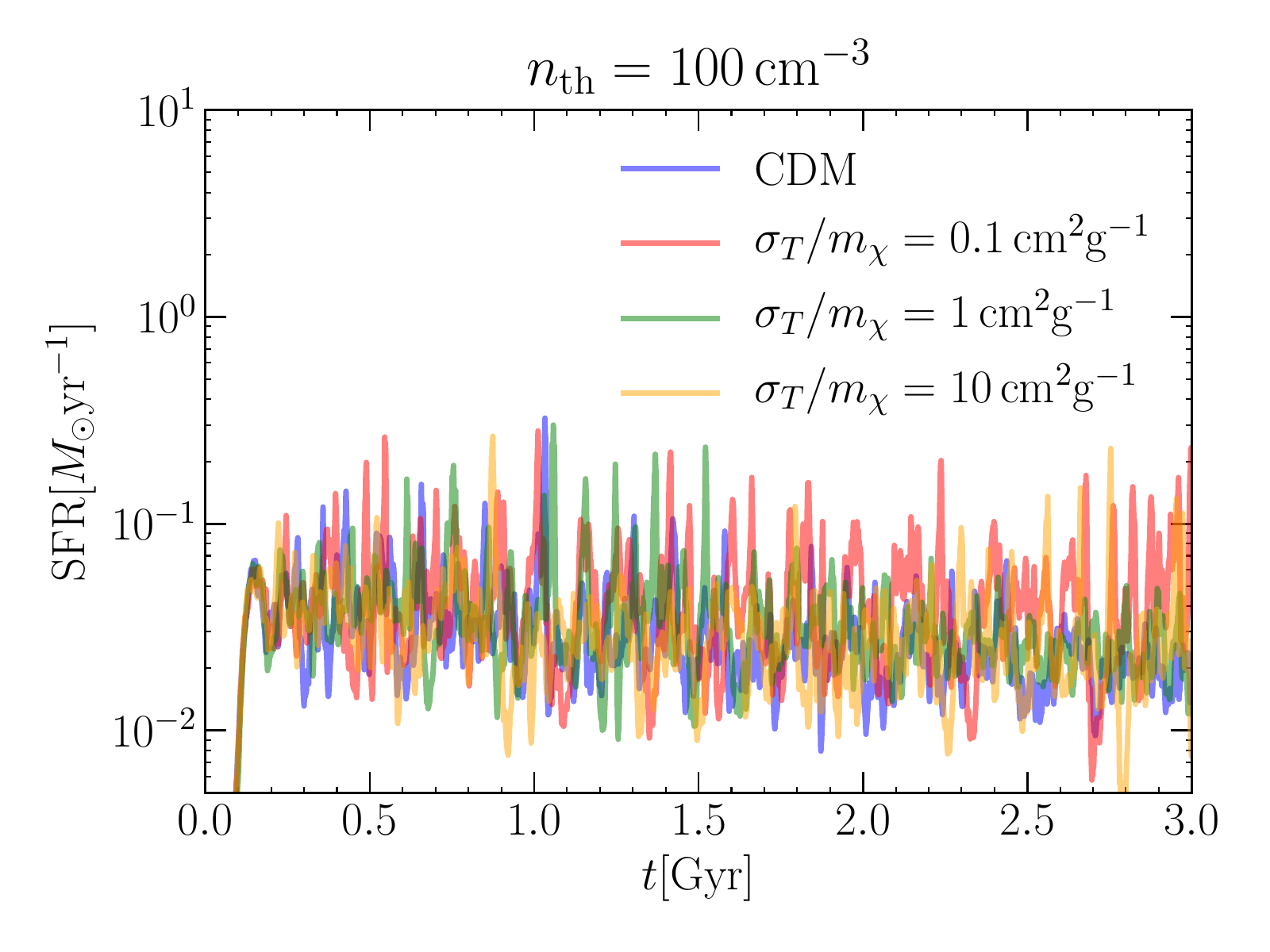}
    \includegraphics[width=0.48\linewidth,trim={0.5cm 0.5cm 0.5cm 0.5cm},clip=true]{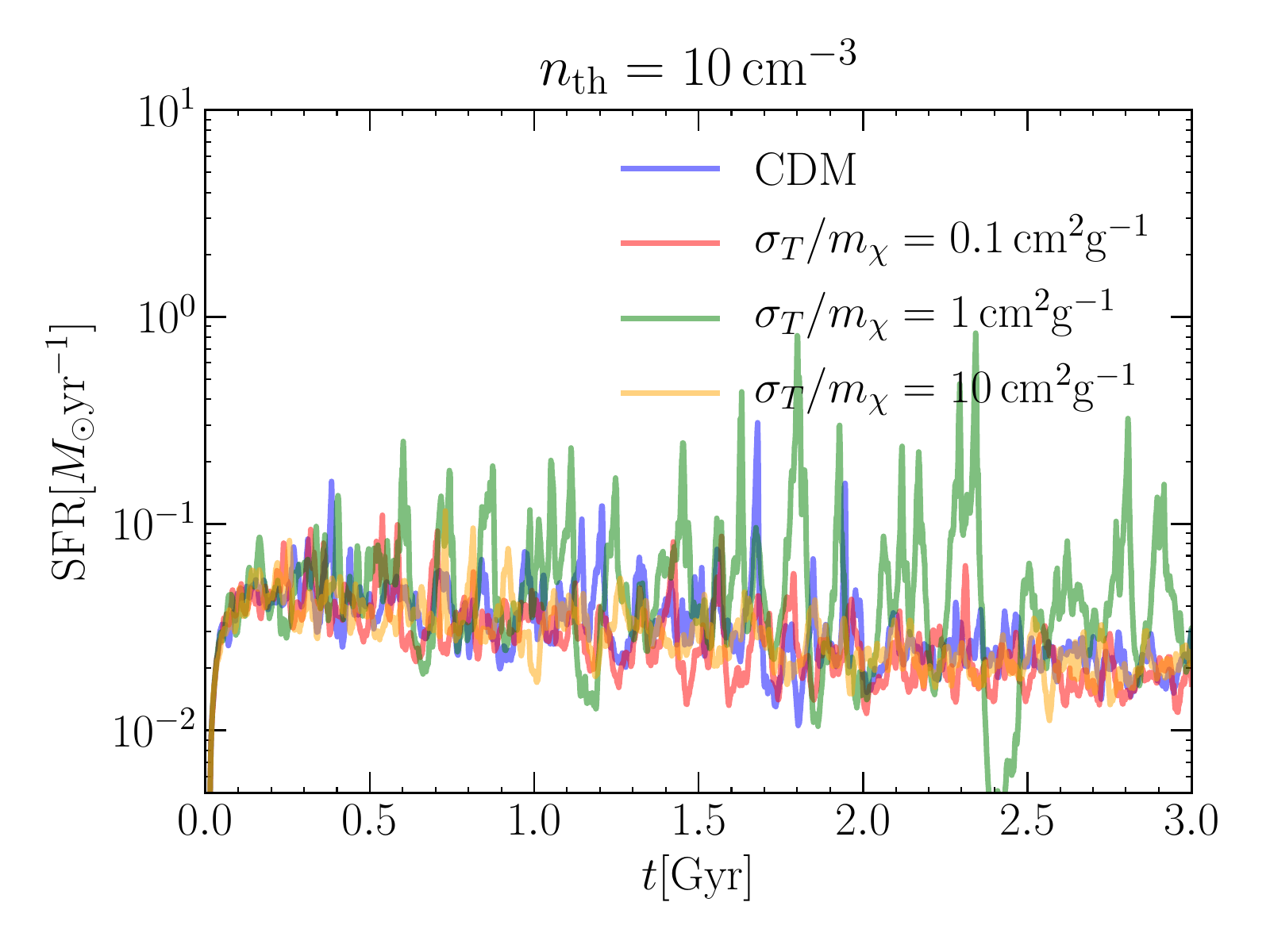}
    \caption{Star formation histories of all 16 simulations for the runs with modified initial conditions (the concentration of the halo is lowered from $c_{200}=18$ to $c_{200}=12$). 
    Clockwise from the upper left, the four panels show the star formation rates of the runs with $n_{\rm th} = 0.1\,{\rm cm^{-3}}$, $n_{\rm th} = 1\,{\rm cm^{-3}}$, $n_{\rm th} = 10\,{\rm cm^{-3}}$, and $n_{\rm th} = 100\,{\rm cm^{-3}}$. Within each panel, each run is colour coded by its SIDM momentum transfer cross section as indicated in the legend. On average, star formation histories are burstier than in the corresponding simulations with $c_{200} = 18$, particularly at lower star formation thresholds. The trend of runs with larger star formation thresholds having burstier star formation histories is preserved, with some stochasticity, mirroring the stochasticity in the star formation histories shown in Figs.~\ref{fig:sfhs} and \ref{fig:sfr} (see also the discussion in appendix \ref{app_ics}.)}
    \label{fig:c12_sfhs}
\end{figure*}

We show the star formation histories of all 16 new simulations in Fig.~\ref{fig:c12_sfhs}. Two clear trends emerge. As in our original simulation suite, star formation histories are, on average, burstier in simulations with larger star formation thresholds. However, even at $n_{\rm th} = 0.1\,{\rm cm^{-3}}$, we now observe significantly burstier star formation than before. This implies that impulsive SNF can now occur in essentially all of our simulations -- and we indeed find that it does. As was the case in our initial simulation suite, we find divergent behaviour between simulations with identical star formation thresholds (see also appendix \ref{appendix_3}). This is due to the stochastical implementation of SNF and star formation (see \citealt{2019MNRAS.489.4233M}). In particular, notice the significantly less bursty star formation histories of the runs with $n_{\rm th} = 10\,{\rm cm^{-3}}\,\sigma_T/m_\chi = 0.1\,{\rm cm^2g^{-1}}$ and $n_{\rm th} = 1\,{\rm cm^{-3}}\,\sigma_T/m_\chi = 1\,{\rm cm^2g^{-1}}$, as they will be reflected in the core size, as well as in some baryonic properties, as we show later on. Moreover, notice the very bursty star formation history of the run with $n_{\rm th} = 10\,{\rm cm^{-3}}\,\sigma_T/m_\chi = 1\,{\rm cm^2g^{-1}}$. Below, we will illustrate on this run how a bursty star formation history perpetuates itself, and how the lower initial DM to baryon ratio in the centre of the dwarf results in burstier star formation histories overall. 

\begin{figure*}
    \centering
    \includegraphics[width=0.48\linewidth,trim={0.5cm 0.5cm 0.5cm 0.5cm},clip=true]{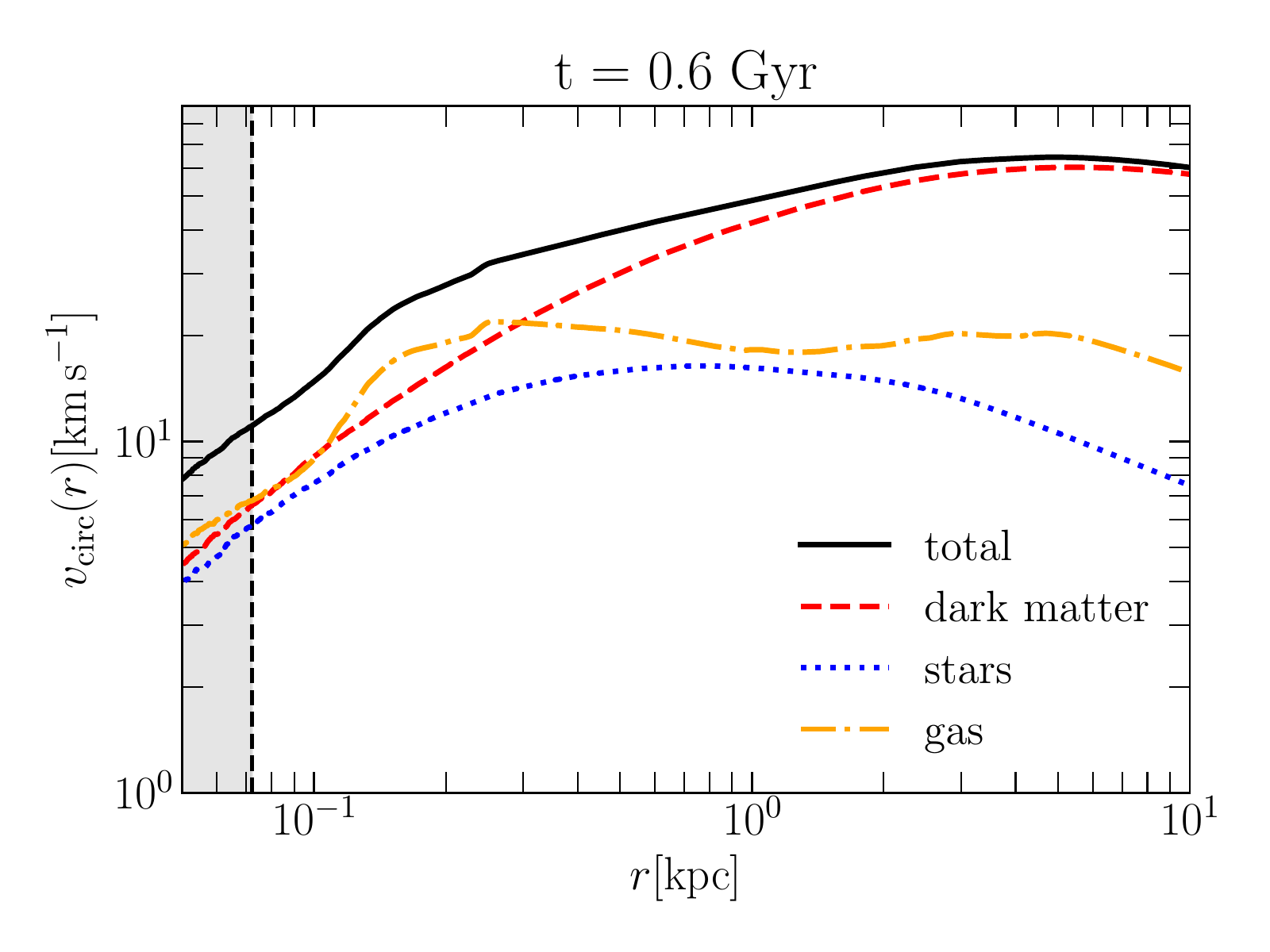}
    \includegraphics[width=0.48\linewidth,trim={0.5cm 0.5cm 0.5cm 0.5cm},clip=true]{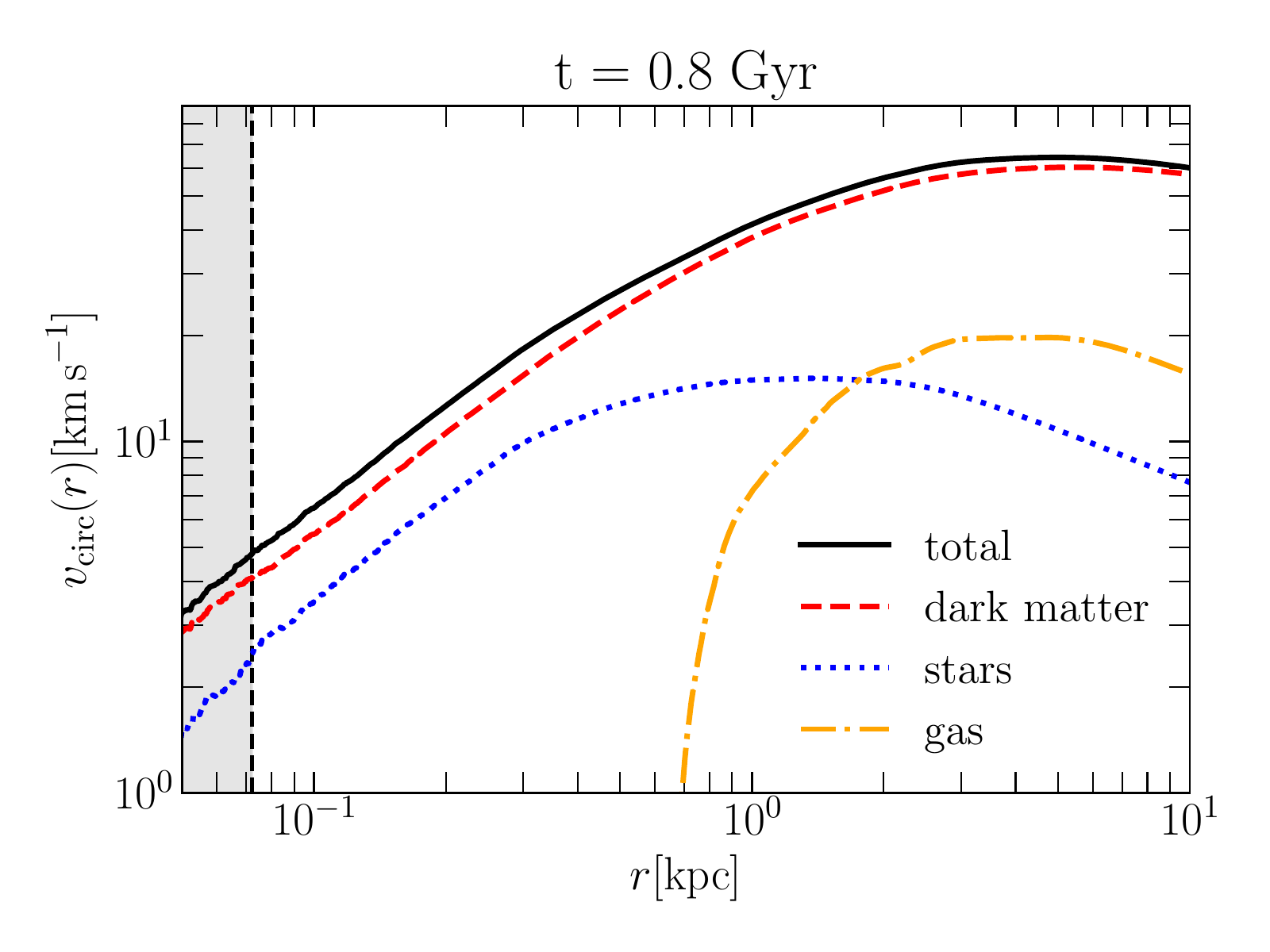}
    \caption{Relative contribution of DM, stars, and gas to the total circular velocity curves of the simulation with $n_{\rm th} = 10\,{\rm cm^{-3}}$ and $\sigma_T/m_\chi = 1\,{\rm cm^2g^{-1}}$ at two times corresponding to two consecutive snapshots, $t = 0.6$ Gyr (left-hand panel) and $t = 0.8$ Gyr (right-hand panel). Cool gas can temporarily dominate the enclosed mass (and hence circular velocity) profile in the centre of the galaxy, before a large burst of star formation occurs and the gas is expelled by the subsequent supernovae, leading to a galaxy centre devoid of gas.}
    \label{fig:cons_rot_curves}
\end{figure*}

In Fig.~\ref{fig:cons_rot_curves}, we show, after $0.6$ Gyr (left-hand panel) and after $0.8$ Gyr of simulation time (right-hand panel), the circular velocity curves derived from the enclosed mass profiles of the simulation with $n_{\rm th} = 10\,{\rm cm^{-3}}$ and $\sigma_T/m_\chi = 1\,{\rm cm^2g^{-1}}$ (and a very bursty star formation history), along with the relative contributions from DM, gas, and stars (as in Fig.~\ref{fig:icrot}). The two consecutive snapshots illustrate how a bursty star formation history arises. In the left-hand panel, the central mass distribution -- and hence the gravitational potential -- is dominated by cold gas. This cold, dense gas is then converted into stars, which subsequently leads to an impulsive episode of SNF. Feedback energy is coupled to the gas, which in turn results in a large-scale galactic wind in which the gas is expelled from the centre of the galaxy. If star formation and subsequent feedback are sufficiently concentrated towards the centre of the galaxy, this will result in a central region devoid of gas -- as is the case in the right-hand panel of Fig.~\ref{fig:cons_rot_curves}. As a consequence, gas that subsequently streams back towards the centre of the galaxy is not affected by ram pressure, and can cool down fast and efficiently, thus perpetuating the impulsive feedback cycle. Essentially, this can continue until the gas is depleted, or -- in a cosmological setup -- removed by mergers, tidal stripping, or tidal heating (see e.g. \citealt{2019Galax...7...81Z} and references therein). 

Since the burstiness of the star formation history relies on the gas repeatedly being expelled from (and then streaming back into) the centre of the dwarf galaxy, how bursty the star formation history depends considerably on 
how centrally concentrated are the stars that form initially. 
Moreover, some interplay between bursty star formation histories and the presence of self-interactions is to be expected, although the resulting effect is a priori unclear. SIDM lowers the central DM mass at a rate that depends on the self-interaction cross section. A lower DM mass means that it is easier for baryons to dominate the central gravitational potential, which favours bursty star formation and impulsive feedback. On the other hand, less central DM means a weaker gravitational potential overall, which could prevent cold gas from accumulating in case the bursty star formation mode has not yet started. The complex interplay between SIDM and star formation, along with the inherent stochasticity of the star formation mechanism, is reflected in the diversity of star formation histories at a fixed star formation threshold seen in Fig.~\ref{fig:c12_sfhs}. The lower initial halo concentration, on the other hand, is clearly reflected in the overall burstier star formation histories, especially at lower star formation thresholds, in agreement with \citet{2016ApJ...819..101G}. 

\subsection{Final dark matter density profiles}

\begin{figure}
    \centering
    \includegraphics[width=\linewidth,trim={0.5cm 0.5cm 0.5cm 0.5cm},clip=true]{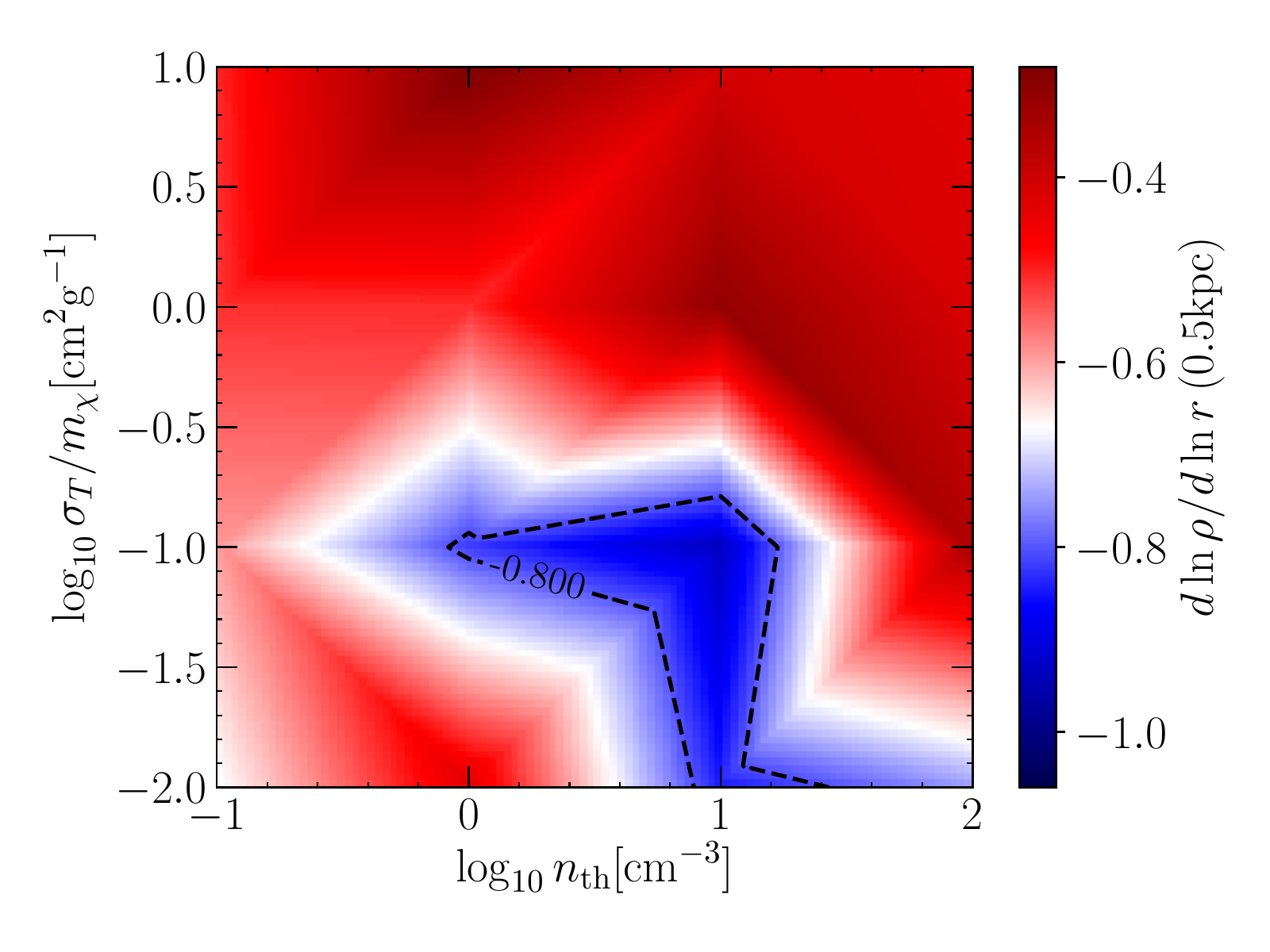}
    \caption{Same as the left-hand panel of Fig.~\ref{fig:app_fig_one}, but for the 16 simulations with $c_{200} = 12$ initially.}
    \label{fig:rv_log_slopes}
\end{figure}

In Fig.~\ref{fig:rv_log_slopes} we show the final (measured after 3 Gyr of simulation time) logarithmic slopes of the spherically averaged DM density profiles at $r = 500$ pc as a function of self-interaction cross section and star formation threshold, and interpolated between all 16 simulations of the simulation suite with $c_{200} = 12$ initially. As before, the divergent colour map is centred around the value measured for the CDM run with $n_{\rm th} = 0.1\,{\rm cm^{-3}}$. Comparison to Fig.~\ref{fig:app_fig_one} reveals that the final DM density profiles are more cored for virtually every combination of $\sigma_T/m_\chi$ and $n_{\rm th}$. Two regimes are particularly noteworthy:
\begin{itemize}
    \item Haloes in CDM runs with low star formation thresholds are now cored, while they remained cuspy in the main simulation suite (with $c_{200} = 18$). This is directly related to the difference in the star formation histories outlined above. 
    \item Runs with $\sigma_T/m_\chi = 10\,{\rm cm^2g^{-1}}$ are universally cored as well. SIDM scattering rates depend on the local DM density, and are thus reduced by decreasing the initial DM concentration. Gravothermal collapse occurs after the SIDM halo progresses through the cored stage, and the timescale on which this happens is directly related to the scattering rate. Hence, the initially lower central DM density slows down gravothermal collapse significantly. 
\end{itemize}

Aside from those two points, we observe that the final cores are less isothermal for the simulations with ($\sigma_T/m_\chi\,[{\rm cm^2g^{-1}}], n_{\rm th}\,[{\rm cm^{-3}}]$) = ([0,10],[0.1,10],[0.1,1]). Notably, two of these simulations correspond to the runs with less bursty star formation overall, as we outlined above. The third one, the CDM run with $n_{\rm th} = 10\,{\rm cm^{-3}}$, initially has a bursty star formation history that settles into a smoother mode towards the end of the simulation (see Fig.~\ref{fig:c12_sfhs}). We conclude that, while for $c_{200} = 12$ star formation histories are burstier on average and essentially all final DM density profiles are cored, the final core size still strongly depends on how impulsive SNF is exactly, as long as $\sigma_T/m_\chi \le 0.1\,{\rm cm^2g^{-1}}$. Thus, the correlation between core size and impulsiveness of SNF/burstiness of star formation history is maintained.

\subsection{Baryonic signatures}\label{app_bar_sig}

\begin{figure*}
    \centering
    \includegraphics[width=0.48\linewidth,trim={0.5cm 0.5cm 0.5cm 0.5cm},clip=true]{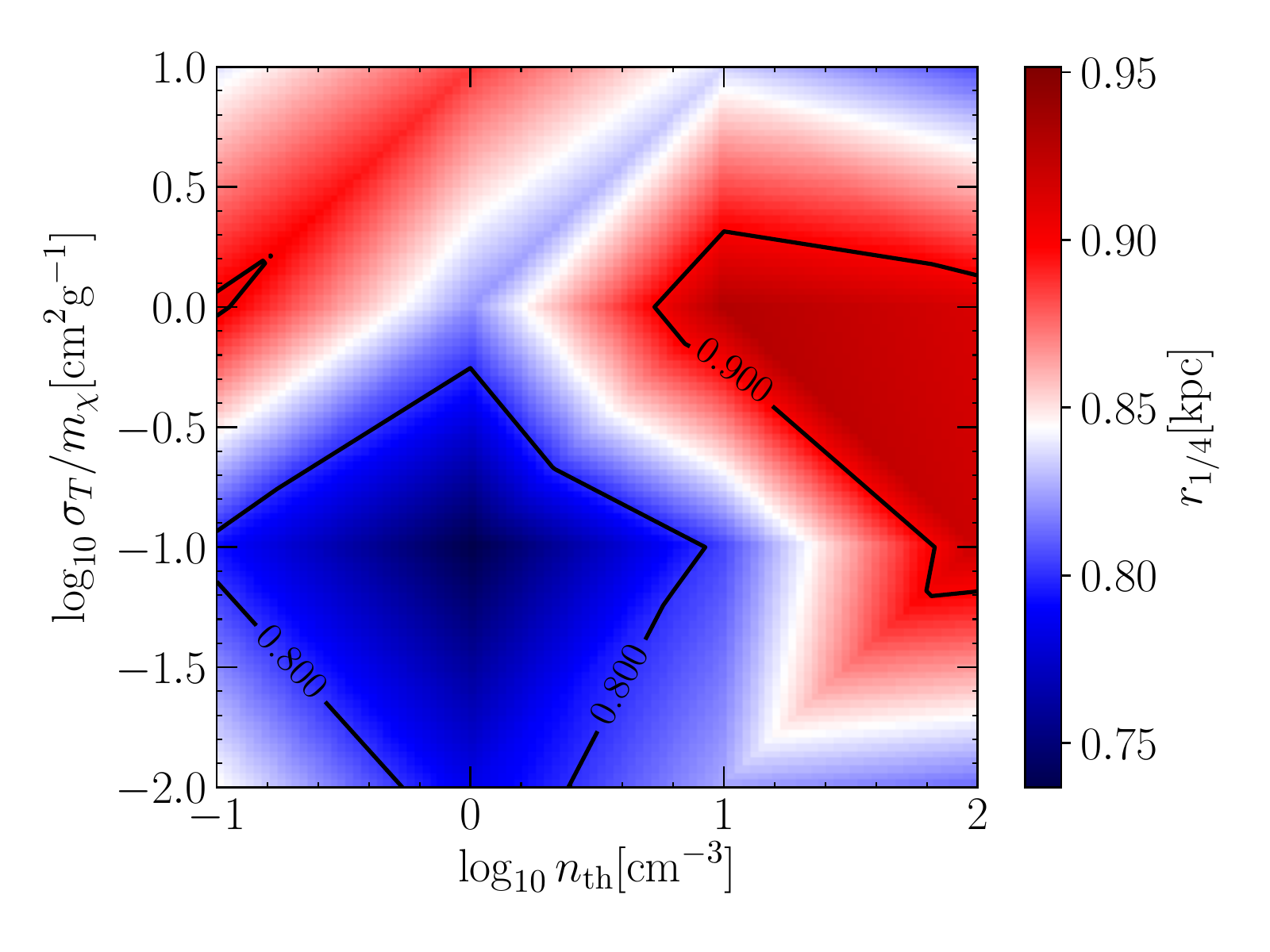}
    \includegraphics[width=0.48\linewidth,trim={0.5cm 0.5cm 0.5cm 0.5cm},clip=true]{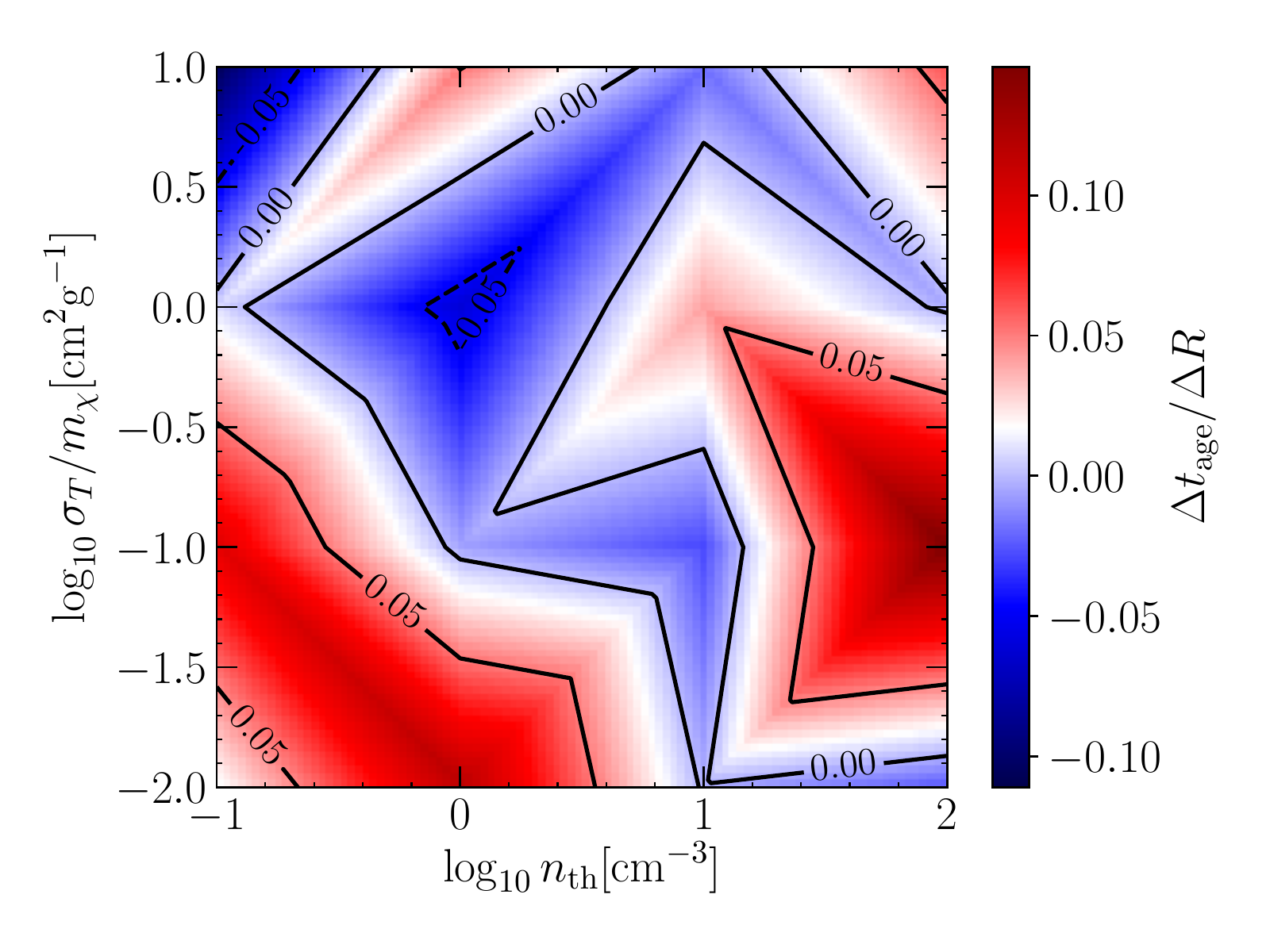}
    \caption{Left-hand panel: Same as bottom panel of Fig.~\ref{fig:stellar_size}, but for $c_{200}=12$ initially. Right-hand panel: Same as top left panel of Fig.~\ref{fig:newgradients}, but for $c_{200}=12$.}
    \label{fig:rv_bar}
\end{figure*}

Finally, in Fig.~\ref{fig:rv_bar} we report how two key baryonic signatures change when changing the initial concentration. On the left-hand panel, we show the stellar quarter mass radius, as a function of star formation threshold and self-interaction cross section, interpolated between simulations in the usual way. On the right-hand panel we show the same for the slope of a linear fit to the stellar ages as a function of cylindrical radius. 

At first glance, it appears that neither panel of Fig.~\ref{fig:rv_bar} looks even remotely similar to its counter part in the main simulation suite. Closer analysis, however, reveals that this does not invalidate our prior conclusions. The key reason for that 
is that the x-axis shows the numerical value of the star formation threshold, while the property of interest is instead the ``burstiness'' of the star formation history, a property that we cannot set a priori in a deterministic way. 

Taking a closer look at the right-hand panel of Fig.~\ref{fig:rv_bar}, it is obvious that -- contrary to the top left panel of Fig.~\ref{fig:newgradients} -- whether or not positive age gradients are present does not correlate with the star formation threshold. However, comparison with Fig.~\ref{fig:c12_sfhs} reveals that the ``burstiness'' of the simulated star formation histories correlates strongly with the observed radial age gradient: the strongest positive gradients appear in simulations with bursty star formation histories, while the weaker -- or negative -- gradients correspond to simulations with smoother star formation histories. Thus, the physical dependence of the age gradients on the star formation histories is unchanged. 

The left-hand panel, on the other hand, remains inconclusive, especially when compared to the clear picture that emerged in Fig.~\ref{fig:stellar_size}. Galaxies are more extended in simulations with larger star formation thresholds (though not necessarily corresponding to burstier star formation histories, as can be seen when comparing to the right-hand panel), as well as in some, though not all, simulations with $\sigma_T/m_\chi \ge 1\,{\rm cm^2g^{-1}}$. 
If the now present additional expansion of the simulated galaxies in simulations with large star formation thresholds is indeed physical, an explanation would have to rely on the larger gas densities that can be achieved in simulations with larger star formation thresholds, and the subsequent impulsive feedback. This means that above some threshold value for the star formation threshold (that also depends on the intial DM to baryon ratio), the effect that initially hinders expansion of the dwarf galaxy (see Section \ref{subsec:galsize}) would eventually be reversed, as the stronger induced fluctuation in the potential begins to outweigh the fact that star formation is more concentrated to the centre of the galaxy.

Note that the key feature observed in Fig.~\ref{fig:stellar_size} cannot possibly be present here, as there is no simulation with purely SIDM-induced core formation and a completely smooth star formation history in the simulation suite with $c_{200} = 12$.

The strongest statement that we can make from the properties of the baryons is obtained when combining the predictions of both panels of Fig.~\ref{fig:rv_bar}. Cores that are predominantly formed by self-interactions, and in the host haloes of galaxies with relatively smooth star formation histories, correlate with extended galaxies and shallow (or negative) age gradients. Cored DM haloes hosting more compact galaxies with steeper age gradients, on the other hand, are more likely to have formed their core through SNF (or, in more quantifiable terms, their galaxies have burstier star formation histories). These statements can be derived directly by comparing Figs.~\ref{fig:stellar_size}, \ref{fig:c12_sfhs}, and \ref{fig:rv_bar} and, while they do not encompass all of the runs we conducted, they apply to both simulation suites alike. This suggests that a statistical comparison of the properties of the baryons in observed nearby dwarfs to the results of a future suite of zoom simulations of the Local Group may indeed hold the key to differentiating between different core formation mechanisms, and thus between different cosmologies. Below, we discuss anticipated additional effects that may be important in such a cosmological setup. 

\subsection{Towards a cosmological setup}\label{app_cosmosetup}
Appendix \ref{appc} has been focused on changing the initial conditions of our simulation suite of a single isolated halo, thus investigating the impact of different initial DM to baryon ratios in the centre of the galaxy, which may arise as a result of different accretion histories. Recent results from cosmological simulations suggest that the star formation threshold plays a key role in regulating the impulsiveness of SNF (\citealt{2019MNRAS.488.2387B}, \citealt{2020MNRAS.499.2648D}), similar to the results of our main simulation suite (with $c_{200} = 18$ initially). This does not mean that our initial conditions chosen in the main paper are more ``correct'' than the choice made in Appendix \ref{appc} from a physical point of view, but it serves to illustrate that the results presented in the main article may be of interest when it comes to comparing the results of simulations to observations. More so, we demonstrated that even with a different initial setup, baryonic signatures can still be useful to identify the dominant core formation mechanism (see Section \ref{app_bar_sig}). 

However, we stress that this work is to be seen as an in-depth theoretical exploration of the differences between SIDM and SNF in an idealized setup. The mass aggregation history (MAH), the merger history, and the local environment can potentially act as perturbers to the signatures presented here. In particular, the effectiveness of SNF will strongly depend on the MAH, simply because it depends on the ratio of DM to gas/baryons -- and thus it should also depend on the evolution of this ratio. Major mergers can alter the inner structure of a galaxy (e.g. \citealt{2021arXiv211013172Z}), and ram pressure can strip the galaxy of its gas as it becomes a satellite of a larger galaxy, thus quenching star formation. Furthermore, the effects of dynamical friction and tidal stripping on the structure of galaxies may even depend on the chosen simulation parameters (\citealt{2018MNRAS.474.3043V}, \citealt{2018MNRAS.475.4066V}). 

The impact of SIDM with a fixed cross section is also expected to vary depending on the environment of the galaxy and its MAH. In particular, whether or not a given cross section leads to gravothermal collapse or not will depend on the merger history (e.g. \citealt{2002ApJ...581..777C}, \citealt{Dave2001}), and on the local environment (e.g. \citealt{2020PhRvD.101f3009N}). Note that mergers and tidal stripping interact with SIDM in opposing ways, though the key mechanism is similar. Mergers can introduce dynamically hot material into the DM halo, and in particular into the annuli surrounding the central core, thus slowing down the runaway collapse. Tidal stripping, on the other hand, removes material primarily from the halo's outer regions, thus promoting inside-out flux of heat and accelerating gravothermal collapse. This mechanism primarily affect subhaloes, and can help to provide an explanation for the diversity of rotation curves seen in the MW satellites (\citealt{2019PhRvD.100f3007Z}). 

The baryonic signatures presented here provide a good starting point towards distinguishing between SNF and SIDM with upcoming observations of nearby dwarf galaxies. In future work, our goal is to investigate whether the properties of baryons in nearby dwarfs will enable us to definitively determine the dominant core formation mechanism, using a suite of cosmological zoom simulations of the Local Group.

\bsp	
\label{lastpage}
\end{document}